\documentclass[aps,pra,groupedaddress,tightenlines, twocolumn]{revtex4}

\usepackage{amsmath}
\usepackage[english]{babel}

\begin{document}

\title{Distinguishability measures between ensembles of quantum states}

\author{Ognyan Oreshkov and John Calsamiglia}

\affiliation{Grup de F\'{i}sica Te\`{o}rica, Universitat Aut\`{o}noma de
Barcelona, 08193 Bellaterra (Barcelona), Spain}
\date{\today}

\begin{abstract}

A quantum ensemble $\{(p_x, \rho_x)\}$ is a set of quantum states
each occurring randomly with a given probability. Quantum ensembles
are necessary to describe situations with incomplete a priori
information, such as the output of a stochastic quantum channel
(generalized measurement), and play a central role in quantum
communication. In this paper, we propose measures of distance and
fidelity between two quantum ensembles. We consider two approaches:
the first one is based on the ability to mimic one ensemble given
the other one as a resource and is closely related to the
Monge-Kantorovich optimal transportation problem, while the second
one uses the idea of extended-Hilbert-space (EHS) representations
which introduce auxiliary pointer (or flag) states. Both types of
measures enjoy a number of desirable properties. The Kantorovich
measures, albeit monotonic under deterministic quantum operations,
are not monotonic under generalized measurements. In contrast, the
EHS measures are. We present operational interpretations for both
types of measures. We also show that the EHS fidelity between
ensembles provides a novel interpretation of the fidelity between
mixed states---the latter is equal to the maximum of the fidelity
between all pure-state ensembles whose averages are equal to the
mixed states being compared. We finally use the new measures to
define distance and fidelity for stochastic quantum channels and
positive operator-valued measures (POVMs). These quantities may be
useful in the context of tomography of stochastic quantum channels
and quantum detectors.

\end{abstract}

\maketitle



\section{Introduction}

A fundamental difference between classical and quantum systems is
that, while classical states can be faithfully distinguished, two
generic quantum states cannot be distinguished with arbitrary
precision by any operational means. A natural measure that
quantifies the similarity of two pure quantum states $|\psi\rangle$
and $|\phi\rangle$ is the transition probability between them, i.e.,
the probability with which the two states would yield the same
outcome under a measurement for which one of the states is the
unique state that yields a particular outcome with certainty. This
quantity is symmetric with respect to the states and is given by the
square of their overlap, $|\langle\psi|\phi\rangle|^2$. In the case
of mixed states, there is no straightforward analogue of the
transition probability since there is no measurement for which a
mixed state is the unique state that yields a particular outcome
with certainty.

A generalization of the concept of transition probability to mixed
states was proposed by Uhlmann \cite{Uhl76} and it is given by the
minimum of the transition probability between two purifications of
the mixed states, over all possible purifications. The square root
of this quantity, which is given by the simple expression
\begin{equation}
F(\rho,\sigma)=\textrm{Tr}\sqrt{\sqrt{\sigma}\rho\sqrt{\sigma}},\label{fidelity}
\end{equation}
is known as the \textit{square root fidelity} between two density
matrices $\sigma$ and $\rho$ and has proven extremely useful in
quantum information theory \cite{NieChu00}. From the square root
fidelity (or fidelity for short), one can define various distances
between states, such as the Bures distance
$B(\rho,\sigma)=\sqrt{1-F(\rho,\sigma)}$ \cite{Bures1, Bures2} or
the Bures angle $A(\rho,\sigma)=\arccos F(\rho,\sigma)$
\cite{Uhlmann92, NieChu00}, which can be regarded as measures of the
difference between two states. In addition to fidelity-based
measures, various other measures of distance have been proposed
(see, e.g., Refs.~\cite{Helstrom, HilbertSchmidt, Chernoffbound,
Chernoffdistance, Ore08, Zyczbook}). The \textit{trace distance}
\cite{Helstrom},
\begin{gather}
\Delta(\rho,\sigma) =
\frac{1}{2}\parallel\rho-\sigma\parallel,\label{tracedistance}
\end{gather}
for example, where $\parallel
O\parallel=\textrm{Tr}\sqrt{O^{\dagger}O}$ is the trace norm of an
operator $O$, is widely used due to its simple form, various useful
properties, and its operational meaning related to the maximum
probability with which the two states $\rho$ and $\sigma$ can be
distinguished by a measurement.

The problem of distinguishing two quantum states bares strong
similarity to the problem of distinguishing two classical
probability distributions by looking at the value of a random
variable sampled from one of them. Unless the supports of the two
distributions have no overlap, the probability of guessing correctly
from which ensemble the variable was drawn is strictly smaller than
unity. In the classical case, however, the two probability
distributions concern the outcomes of only a single observable---the
one corresponding to the random variable. In the quantum case, there
is a continuum of possible observations that one can perform on the
systems and a continuum of corresponding random variables.

Different quantum measurements establish different correspondence
between quantum states and probability distributions. This suggests
a natural approach to defining distinguishability measures between
states. For instance, the fidelity between two quantum states is
equal to the minimum statistical overlap between the probability
distributions generated by all possible measurements performed on
the states \cite{Fuchs96}. The statistical overlap in question is
the Bhattacharyya coefficient $\sum_x\sqrt{P(x)Q(x)}$ between
classical probability distributions $P(x)$ and $Q(x)$ (here $x$ is a
\textit{classical} random variable). Similarly, the trace distance
\eqref{tracedistance} can be obtained  by maximizing  over all
possible measurements the Kolmogorov distance
$\sum_x\frac{1}{2}|P(x)-Q(x)|$ between the corresponding outcome
probability distributions. As expected, in the limit of commuting
density matrices, both the fidelity and the trace distance reduce to
their classical counterparts, i.e., to the Bhattacharyya overlap and
the Kolmogorov distance, respectively.

As is manifested in these examples, density matrices can be thought
of as generalizations of classical probability distributions, which
include the latter as a special case. However, in many quantum
information scenarios, one often deals with an even more general
concept, which is a hybrid between the quantum and classical cases.
This is the concept of a probabilistic ensemble of quantum states,
i.e., a classical probability distribution of density matrices.
Ensembles of quantum states describe situations in which a quantum
system can take a number of different states at random according to
some probability distribution. Such a situation is, for example, the
outcome of a quantum measurement. A quantum measurement can be
regarded as a stochastic quantum channel that outputs different
quantum states with probabilities that depend on the input state
according to the generalized Born rule \cite{NieChu00}. When the
measurement is projective, the possible output states are orthogonal
and the output ensemble can be regarded as a classical one. However,
in the case of generalized measurements the states need not be
orthogonal, and the output of the channel is a genuine quantum
ensemble.

A quantum state is said to ``...capture the best information
available about how a quantum system will react in this or that
experimental situation'' \cite{braunstein_criteria_2000}.
Accordingly, a quantum ensemble gives the best information available
about how a quantum system will react  in this or that experimental
situation when the choice of experiment can be made conditional on
some classical side information. The uses or applications of the
quantum system will depend strongly on the particular quantum states
that appear in the ensemble and on their probabilities.

It should be noted that in the context of resource theory
\cite{resourcetheory}, a protocol consisting of allowed
transformations generally involves measurements, and the resource
available after a measurement is given by the average resource of
the resulting ensemble. For example, the restriction to local
operations and classical communication (LOCC) naturally gives rise
to entanglement as a resource, which is quantified by an
entanglement monotone---a function which does not increase
\textit{on average} under LOCC \cite{Vidal00b, OB06}. In this sense,
entanglement can be thought of as a function defined on ensembles.
Ensembles of quantum states have various other applications in
quantum information theory, with particularly notable ones in
quantum communication, e.g., for representing sources of quantum
states used for communication \cite{ensforcom1, ensforcom2}, or for
describing ``static resources'' of shared classical-quantum
correlations in multi-partite systems \cite{DevWin03}.

Even though various measures of distance and fidelity between
quantum states have been studied, similar measures for ensembles of
states have been lacking. With the development of quantum
technology, it becomes important to be able to rigorously compare
different experimental schemes and assess the degree to which they
differ from ideal ones. The existing measures of distance and
fidelity between quantum states are sufficient for this purpose when
the system of interest at a given stage of the experiment is
described by a single quantum state. These measures can also be used
to define distance and fidelity between deterministic quantum
operations, i.e., completely positive trace-preserving (CPTP) maps
\cite{GLN05}. However, in many situations an experiment may involve
states obtained randomly according to some probability distribution,
such as the states obtained during the process of entanglement
concentration \cite{concentration}, or the states resulting from the
measurement of an error syndrome during and error-correction
protocol \cite{Got97}, or simply a source of quantum states used for
communication. It is therefore important to have a
distinguishability measure between two ensembles of states.
Furthermore, the tools of quantum information involve not only CPTP
maps but also stochastic quantum operations (generalized
measurements), and a figure of merit comparing two such operations
(e.g., a real one with an ideal one) would require a quantitative
comparison between their output ensembles. Rigorous measures that
compare generalized measurements would be useful, in particular, for
assessing the performance of quantum detectors, which can now be
characterized experimentally \cite{TQD} through quantum detector
tomography \cite{TQD1,TQD2,TQD3}.

The purpose of this paper is to propose measures of distance and
fidelity between ensembles of quantum states and use them to define
distance and fidelity between generalized measurements. The rest of
the paper is organized as follows. In Sec.~II, we review the concept
of an ensemble of quantum states and establish nomenclature. In
Sec.~III, we discuss some basic properties that we expect a measure
of distinguishability between ensembles to have, and rule out
several naive candidates. In Sec.~IV, we propose measures of
distance and fidelity of a Kantorovich type and study their
properties. We first introduce the measure of distance on the basis
of intuitive considerations concerning the ability of states
obtained randomly from one ensemble to mimic states obtained
randomly from the other ensemble. The measure is based on the trace
distance between states and satisfies a number of desirable
properties. In addition to the standard distance properties, it is
jointly convex, monotonic under averaging of the ensembles and under
CPTP maps. When the ensembles are discrete, the measure is
equivalent to a linear program and can be computed efficiently in
the size of the set of states participating in the ensembles. We
show that for simple limiting cases, the distance between ensembles
reduces to intuitive expressions involving the trace distance
between states. We introduce a measure of fidelity between ensembles
in a similar fashion. The fidelity satisfies properties analogous to
those of the distance and also can be computed as a linear program.
We provide operational interpretations of both quantities. We show
that for the case when the measures are based on the trace distance
and the standard fidelity, the measures are not monotonic under
generalized measurements. We explain why this is natural considering
the operational interpretations of the quantities and derive
necessary and sufficient conditions which the basic measures of
distance or fidelity between states have to satisfy in order for the
corresponding Kantorovich measures to be monotonic under
measurements. In Sec.~V, we propose measures of distance and
fidelity which make use of the extended-Hilbert-space (EHS)
representation of ensembles \cite{DevWin03}. We argue that to every
ensemble of quantum states there is a corresponding class of valid
EHS representations and provide a rigorous definition of this class.
We then define the measures as a minimum (maximum) of the distance
(fidelity) between all possible EHS representations of the ensembles
being compared. We show that these definitions can be simplified and
are equivalent to convex optimization problems. We also provide
equivalent formulations without reference to an extended Hilbert
space. These quantities are based on the trace distance and the
square root fidelity and inherit all their celebrated properties
such as joint convexity in the case of the trace distance or strong
concavity in the case of the fidelity. In addition, they are
monotonic under averaging of the ensembles, as well as under
generalized measurements. The latter property can be regarded as a
generalization of the monotonicity under CPTP maps of the trace
distance and the square root fidelity. The EHS measures are upper
(lower) bounded by the Kantorovich distance (fidelity). We provide
operational interpretations for the EHS measures too. In Sec.~VI, we
present a novel interpretation of the standard fidelity between
mixed states as a maximum of the fidelity between all pure-state
ensembles from which the mixed states being compared can be
constructed. The fidelity between pure-state ensembles used in this
definition is of the EHS type but can be expressed without any
reference to fidelity between mixed states and has a form which can
be regarded as a generalization of the Bhattacharyya overlap. In
Sec.~VII, we use the measures between ensembles of quantum states to
define distance and fidelity between generalized measurements. We
consider two definitions---one based on the Jamio{\l}kowski
isomorphism \cite{Jamiolkowski} and another based on worst-case
comparison---and discuss their properties. We also propose distance
and fidelity between positive operator-valued measures (POVMs). In
Sec.~VIII, we conclude.

\section{Ensembles of quantum states}

Let $\mathcal{B}(\mathcal{H})$ denote the set of linear operators on
a finite-dimensional Hilbert space $\mathcal{H}$. For the purposes
of this paper, a (probabilistic) ensemble of quantum states is a set
of pairs $\{(p_x,\rho_x)\}$ of probabilities $p_x$ ($p_x\geq 0$,
$\sum_xp_x=1$) and distinct density matrices
$\rho_x\in\mathcal{B}(\mathcal{H})$ ($\rho_x>0$,
$\textrm{Tr}(\rho_x)=1$), $\rho_x\neq\rho_y$ for $x\neq y$. For
simplicity, we will assume that the set of states participating in
an ensemble is discrete (i.e., the index $x$ runs over a countable
set), although we expect that our considerations extend to
non-discrete ensembles as well. We will use the concept of ensemble
of states to describe situations in which a system takes a state
$\rho_x$ at random with probability $p_x$. The statement that a
system takes the state $\rho_x$ means that there exists {classical}
information about the identity of the state. This is to be
distinguished from the situation in which no information about the
identity of the state exists or can be obtained. In the latter case,
for all practical purposes, the average density matrix of the
ensemble, $\overline{\rho}=\sum_xp_x\rho_x$, provides a complete
description of the state of the system.

An example of an ensemble of states is the output of a
non-destructive generalized measurement. Under the most general type
of quantum measurement, a density matrix $\rho$ transforms as
\begin{equation}
\rho \rightarrow \rho_i=\frac{\mathcal{M}_i(\rho)}{\textrm{Tr}
\mathcal{M}_i(\rho)}, \hspace{0.2cm} \textrm{with probability}
\hspace{0.2cm}p_i= {\textrm{Tr} \mathcal{M}_i(\rho)},\label{genmeas}
\end{equation}
where $\mathcal{M}_i(\cdot)=\sum_j M_{ij}(\cdot) M_{ij}^{\dagger}$
is the measurement superoperator corresponding to measurement
outcome $i$. (The operators $M_{ij}$ satisfy the completeness
relation $\sum_{i,j}M_{ij}^{\dagger}M_{ij}=I$.) Note that different
measurement outcomes do not necessarily yield different output
states. For example, both outcomes of a measurement on a qubit
system with measurement superoperators
$\mathcal{M}_1(\cdot)=|0\rangle\langle 0|(\cdot)|0\rangle\langle 0|$
and $\mathcal{M}_2(\cdot)=|0\rangle\langle 1|(\cdot)|1\rangle\langle
0|$ leave the system in the state $|0\rangle\langle 0|$, although
they provide information about the input state. If $\{\rho_x\}$ is
the set of distinct output states, each occurring with probability
$p_x=\sum_{i: \textrm{ }\rho_i=\rho_x}p_i$, the ensemble of
post-measurement states resulting from the stochastic transformation
\eqref{genmeas} is $\{(p_x,\rho_x)\}$.

Let $\{(p_x, \rho_x)\}$ be an ensemble of density matrices over a
Hilbert space $\mathcal{H}$. If $\Omega_1$ is the set of all density
matrices $\rho_x$ that participate in the ensemble, we can
equivalently represent the ensemble as a probability distribution
$P(\rho)$, $\rho\in \Omega_1$ ($P(\rho_x)\equiv p_x$), over the set
$\Omega_1$ . Consider a second ensemble, $Q(\sigma)$,
$\sigma\in\Omega_2$, where the set $\Omega_2$ is not necessarily
equal to $\Omega_1$. We can think of the two ensembles as
corresponding to probability distributions over the same set, by
extending the definitions of $P(\rho)$ and $Q(\sigma)$ to the larger
set $\Omega=\Omega_1\cup\Omega_2$ through assigning zero
probabilities to those states that do not participate in the
respective ensembles. Therefore, without loss of generality, we will
treat the ensembles that we compare as probability distributions
$P(\rho)$ and $Q(\rho)$ over the same set $\Omega$. (Sometimes, when
it is clear from the context, we will denote the ensembles we
compare simply by $P$ and $Q$.) Most generally, the set $\Omega$ can
be taken to be the full set of density matrices over $\mathcal{H}$,
but in this paper we will assume that $\Omega$ is discrete.

The fact that $P(\rho)$ and $Q(\rho)$ are valid probability distributions is expressed in the conditions
\begin{gather}
\underset{\rho'\in\Omega}{\sum}P(\rho')=1,\hspace{0.3cm} P(\rho)\geq 0, \hspace{0.2cm} \forall  \rho\in\Omega,\\
\underset{\rho'\in\Omega}{\sum}Q(\rho')=1,\hspace{0.3cm}
Q(\rho)\geq 0, \hspace{0.2cm} \forall  \rho\in\Omega.
\end{gather}

If our world is ultimately quantum, it is natural to expect that an
ensemble of quantum states must have a description in terms of the
state of a (possibly larger) quantum system. Indeed, there is a
correspondence between an ensemble of the form $\{(p_x, \rho_x)\}$
and a state of the form
\begin{gather}
\widehat{\rho}=\sum_x p_x \rho_x\otimes |x\rangle\langle x|,\label{state1}
\end{gather}
where the pointer (or flag) states $\{|x\rangle\}$ are an
orthonormal set in the Hilbert space of an auxiliary system of a
sufficiently large dimension \cite{DevWin03}. The pointer states can
be thought of as carrying the classical information about which
particular state from the ensemble we are given---a measurement of
the classical system yields the quantum state $\rho_x$ with
probability $p_x$, which is equivalent to drawing a state randomly
from the ensemble. Reversely, if we are given a state drawn randomly
from the ensemble, we can record our knowledge about the identity of
the state in a `classical' pointer attached to it and forget the
information about the state since this information is stored in the
pointer and can always be retrieved. After the latter operation, the
state of the original system plus the pointer system is described by
$\sum_x p_x \rho_x\otimes |x\rangle\langle x|$. This representation
is referred to as an extended-Hilbert-space (EHS) representation of
an ensemble \cite{DevWin03}. For simplicity and in order to
distinguish the system storing the classical memory from the quantum
system, we will use the following notation for the pointers:
 \begin{equation}
 [x]\equiv|x\rangle\langle x|.
 \end{equation}
In this notation, the state \eqref{state1} reads
 \begin{gather}
\widehat{\rho}=\sum_x p_x \rho_x\otimes [x].\label{state2}
\end{gather}
In terms of the description of an ensemble as a probability
distribution $P(\rho)$ over a set of states $\Omega$, an EHS
representation of this type can be written as
 \begin{gather}
 \widehat{\rho}_P=\sum_{\rho\in\Omega}P(\rho)\rho\otimes [\rho],\label{EHSsimple}
 \end{gather}
where $\{[\rho]\}$ is an orthonormal set of pure pointer states
$|\rho\rangle\langle\rho|$, each of which is associated with a
unique density matrix $\rho \in \Omega$. We will develop this
concept further in Sec.~V.

\section{Naive candidates}

Before we propose distinguishability measures between two ensembles
of quantum states, let us first consider what properties we expect
such measures to have. The answer to this question will depend on
the operational context in which we want to compare the ensembles.

We could ask, for example, how different on average two states drawn
randomly from the two ensembles are. Given a measure of distance
$d(\rho,\sigma)$ between states, the average distance in that sense
would be
$\underset{\rho\in\Omega}{\sum}\underset{\sigma\in\Omega}{\sum}P(\rho)Q(\sigma)d(\rho,\sigma)$.
This quantity obviously could be non-zero even when the two
ensembles are identical. Similarly, we could look at the average
fidelity which can be smaller than 1 for identical ensembles. Thus
even though these quantities have a well defined meaning, they are
not good measures of distinguishability.

Another possibility is to look at a distance $d
(\overline{\rho}_P,\overline{\rho}_Q)$ between the average density
matrices
$\overline{\rho}_P=\underset{\rho\in\Omega}{\sum}P(\rho)\rho$ and
$\overline{\rho}_Q=\underset{\rho\in\Omega}{\sum}Q(\rho)\rho$ of the
two ensembles, or the fidelity
$F(\overline{\rho}_P,\overline{\rho}_Q)$ between them. Obviously,
for identical ensembles the distance is equal to 0 and the fidelity
is equal to 1. However, these quantities cannot discriminate between
different ensembles that have the same average density matrices.
Imagine, for example, that an experimentalist has at her disposal
two devices. The first one produces the two-qubit Bell states
$\frac{|00\rangle+|11\rangle}{\sqrt{2}}$,
$\frac{|00\rangle-|11\rangle}{\sqrt{2}}$,
$\frac{|01\rangle+|10\rangle}{\sqrt{2}}$,
$\frac{|01\rangle-|10\rangle}{\sqrt{2}}$, each occurring with
probability $1/4$, together with a classical indicator specifying
which state is produced. The second device produces the two-qubit
product states $|00\rangle$, $|01\rangle$, $|10\rangle$,
$|11\rangle$, each occurring with probability $1/4$, again with an
indicator of the identity of the state. Although the average states
in the two cases are the same, the ensembles produced by the two
devices have very different properties. In the first case, the
average entanglement between the two qubits is maximal, whereas in
the second case it is zero. Therefore, in order to capture the
difference between two ensembles, we would like our measure of
distance (fidelity) to be 0 (1) if and only if $P(\rho)=Q(\rho)$,
$\forall \rho \in \Omega$.

Measures of distance and fidelity which satisfy the latter
requirement could be any measures of distance and fidelity between
probability distributions which treat $\rho$ as a classical
variable. Consider, for example, the Kolmogorov distance
$\frac{1}{2}\underset{\rho\in\Omega}{\sum}|P(\rho)-Q(\rho)|$. Note
that this distance is precisely equal to the trace distance between
the EHS representations of the two ensembles of type
\eqref{EHSsimple},
\begin{gather}
\Delta(\widehat{\rho}_P, \widehat{\rho}_Q)=
\frac{1}{2}\parallel \sum_{\rho\in\Omega}P(\rho)\rho\otimes [\rho]- \sum_{\sigma\in\Omega}Q(\sigma)\sigma\otimes [\sigma]\parallel\notag\\
=\frac{1}{2}\underset{\rho\in\Omega}{\sum}|P(\rho)-Q(\rho)|.
\end{gather}
In a similar manner, we could look at the Bhattacharyya overlap
$\underset{\rho\in\Omega}{\sum}\sqrt{P(\rho)Q(\rho)}$, which is
equal to the fidelity between the two EHS representations of type
\eqref{EHSsimple}. Such measures, however, do not take into account
the quantum-mechanical aspect of the variables $\rho$. If the two
distributions $P$ and $Q$ have supports on non-overlapping subsets
of $\Omega$, the above distance (fidelity) would be maximal
(minimal), but as we mentioned earlier, two distinct density
matrices are not necessarily distinguishable (they often behave as
if they are the same state) and we would like our distance and
fidelity to capture this property. In particular, in the special
case where each of the two ensembles consists of a single state, we
would like the measures between the two ensembles to be equal to the
distance or fidelity between the respective states. If we used the
above distance (fidelity) between classical probability
distributions in this case, we would obtain a maximum (minimum)
value even if the two states are very similar. At the same time, it
is natural to expect that a distance between ensembles would reduce
to a distance between classical probability distributions when the
states participating in the ensembles are orthogonal.

\section{Distance and fidelity of a Kantorovich type}

\subsection{Motivating the definitions}

The above examples suggest that distinguishability measures with the
desired properties may have to be non-trivial functions of the
probability distributions and the set of states participating in the
ensembles. Heuristically, a distance (fidelity) between two quantum
states can be regarded as a measure of the extent to which the two
states do not (do) behave as if they are the same state (the precise
meaning of this statement depends on the operational meaning of the
distance (fidelity) in question). In a similar manner, we would
expect a distance (fidelity) between two ensembles of quantum states
to compare the extent to which the two ensembles do not (do)
``behave'' as if they are the same ensemble. Since the ensemble is a
statistical concept which describes the situation of having
particular states with particular probabilities, we would like to
compare the extent to which states drawn randomly from one ensemble
can be used to mimic states drawn randomly from the other ensemble.

When states drawn randomly from the ensemble
$\{(Q(\sigma),\sigma)\}$ are used to mimic states drawn from the
ensemble $\{(P(\rho),\rho)\}$, a given state $\sigma$ obtained
according to the distribution $Q(\sigma)$ most generally can be
taken with different probabilities to pass off as different states
$\rho$ from $\{(P(\rho),\rho)\}$. In other words, the process of
mimicking one ensemble using the other one as a resource can be
described by a transition probability matrix whose elements
$T(\rho|\sigma)$, $\rho,\sigma\in\Omega$, describe the probabilities
with which the state $\sigma$ sampled from the distribution
$Q(\sigma)$ is taken to pass off as the state $\rho$ sampled from
$P(\rho)$. The requirement that under this simulation the
probabilities are consistent with the probabilities $P(\rho)$ and
$Q(\sigma)$, respectively, is expressed in the condition
$\underset{\sigma\in\Omega}{\sum}T(\rho|\sigma)Q(\sigma)=P(\rho)$.
The fact that $T(\rho|\sigma)$ describe valid transition
probabilities imposes the conditions $T(\rho|\sigma)\geq0$,
$\forall\rho,\sigma\in\Omega$, and
$\underset{\rho\in\Omega}{\sum}T(\rho|\sigma)=1$,
$\forall\sigma\in\Omega$.

In order to measure how much the state $\sigma$ fails to mimic the
state $\rho$, we can use any measure of distance between states. In
this paper, we will concentrate on the case of the trace distance,
$\Delta(\rho,\sigma)$ (Eq.~\eqref{tracedistance}). To measure the
degree to which a map $T(\rho|\sigma)$ from one ensemble to the
other fails to mimic the latter, we propose to use the average
distance between the actual states and those that they mimic:
$\sum_{\rho,\sigma\in\Omega}T(\rho|\sigma)Q(\sigma)\Delta(\rho,\sigma)$.
We can write this expression in an explicitly symmetric form by
introducing the joint probability distribution
$\Pi(\rho,\sigma)\equiv T(\rho|\sigma)Q(\sigma)$ which satisfies the
marginal conditions $\sum_{\sigma\in\Omega}\Pi(\rho,\sigma)=P(\rho),
\hspace{0.2cm}\forall \rho \in \Omega$, and
$\sum_{\rho\in\Omega}\Pi(\rho,\sigma)=Q(\sigma),
\hspace{0.2cm}\forall \sigma \in \Omega$:
\begin{equation}
D_{
\Pi}(P,Q)=\underset{\rho,\sigma\in\Omega}{\sum}\Pi(\rho,\sigma)\Delta(\rho,\sigma).\label{quantity}
\end{equation}
Clearly, different choices of the map $T(\rho|\sigma)$ (or
equivalently, of $\Pi(\rho,\sigma)$) can yield different values for
the quantity \eqref{quantity}. Therefore, we define the distance between the two
ensembles as the minimum of the quantity \eqref{quantity} over all
possible choices of $\Pi(\rho,\sigma)$, i.e., we choose the optimal mimicking strategy.

\textbf{Definition 1 (Kantorovich distance).} Let $P(\rho)$ and
$Q(\rho)$, $\rho\in \Omega$, be two ensembles (probability
distributions over $\Omega$), which we denote by $P$ and $Q$ for
short. Then
\begin{equation}
D^K(P,Q)=\min_{\Pi(\rho,\sigma)}\sum_{\rho,\sigma\in\Omega}\Pi(\rho,\sigma)\Delta(\rho,\sigma),\label{definition}
\end{equation}
where minimum is taken over all joint probability distributions
$\Pi(\rho,\sigma)$ with marginals
$\sum_{\sigma\in\Omega}\Pi(\rho,\sigma)=P(\rho),
\hspace{0.2cm}\forall \rho \in \Omega$, and
$\sum_{\rho\in\Omega}\Pi(\rho,\sigma)=Q(\sigma),
\hspace{0.2cm}\forall \sigma \in \Omega$.

The quantity \eqref{definition} is of the same form as the
Kantorovich formulation of the optimal transportation problem
\cite{Kantorovich}, which is a relaxation of a problem studied in
1781 by  Monge. In 1975, Kantorovich received the Nobel Prize in
Economics, together with Koopmans, for their contributions to the
theory of optimum allocation of resources, and he is considered to
be one of the fathers of linear programming. The optimal
transportation problem can be cast in the spirit of its original
formulations as follows:

\emph{Assume you have to transport the coal produced in some mines
$X$ to the factories $Y$. The amounts produced in each mine
$\{P_{1},P_{2},\ldots\}$ as well as the needs for each factory
$\{Q_{1},Q_{2},\ldots\}$ are given. There is a cost per unit of mass
$c(x,y)$ to move coal from mine $x$ to factory $y$. The problem is
to find the optimal transportation plan or transportation map
$T(y|x)$, i.e., for every mine $x$ determine how much material has
to be carried to every factory $y$ so as to minimize the overall
cost.}

The analogy with the above definition \eqref{definition} is
straightforward: mines and factories play the role of the quantum
states  $\rho$ and $\sigma$ in each ensemble respectively, and the
cost function is given by the trace distance.  Kantorovich's
formulation extended also to non-discrete probability measures
\cite{Villani} and was one of the first infinite-dimensional linear
programming problems to be considered. If the probability measures
are defined over a metric space and the cost function is taken to be
the corresponding distance function, the optimal average cost is
known as the Kantorovich distance (also referred to as Wasserstein
distance \cite{Vasershtein69}). The optimal transportation problem
is now an active field of research with tight connections with
problems in geometry, probability theory, differential equations,
fluid mechanics, economics and image or data processing.

Based on the same idea we can define a fidelity between two
ensembles, which we will refer to as the Kantorovich fidelity.

\textbf{Definition 2 (Kantorovich fidelity).} The Kantorovich
fidelity between the ensembles $P(\rho)$ and $Q(\rho)$, $\rho\in
\Omega$, is
\begin{equation}
F^K(P,Q)=\max_{\Pi(\rho,\sigma)}\sum_{\rho,\sigma\in\Omega}\Pi(\rho,\sigma)F(\rho,\sigma),\label{definition2}
\end{equation}
where $F(\rho,\sigma)$ is the square root fidelity between $\rho$
and $\sigma$ (Eq.~\eqref{fidelity}), and maximum is taken over all
joint probability distributions $\Pi(\rho,\sigma)$ that satisfy
$\sum_{\sigma\in\Omega}\Pi(\rho,\sigma)=P(\rho),
\hspace{0.2cm}\forall \rho \in \Omega$, and
$\sum_{\rho\in\Omega}\Pi(\rho,\sigma)=Q(\sigma),
\hspace{0.2cm}\forall \sigma \in \Omega$.

\subsection{Properties of the Kantorovich distance}

Let $\mathcal{P}_{\Omega}$ denote the set of probability
distributions over a set of density matrices $\Omega$.

\textbf{Property 1 (Positivity).}
\begin{gather}
D^K(P,Q)\geq 0, \\ \forall \hspace{0.1cm}P,Q \in \mathcal{P}_{\Omega},   \notag
\end{gather}
with equality
\begin{gather}
D^K(P,Q)=0 \hspace{0.2cm} \textrm{iff} \hspace{0.2cm}
P(\rho)=Q(\rho), \hspace{0.2cm} \forall \rho \in \Omega.
\end{gather}

\textbf{Proof.} Since all terms in Eq.~\eqref{definition} are
non-negative, the distance $D^K(P,Q)$ is also non-negative.
Obviously, if $P(\rho)=Q(\rho)$, $\forall \rho \in \Omega$, we
obtain $D^K(P,Q)=0$ by choosing the joint probability distribution
$\Pi(\rho,\sigma)=\delta_{\rho,\sigma}P(\rho)$. Inversely, assume
that $D^K(P,Q)=0$. This means that all terms in
Eq.~\eqref{definition} must be zero, which can happen only if
$\Pi(\rho, \sigma)\propto\delta_{\rho,\sigma}$. From the condition
for the marginal probability distributions, we see that
$\Pi(\rho,\sigma)=\delta_{\rho,\sigma}P(\rho)$ and
$P(\rho)=Q(\rho)$.

\textbf{Property 2 (Normalization).}
\begin{gather}
D^K(P,Q)\leq 1,  \\ \forall \hspace{0.1cm}P,Q \in
\mathcal{P}_{\Omega},  \notag
\end{gather}
with equality
\begin{gather}
D^K(P,Q)=1
\end{gather}
if and only if the supports of P and Q are orthogonal sets of
states.

\textbf{Proof.} Since $\Delta(\rho,\sigma)\leq 1$, then for any
given $\Pi(\rho,\sigma)$ we have
$\sum_{\rho,\sigma\in\Omega}\Pi(\rho,\sigma)\Delta(\rho,\sigma) \leq
\sum_{\rho,\sigma\in\Omega}\Pi(\rho,\sigma)=1$. Furthermore,
$\Delta(\rho,\sigma)= 1$ if and only if $\rho$ and $\sigma$ are
orthogonal. Observe that the only non-zero values $\Pi(\rho,\sigma)$
of a joint probability distribution that respects the marginal
distributions $P$ and $Q$ are those for which $\rho$ is in the
support of $P$ and $\sigma$ is in the support of $Q$. Therefore, if
$P$ and $Q$ have supports on sets of density matrices which are
orthogonal, every non-zero component $\Pi(\rho,\sigma)$ in the sum
on the right-hand side of Eq.~\eqref{definition} will be multiplied
by $\Delta(\rho,\sigma)=1$, which implies that $D^K(P,Q)=1$.
Inversely, since $\sum_{\rho,\sigma\in\Omega}\Pi(\rho,\sigma)=1$ if
$D^K(P,Q)=1$, then every non-zero $\Pi(\rho,\sigma)$ on the
right-hand side of Eq.~\eqref{definition} must be multiplied by $1$,
which implies that $P$ and $Q$ must have supports on orthogonal
sets.

\textbf{Property 3 (Symmetry).}
\begin{gather}
D^K(P,Q)=D^K(Q,P),
\\ \forall \hspace{0.2cm}P,Q \in \mathcal{P}_{\Omega}\notag.
\end{gather}

\textbf{Proof.} The symmetry follows from the definition
\eqref{definition} and the symmetry of $\Delta(\rho,\sigma)$.

\textbf{Property 4 (Triangle inequality).}
\begin{gather}
D^K(P,R)\leq D^K(P,Q)+D^K(Q,R),\\ \forall \hspace{0.1cm}P,Q,R \in \mathcal{P}_{\Omega}\notag.
\end{gather}

\textbf{Proof.} Let $\Pi^{PQ}(\rho,\sigma)$ and
$\Pi^{QR}(\rho,\sigma)$ be the two joint probability distributions
which achieve the minimum in Eq.~\eqref{definition} for the pairs of
distributions $(P,Q)$ and $(Q,R)$, respectively. Consider the
quantity
\begin{gather}
\tilde{\Pi}^{PR}(\rho,\sigma)=\sum_{\kappa\in\Omega}\Pi^{PQ}(\rho,\kappa)
\frac{1}{Q(\kappa)}\Pi^{QR}(\kappa,\sigma),\hspace{0.3cm}\rho,\sigma\in\Omega
\end{gather}
where for $Q(\kappa)=0$, we define $\Pi^{PQ}(\rho,\kappa)
\frac{1}{Q(\kappa)}\Pi^{QR}(\kappa,\sigma)=0$ (note that if
$Q(\kappa)=0$, then
$\Pi^{PQ}(\rho,\kappa)=\Pi^{QR}(\kappa,\sigma)=0$, $\forall\rho,
\sigma \in \Omega$). One can readily verify that this is a valid
joint probability distribution with marginals $P$ and $R$.
Therefore, we have
\begin{gather}
D^K(P,R)\leq \sum_{\rho,\sigma\in\Omega}\tilde{\Pi}^{PR}(\rho,\sigma)\Delta(\rho,\sigma)\notag\\
=\sum_{\rho,\sigma,\kappa\in\Omega}\Pi^{PQ}(\rho,\kappa)
\frac{1}{Q(\kappa)}\Pi^{QR}(\kappa,\sigma)\Delta(\rho,\sigma)\notag\\
\leq \sum_{\rho,\sigma,\kappa\in\Omega}\Pi^{PQ}(\rho,\kappa)
\frac{1}{Q(\kappa)}\Pi^{QR}(\kappa,\sigma)\Delta(\rho,\kappa)\notag\\
+\sum_{\rho,\sigma,\kappa\in\Omega}\Pi^{PQ}(\rho,\kappa)
\frac{1}{Q(\kappa)}\Pi^{QR}(\kappa,\sigma)\Delta(\kappa,\sigma)\notag\\
=\sum_{\rho,\kappa\in\Omega}\Pi^{PQ}(\rho,\kappa)
\Delta(\rho,\kappa)
+\sum_{\sigma,\kappa\in\Omega}\Pi^{QR}(\kappa,\sigma)\Delta(\kappa,\sigma)\notag\\
=D^K(P,Q)+D^K(Q,R),
\end{gather}
where in the second inequality we have used the triangle inequality
for $\Delta$.

\textbf{Property 5 (Joint convexity).}
\begin{gather}
D^K(pP_1+(1-p)P_2,pQ_1+(1-p)Q_2)\\\leq pD^K(P_1,Q_1)+(1-p)D^K(P_2,Q_2),\notag\\
\forall \hspace{0.1cm}P_1,P_2,Q_1,Q_2 \in \mathcal{P}_{\Omega},
\hspace{0.2cm} \forall \hspace{0.1cm}p\in[0,1] \notag.
\end{gather}

\textbf{Proof.} Let $\Pi^{1}(\rho,\sigma)$ and
$\Pi^{2}(\rho,\sigma)$ be two joint probability distributions which
achieve the minimum in Eq.~\eqref{definition} for the pairs of
distributions $(P_1,Q_1)$ and $(P_2,Q_2)$, respectively. It is
immediately seen that
\begin{gather}
\tilde{\Pi}^{12}(\rho,\sigma)={p}\Pi^{1}(\rho,\sigma)+(1-p)\Pi^{2}(\rho,\sigma)
\end{gather}
is a joint probability distribution with marginals $pP_1+(1-p)P_2$
and $pQ_1+(1-p)Q_2$. Therefore,
\begin{gather}
D^K(pP_1+(1-p)P_2,pQ_1+(1-p)Q_2)\notag\\\leq \sum_{\rho,\sigma\in\Omega}\tilde{\Pi}^{12}(\rho,\sigma)\Delta(\rho,\sigma)\notag\\
=p\sum_{\rho,\sigma\in\Omega}{\Pi}^{1}(\rho,\sigma)\Delta(\rho,\sigma)\notag
+(1-p)\sum_{\rho,\sigma\in\Omega}{\Pi}^{2}(\rho,\sigma)\Delta(\rho,\sigma)\notag\\
=pD^K(P_1,Q_1)+(1-p)D^K(P_2,Q_2).
\end{gather}

\textbf{Property 6 (Monotonicity under CPTP maps).}

Let $\mathcal{E}: \mathcal{B}(\mathcal{H})\rightarrow
\mathcal{B}(\mathcal{H'})$, where $\mathcal{H}$ and $\mathcal{H'}$
generally can have different dimensions, be a completely positive
trace-preserving (CPTP) map. (Any such map can be written in the
Kraus form  $\mathcal{E}(\rho)=\sum_iM_i\rho M_i^{\dagger}$,
$\forall \rho \in \mathcal{B}(\mathcal{H})$ \cite{Kraus83}). Denote
the set of density matrices consisting of $\mathcal{E}(\rho)$, with $\rho
\in \Omega$, by $\Omega_{\mathcal{E}}$. If we apply the same CPTP
map $\mathcal{E}$ to every state in an ensemble $P(\rho)$, $\rho\in
\Omega$, we obtain another ensemble $P'(\rho')$, $\rho'\in
\Omega_{\mathcal{E}}$. Note that generally $P(\rho)\neq
P'(\mathcal{E}(\rho))$, because the map $\mathcal{E}$ may be such
that it takes two or more different states from $\Omega$ to one and
the same state in $\Omega_{\mathcal{E}}$, e.g.,
$\mathcal{E}(\rho_1)=\mathcal{E}(\rho_2)$, $\rho_1\in\Omega$,
$\rho_2\in\Omega$, $\rho_1\neq \rho_2$. (The opposite obviously
cannot happen because every state $\rho$ in $\Omega$ is mapped to a
unique state $\mathcal{E}(\rho)\in \Omega_{\mathcal{E}}$.) Thus the
operation $\mathcal{E}$ induces a map from the set of probability
distributions over $\Omega$ to the set of probability distributions
over $\Omega_{\mathcal{E}}$. Denote this map by
$M_{\mathcal{E}}:\mathcal{P}_{\Omega} \rightarrow
\mathcal{P}_{\Omega_{\mathcal{E}}}$.

Now we can state the property of monotonicity under CPTP maps as
follows: For all CPTP maps $\mathcal{E}$,
\begin{gather}
D^K(P,Q)\geq D^K(M_{\mathcal{E}}(P),M_{\mathcal{E}}(Q)),
\end{gather}
where $M_{\mathcal{E}}:\mathcal{P}_{\Omega} \rightarrow
\mathcal{P}_{\Omega_{\mathcal{E}}}$ is the map induced by
$\mathcal{E}$.

\textbf{Proof.} Let $\Pi(\rho,\sigma)$ be a joint probability
distribution for which the minimum in the definition
\eqref{definition} of $D^K(P,Q)$ is attained. Observe that
$\sum_{\rho,\sigma\in \Omega}
\Pi(\rho,\sigma)\Delta(\mathcal{E}(\rho),\mathcal{E}(\sigma))=\sum_{\rho',\sigma'\in
\Omega_{\mathcal{E}}} \Pi'(\rho',\sigma')\Delta(\rho',\sigma')$,
where $\Pi'(\rho',\sigma')$ is a joint probability distribution over
$\Omega_{\mathcal{E}}\times \Omega_{\mathcal{E}}$ with marginals
$P'(\rho')$ and $Q(\rho')$. This can be seen from the fact that
$P'(\rho')=\sum_xP(\rho_x)$, where the sum is over all
$\rho_x\in\Omega$ such that $\rho'=\mathcal{E}(\rho_x)$. Similarly,
$Q'(\sigma')=\sum_yP(\sigma_y)$, where the sum is over all
$\sigma_y\in\Omega$ such that $\sigma'=\mathcal{E}(\sigma_y)$.
Therefore, we have that
\begin{gather}
D^K(M_{\mathcal{E}}(P),M_{\mathcal{E}}(Q))\leq
\sum_{\rho',\sigma'\in \Omega_{\mathcal{E}}}\Pi'(\rho',\sigma')\Delta(\rho',\sigma')\notag\\
= \sum_{\rho,\sigma\in \Omega} \Pi(\rho,\sigma)\Delta(\mathcal{E}(\rho),\mathcal{E}(\sigma))\leq\sum_{\rho,\sigma\in \Omega} \Pi(\rho,\sigma)\Delta(\rho,\sigma)
\notag\\
=D^K(P,Q),
\end{gather}
where the last inequality follows from the monotonicity of
$\Delta(\rho,\sigma)$ under CPTP maps \cite{ruskaicontractivity}.

\textbf{Corollary (Invariance under unitary maps).}

For all unitary maps $\mathcal{U}$,
\begin{gather}
D^K(P,Q)=D^K(M_{\mathcal{U}}(P),M_{\mathcal{U}}(Q)).
\end{gather}
The property follows from the fact that unitary maps are reversible
CPTP maps.

\textbf{Property 7 (Monotonicity under averaging).} Let
$\overline{P}$ denote the singleton ensemble consisting of the
average state of $P(\rho)$,
$\overline{\rho}_P=\underset{\rho\in\Omega}{\sum}P(\rho)\rho$. Then
\begin{equation}
D^K(P,Q)\geq D^K(\overline{P},\overline{Q}).
\end{equation}

\textbf{Proof.} Let $\Pi(\sigma,\rho)$ be a joint probability
distribution for which the minimum in the definition
\eqref{definition} of $D(P,Q)$ is attained. Since
$\Delta(\rho,\sigma)$ is jointly convex \cite{NieChu00}, we have
\begin{gather}
D^K(P,Q)=\sum_{\rho,\sigma\in \Omega} \Pi(\rho,\sigma)\Delta(\rho,\sigma)\notag\\
\geq \Delta( \sum_{\rho,\sigma\in \Omega}\Pi(\rho,\sigma)\rho,\sum_{\rho,\sigma\in \Omega}\Pi(\rho,\sigma)\sigma)\notag\\
=\Delta(\sum_{\rho\in\Omega}P(\rho)\rho,\sum_{\sigma\in
\Omega}Q(\sigma)\sigma)\notag\\=\Delta(\overline{\rho}_P,
\overline{\rho}_Q)=D^K(\overline{P},\overline{Q}).
\end{gather}
(For the last equality, see Eq.~\eqref{2sinK} below.)

\textbf{Corollary.} If two distributions are close, their average
states are also close, i.e.,
\begin{equation}
\textrm {if}\hspace{0.2cm} D^K(P,Q)\leq
\varepsilon,\hspace{0.2cm}\textrm{then}\hspace{0.2cm}
\Delta(\overline{\rho}_P, \overline{\rho}_Q)\leq
D^K(P,Q)\leq\varepsilon.
\end{equation}

\textbf{Property 8 (Continuity of the average of a continuous
function)}. Let $h(\rho)$ be a bounded function, which is continuous
with respect to the distance $\Delta$. Then the ensemble average of
$h(\rho)$,
$\overline{h}_P=\underset{\rho\in\Omega}{\sum}P(\rho)h(\rho)$, is
continuous with respect to $D^{K}$.

\textbf{Proof.} The proof is presented in Appendix A.

\textbf{Comment.} Property 8 naturally reflects the idea of states
as resources. Assuming that a resource is a continuous function of
the state, if two ensembles are close, their corresponding average
resources must also be close.

\textbf{Example (Continuity of the Holevo information).} A function
of ensembles, which is of great significance in quantum information
theory, is the Holevo information \cite{ensforcom2}
\begin{equation}
\chi(P)=S(\overline{\rho})-\sum_xp_xS(\rho_x).\label{Holevo}
\end{equation}
Here $\overline{\rho}=\sum_xp_x\rho_x$ is the average density matrix
of the ensemble $\{(p_x,\rho_x)\}$ which we denote by $P$ for short,
and $S(\rho)=-\textrm{Tr}(\rho \log \rho)$ is the von Neumann
entropy. This function gives an upper bound to the amount of
information about the index $x$ extractable through measurements on
a state obtained randomly from the ensemble and is used to define
the classical capacity of a quantum channel under independent uses
of the channel \cite{Holevo2, SchumWest}. The second term in the
expression \eqref{Holevo} is the average of the von Neumann entropy
over the ensemble, while the first term is the von Neumann entropy
of the average. Since $S(\rho)$ is a continuous function, from
Property 8 and the Corollary of Property 7 one can easily see that
the Holevo information is a continuous function of the ensemble with
respect to the Kantorovich distance. It would be interesting,
however, to obtain an explicit bound of that continuity. For this
purpose, we will need the following lemma.

\textbf{Lemma 1}. If a function $h(\rho)$ satisfies the continuity
property
\begin{equation}
|h(\rho)-h(\sigma)|\leq g[\Delta(\rho, \sigma)]\label{con4}
\end{equation}
for some function $g[x]$ that is concave in $x\in [0,1]$, then the
ensemble average of  $h(\rho)$ satisfies
\begin{equation}
|\overline{h}_P-\overline{h}_Q|\leq g[D^K(P,Q)].
\end{equation}

\textbf{Proof.} Let $\Pi(\rho,\sigma)$ be a joint probability
distribution which attains the minimum in Eq.~\eqref{definition} for
the distributions $P$ and $Q$. Then,
\begin{gather}
|\overline{h}_P-\overline{h}_Q|=|\underset{\rho\in\Omega}{\sum}P(\rho)h(\rho) -\underset{\sigma\in\Omega}{\sum}Q(\sigma)h(\sigma)|\notag\\
=|\sum_{\rho,\sigma\in \Omega}\Pi(\rho,\sigma)h(\rho)
-\sum_{\rho,\sigma\in \Omega}\Pi(\rho,\sigma)h(\sigma)|\notag\\
\leq
\sum_{\rho,\sigma\in\Omega}\Pi(\rho,\sigma)|h(\rho)-h(\sigma)|\leq
\sum_{\rho,\sigma\in\Omega}\Pi(\rho,\sigma)g[\Delta(\rho,\sigma)]\notag\\
\leq
g\left[\sum_{\rho,\sigma\in\Omega}\Pi(\rho,\sigma)\Delta(\rho,\sigma)\right]=
g[D^K(P,Q)].
\end{gather}

\textbf{Theorem 1 (A Fannes-type inequality for the ensemble average
of the von Neumann entropy).} For any two ensembles $P$ and $Q$ of
density matrices over a $d$-dimensional Hilbert space,
\begin{equation}
|\overline{S}_P-\overline{S}_Q|\leq
D^K\log_2(d-1)+H((D^K,1-D^K)),\label{Fannes}
\end{equation}
where $D^K$ is the Kantorovich distance between the ensembles $P$
and $Q$, and $H((D^K,1-D^K))=-D^K\log_2(D^K)-(1-D^K)\log_2(1-D^K)$
is the Shannon entropy of the binary probability distribution
$(D^K,1-D^K)$.

\textbf{Comment.} This inequality is based on a Fannes-type
inequality for the von Neumann entropy due to Audenaert
\cite{Audenaert}, which is stronger than the original inequality by
Fannes \cite{Fannes} and provides the sharpest continuity bound for
the von Neumann entropy based on $\Delta$ and $d$.

\textbf{Proof.} In Ref.~\cite{Audenaert}, it was shown that
\begin{gather}
|{S}(\rho)-{S}(\sigma)|\leq
\Delta\log_2(d-1)+H((\Delta,1-\Delta)).\label{Auden}
\end{gather}
The theorem follows from Lemma 1 and the fact that the right-hand
side of Eq.~\eqref{Auden} is a concave function of $\Delta$.

\textbf{Corollary (Continuity bound for the Holevo information).}
The term $S(\overline{\rho})$ in the expression \eqref{Holevo} for
the Holevo information is not an average of a function, but
according to the Corollary of Property 7,
$\Delta(\overline{\sigma},\overline{\rho})\leq D^K(P,Q)$. The
right-hand side of Eq.~\eqref{Auden} is monotonically increasing in
the interval $0\leq\Delta\leq(d-1)/d$ and monotonically decreasing
in the interval $(d-1)/d <~\Delta\leq~1$. Therefore, we can write
\begin{gather}
|S(\overline{\sigma})-S(\overline{\rho})|\leq
D^K\log_2(d-1)+H((D^K,1-D^K))\notag\\
\textrm{for}\hspace{0.2cm} 0\leq D^K\leq (d-1)/d.\label{Hol2}
\end{gather}
Combining Eq.~\eqref{Fannes} and Eq.~\eqref{Hol2}, we obtain
\begin{gather}
|\chi(Q)-\chi(P)|\leq 2 D^K\log_2(d-1)+2H((D^K,1-D^K))\notag\\
\textrm{for}\hspace{0.2cm} 0\leq D^K\leq (d-1)/d.
\end{gather}
For the interval $(d-1)/d < D^K\leq 1$, we can upper bound
$|S(\overline{\sigma})-S(\overline{\rho})|$ by its maximum value
$\log_2(d)$, and we can write the weaker inequality
\begin{gather}
|\chi(Q)-\chi(P)|\leq \notag\\
\log_2(d)+ D^K\log_2(d-1)+H((D^K,1-D^K))\notag\\
\textrm{for}\hspace{0.2cm} (d-1)/d < D^K\leq 1.
\end{gather}

\textbf{Property 9 (Stability).} Let $P(\rho)$, $\rho\in\Omega$, and
$R(\sigma')$, $\sigma'\in\Omega'$, be two ensembles of quantum
states, where $\Omega$ and $\Omega'$ are sets of states of two
different systems. Define the tensor product of the two ensembles as
the ensemble $\{(P(\rho)R(\sigma'),\rho\otimes\sigma')\}$, which we
will denote by $P\otimes R$ for short. Let $P(\rho)$ and $Q(\rho)$
be two ensembles of states in $\Omega$ and $R(\sigma')$ be an
ensemble of states in $\Omega'$. Then,
\begin{gather}
D^K(P\otimes R,Q\otimes R)=D^K(P,Q).
\end{gather}

\textbf{Comment.} The physical meaning of this property is that
unrelated ensembles do not affect the value of $D^K(P,Q)$. Even
though this may seem as a natural property to expect from a
distance, it does not hold in general even for distance measures
between states. For example, the Hilbert-Schmidt distance
$\sqrt{\textrm{Tr}(\rho-\sigma)^2}$, which has a well-defined
operational meaning \cite{HilbertSchmidt}, is not stable.

\textbf{Proof.} Let
\begin{gather}
D^K(P\otimes R,Q\otimes
R)=\notag\\
\sum_{\rho,\sigma\in\Omega; \tau',\kappa'\in\Omega'}\Pi(\rho\otimes
\tau',\sigma \otimes \kappa')\Delta(\rho\otimes \tau',\sigma \otimes
\kappa'),
\end{gather}
where $\Pi(\rho\otimes \tau',\sigma \otimes \kappa')$ has left and
right marginals $P(\rho)R(\tau')$ and $Q(\sigma)R(\kappa')$,
respectively. From the monotonicity of $\Delta$ under partial
tracing it follows that
\begin{gather}
D^K(P\otimes R,Q\otimes R)\geq
\sum_{\rho,\sigma\in\Omega}\Pi'(\rho,\sigma)\Delta(\rho,\sigma),
\end{gather}
where
\begin{gather}
\Pi'(\rho,\sigma)=\sum_{\tau',\kappa'\in\Omega'}\Pi(\rho\otimes
\tau',\sigma \otimes \kappa')
\end{gather}
is a joint probability distribution with left and right marginals
$P(\rho)$ and $Q(\sigma)$, respectively. Therefore,
\begin{gather}
D^K(P\otimes R,Q\otimes R)\geq D^K(P,Q).\label{stabin}
\end{gather}
But by choosing $\Pi(\rho\otimes \tau',\sigma \otimes
\kappa')=\Pi(\rho,\sigma)R(\tau')\delta_{\tau'\kappa'}$, where
$\Pi(\rho,\sigma)$ is a joint distribution which attains the minimum
in the definition \eqref{definition} of $D^K(P,Q)$, and using the
stability of $\Delta$, the equality in Eq.~\eqref{stabin} is
attained. This completes the proof.

\textbf{Property 10 (Linear programming).} The task of finding the
optimal $\Pi(\sigma,\rho)$ in Eq.\eqref{definition} is a linear
program and can be solved efficiently in the cardinality of
$\Omega$.

\textbf{Proof.} If the cardinality of $\Omega$ is N, we can think of
$\Delta(\rho,\sigma)$, $\rho, \sigma \in \Omega$ as the components
$c_{\mu}$, $\mu=(\rho,\sigma)$, of an $N^2$-component vector which
we will denote by $c$. The joint probability distribution
$\Pi(\rho,\sigma)$ over which we want to minimize the expression on
the right-hand side of Eq.~\eqref{definition} can similarly be
thought of as an $N^2$-component vector $x$ with components
$x_{\mu}$, $\mu=(\rho,\sigma)$. Thus the task of finding the optimal
$\Pi(\rho,\sigma)$ can be expressed in the compact form
\begin{equation}
\textrm{Minimize} \hspace{0.4cm}c^T x. \label{linprog0}
\end{equation}
The constraints $\sum_{\sigma\in\Omega}\Pi(\rho,\sigma)=P(\rho),
\hspace{0.2cm}\forall \rho \in \Omega$, and
$\sum_{\rho\in\Omega}\Pi(\rho,\sigma)=Q(\sigma),
\hspace{0.2cm}\forall \sigma \in \Omega$, can also be expressed in a
compact matrix forms as
\begin{gather}
{A}x=a,\notag\\
{B}x=b,\label{linprog1}
\end{gather}
where ${A}$ is an $N\times N^2$ matrix with components $A_{\kappa
\mu}=\delta_{\kappa\rho}$ where $\mu=(\rho,\sigma)$ is a double
index, ${B}$ is an $N\times N^2$ matrix with components
$B(\kappa,\mu)=\delta_{\kappa\sigma}$ ($\mu=(\rho,\sigma)$), and $a$
and $b$ are $N$-component vectors with elements
$a_{\kappa}=P(\kappa)$, $\kappa\in\Omega$, and
$b_{\kappa}=Q(\kappa)$, $\kappa\in\Omega$, respectively. In
addition, the positivity of the quantities $\Pi(\rho,\sigma)$
amounts to the constraint
\begin{equation}
x\geq 0.\label{linprog2}
\end{equation}
Eqs.~\eqref{linprog0}-\eqref{linprog2} are the canonical form of a
linear program, which can be solved efficiently in the length $N^2$
of the vector $x$. This completes the proof.

It is natural to ask about the properties of the distance in certain
simple limiting cases. We consider the following three cases.

\textbf{Limiting case 1 (Two singleton ensembles).} If
$P(\rho)=\delta_{\rho\tau}$, $\rho, \tau\in \Omega$ and
$Q(\rho)=\delta_{\rho\sigma}$, $\rho, \sigma\in \Omega$, i.e., each
of the ensembles $P$ and $Q$ consists of only a single state, then
the distance between the ensembles is equal to the distance between
the respective states,
\begin{equation}
D^K(P,Q)=\Delta(\tau,\sigma).\label{2sinK}
\end{equation}

\textbf{Proof.} Obviously, the only joint probability distribution
with marginals $P$ and $Q$ in this case is
$\Pi(\kappa,\tau)=\delta_{\kappa\sigma}\delta_{\tau\rho}$, so the
property follows.

\textbf{Limiting case 2 (One singleton ensemble)}. If the ensemble
$Q(\rho)$ consists of only one state $\sigma$, i.e.,
$Q(\rho)=\delta_{\rho\sigma}$, $\rho, \sigma\in \Omega$, then the
distance between $P(\rho)$ and $Q(\rho)$ is equal to the average
distance between a state drawn from the ensemble $P(\rho)$ and the
state $\sigma$,
\begin{equation}
D^K(P,Q)=\underset{\rho\in\Omega}{\sum}P(\rho)\Delta(\rho,\sigma).
\end{equation}

\textbf{Proof.} The property follows from the fact that the only
joint probability distribution with marginals $P$ and $Q$ in this
case is $\Pi(\kappa,\rho)= \delta_{\sigma\kappa}P(\rho)$.

\textbf{Limiting case 3 (Classical distributions).} If the set
$\Omega$ consists of perfectly distinguishable density matrices,
i.e., $\Delta (\rho, \sigma)=1-\delta_{\rho\sigma}$, $\forall \rho,
\sigma \in \Omega$, then $D^K(P,Q)$ reduces to the Kolmogorov
distance between the classical probability distributions $P$ and
$Q$,
\begin{equation}
D^K(P,Q)=\frac{1}{2}\underset{\rho\in\Omega}{\sum}|P(\rho)-Q(\rho)|.
\end{equation}

\textbf{Proof.} Since in this case the set $\Omega$ consists of orthogonal states, we can write the right-hand side of Eq.~\eqref{definition} as
\begin{gather}
\min_{\Pi(\rho,\sigma)}\sum_{\rho,\sigma\in\Omega,
\rho\neq\sigma}\Pi(\rho,\sigma)\times
1+\sum_{\rho\in\Omega}\Pi(\rho,\rho)\times 0\notag\\
=\min_{\Pi(\rho,\sigma)}(1-\sum_{\rho\in\Omega}\Pi(\rho,\rho)),\label{a}
\end{gather}
where the equality follows from the fact that
\begin{equation}
\sum_{\rho,\sigma\in\Omega,
\rho\neq\sigma}\Pi(\rho,\sigma)+\sum_{\rho\in\Omega}\Pi(\rho,\rho)=1.
\end{equation}
The minimum in Eq.~\eqref{a} is achieved when
$\sum_{\rho\in\Omega}\Pi(\rho,\rho)$ is maximal, which in turn is
achieved when each of the terms $\Pi(\rho,\rho)$ is maximal. Since
the maximum value of $\Pi(\rho,\rho)$ is
$\textrm{min}(P(\rho),Q(\rho))$, we obtain
\begin{gather}
D^K(Q,P)=(1-\sum_{\rho\in\Omega}\textrm{min}(Q(\rho),P(\rho)))\notag\\
=\frac{1}{2}\sum_{\rho\in\Omega}|Q(\rho)-P(\rho)|.
\end{gather}

\textbf{Comment.} Note that we can distinguish two limits which can
be interpreted as comparing classical probability distributions. One
is Limiting case 3---probability distributions over a set of
orthogonal states. The other is the case where each of the two
ensembles consists of a single state (two singleton ensembles) and
the two states are diagonal in the same basis. In both limits, the
distance $D^K(Q,P)$ reduces to the Kolmogorov distance between
classical distributions.

\subsection{Properties of the Kantorovich fidelity}

The following properties of the Kantorovich fidelity
\eqref{definition2} can be proven similarly to the corresponding
properties of the Kantorovich distance, which is why we present them
without proof.

\textbf{Property 1 (Positivity and normalization).}
\begin{gather}
0\leq F^K(P,Q)\leq 1, \\ \forall \hspace{0.1cm}P,Q \in \mathcal{P}_{\Omega},   \notag
\end{gather}
with
\begin{gather}
F^K(P,Q)=1 \hspace{0.2cm} \textrm{iff} \hspace{0.2cm}
P(\rho)=Q(\rho), \hspace{0.2cm} \forall \rho \in \Omega,
\end{gather}
and
\begin{gather}
F^K(P,Q)=0
\end{gather}
if and only if the supports of P and Q are orthogonal sets of
states.

\textbf{Property 2 (Symmetry).}
\begin{gather}
F^K(P,Q)=F^K(Q,P),
\\ \forall \hspace{0.2cm}P,Q \in \mathcal{P}_{\Omega}\notag.
\end{gather}

\textbf{Property 3 (Joint concavity).}
\begin{gather}
F^K(pP_1+(1-p)P_2,pQ_1+(1-p)Q_2)\\\geq pF^K(P_1,Q_1)+(1-p)F^K(P_2,Q_2),\notag\\
\forall \hspace{0.1cm}P_1,P_2,Q_1,Q_2 \in \mathcal{P}_{\Omega},
\hspace{0.2cm} \forall \hspace{0.1cm}p\in[0,1] \notag.
\end{gather}

\textbf{Property 4 (Monotonicity under CPTP maps).} For all CPTP
maps $\mathcal{E}$,
\begin{gather}
F^K(P,Q)\leq F^K(M_{\mathcal{E}}(P),M_{\mathcal{E}}(Q)),
\end{gather}
where $M_{\mathcal{E}}:\mathcal{P}_{\Omega} \rightarrow
\mathcal{P}_{\Omega_{\mathcal{E}}}$ is the map induced by
$\mathcal{E}$.

\textbf{Corollary (Invariance under unitary maps).} For all unitary
maps $\mathcal{U}$,
\begin{gather}
F^K(P,Q)=F^K(M_{\mathcal{U}}(P),M_{\mathcal{U}}(Q)),
\end{gather}
where $M_{\mathcal{U}}:\mathcal{P}_{\Omega} \rightarrow
\mathcal{P}_{\Omega_{\mathcal{U}}}$ is the map induced by
$\mathcal{U}$.

\textbf{Property 5 (Monotonicity under averaging).} Let
$\overline{P}$ denote the singleton ensemble consisting of the
average state of $P(\rho)$,
$\overline{\rho}_P=\underset{\rho\in\Omega}{\sum}P(\rho)\rho$. Then
\begin{equation}
F^K(P,Q)\leq F^K(\overline{P}, \overline{Q}).
\end{equation}

\textbf{Corollary.} If two distributions are close, their average
states are also close, i.e.,
\begin{equation}
\textrm{if}\hspace{0.2cm}F^K(P,Q)\geq
1-\varepsilon,\hspace{0.2cm}\textrm{then}\hspace{0.2cm}
F(\overline{\rho}_P, \overline{\rho}_Q)\geq 1-\varepsilon.
\end{equation}

\textbf{Property 6 (Stability).} Let $P(\rho)$ and $Q(\rho)$ be two
ensembles of states in $\Omega$ and $R(\sigma')$ be an ensemble of
states in $\Omega'$. Then,
\begin{gather}
F^K(P\otimes R,Q\otimes R)=F^K(P,Q).
\end{gather}

\textbf{Property 7 (Linear programming).} The task of finding the
optimal $\Pi(\rho,\sigma)$ in Eq.\eqref{definition2} is a linear
program and can be solved efficiently in the cardinality of
$\Omega$.

\textbf{Limiting case 1 (Two singleton ensembles).} If
$P(\rho)=\delta_{\rho\tau}$, $\rho, \tau\in \Omega$ and
$Q(\rho)=\delta_{\rho\sigma}$, $\rho, \sigma\in \Omega$, i.e., each
of the ensembles $P$ and $Q$ consists of only a single state, then
the fidelity between the ensembles is equal to the fidelity between
the respective states,
\begin{equation}
F^K(P,Q)=F(\tau,\sigma).
\end{equation}

\textbf{Limiting case 2 (One singleton ensemble)}. If the ensemble
$Q(\rho)$ consists of only one state $\sigma$, i.e.,
$Q(\rho)=\delta_{\rho\sigma}$, $\rho, \sigma\in \Omega$, then the
fidelity between $P(\rho)$ and $Q(\rho)$ is equal to the average
fidelity between a state drawn from the ensemble $P(\rho)$ and the
state $\sigma$,
\begin{equation}
F^K(P,Q)=\underset{\rho\in\Omega}{\sum}P(\rho)F(\rho,\sigma).
\end{equation}

\textbf{Limiting case 3 (Classical distributions).} If the set
$\Omega$ consists of perfectly distinguishable density matrices, i.e.,
$F(\rho, \sigma)=\delta_{\rho\sigma}$, $\forall \rho, \sigma \in
\Omega$, then $F^K(P,Q)$ reduces to the following overlap between
the classical probability distributions over the set $\Omega$:
\begin{equation}
F^K(P,Q)=\underset{\rho\in\Omega}{\sum}\textrm{min}(P(\rho),Q(\rho))=1-\frac{1}{2}\underset{\rho\in\Omega}{\sum}|P(\rho)-Q(\rho)|.\label{lc3fK}
\end{equation}

\textbf{Comment.} As pointed out earlier, there are two limits which
can be interpreted as corresponding to classical probability
distributions---Limiting case 3 (probability distributions over a
set of orthogonal states), and the limit of two singleton ensembles
where the two states are diagonal in the same basis. Here, these two
limits yield different results. In the first case, we obtain
Eq.~\eqref{lc3fK} which is a particular type of overlap between
classical probability distributions. In the second case, if
$P(\rho)$ and $Q(\rho)$ are the spectra of the two density matrices,
the fidelity reduces to the Bhattacharyya overlap
$\underset{\rho\in\Omega}{\sum}\sqrt{P(\rho),Q(\rho)}$ which upper
bounds expression \eqref{lc3fK}. This reflects the fact that the way
$F^K$ treats the overlap between the `classical aspect' of the
probability distribution $P(\rho)$ is not a special case of the way
it treats the overlap between two quantum states. We will show in
subsection E, that this property is intimately related to the fact
that $F^K$ is not monotonic under measurements.  The fidelity which
we propose in Sec.~V is monotonic under measurements and both its
classical limits coincide.

\subsection{Operational interpretations of the Kantorovich measures}

To further develop our understanding of the meaning of the
Kantorovich measures, it is useful to illustrate their
interpretation in the spirit of game theory. Let us consider the
Kantorovich distance first.

The trace distance is related to the maximum average probability
$p_{\mathrm{max}}(\rho,\sigma)$ with which two equally probable
states $\rho$ and $\sigma$ can be distinguished by a measurement as
follows:
$p_{\mathrm{max}}(\rho,\sigma)=\frac{1}{2}+\frac{1}{2}\parallel\rho-\sigma
\parallel$ \cite{Helstrom}. This naturally suggests the following
game scenario. Imagine that Alice has access to two ensembles of
quantum states $P(\rho)$ and $Q(\rho)$, $\rho\in\Omega$. More
precisely, we will assume that she has at her disposal two
sufficiently large pools of states in which the relative frequencies
of different states are approximately equal to the corresponding
probabilities for these states within a satisfactory precision.
Alice has to pick one state from one pool and another state from the
other pool and choose randomly (with equal probability) whether to
send the first state to Bob and throw the other away, or vice versa.
She has to tell Bob which is the pair of states drawn from the two
ensembles. Bob's task is to distinguish, by performing any operation
on the received state, from which ensemble the state he receives has
been drawn. This is repeated until the two pools are depleted (the
two pools are assumed to have equal numbers of states). Bob's
success is measured in terms of the average number of times he
guesses correctly the ensemble from which the state he receives has
been drawn. Alice's goal, on the other hand, is to choose the pairs
of states from the two ensembles in such a way as to make Bob's task
as difficult as possible.

If every time Bob employs the optimal measurement
strategy for distinguishing which state he has been sent, it is
obvious that the optimal strategy for Alice is to pair the states
according to the joint probability distribution $\Pi(\sigma,\rho)$
which minimizes the right-hand side of Eq.~\eqref{definition}, that
is, minimizes the average probability of correctly distinguishing
the two states in each pair by an optimal measurement. The
Kantorovich distance can then be understood as
\begin{gather}
D^K(P,Q)=2p^{\mathrm{Bob}}_{\mathrm{max}}(P,Q)-1,
\end{gather}
where $p^{\mathrm{Bob}}_{\mathrm{\mathrm{max}}}(P,Q)$ is Bob's maximal probability of success
when Alice chooses her strategy optimally.

The fidelity $F^K(P,Q)$ can be given a similar operational
interpretation, although a bit more artificial. The difference is
that Bob's task and corresponding measure of success have to be
chosen so that they are given by the fidelity between the two states
which Bob wants to distinguish at every round. For this purpose, we
can use Fuchs' operational interpretation of the fidelity
\cite{Fuchs96} as the minimum Bhattacharyya overlap between the
statistical distributions generated by all possible measurements on
the states,
\begin{gather}
F(\tau,
\upsilon)=\underset{\{E_i\}}{\textrm{min}}\underset{i}{\sum}\sqrt{\textrm{Tr}(
E_i\tau)}\sqrt{\textrm{Tr}( E_i\upsilon)},
\label{fidelityoper}
\end{gather}
where minimum is taken over all positive operators $\{E_i\}$ that
form a positive operator-valued measure ($\underset{i}{\sum}E_i=I$).
Then we can modify the game as follows. After sending one of the two
states to Bob, Alice does not throw away the other state, but waits
for Bob to tell her the type of measurement he performs on his
state, and she performs the same measurement on her state. They
record their results under many repetitions, and at the end they
calculate the average of the statistical overlap between the
resulting distributions of measurement outcomes for every pair of
states. Bob's task is to minimize this quantity by appropriately
choosing his measurements for every pair of states, while Alice's
goal is again to make Bob's task as difficult as possible by
choosing the pairs of states in a suitable manner.

\subsection{Non-monotonicity under generalized measurements}

The trace distance and the fidelity (as well as all fidelity-based
distance measures between states) are monotonic under CPTP maps
\cite{ruskaicontractivity, NieChu00, GLN05}. This property, also
known as \textit{contractivity}, can be understood as an expression
of the fact that the distinguishability between states described by
these measures cannot be increased by performing any operation on
the states. One may wonder if, when going to the realm of ensembles,
we should expect a measure of distinguishability between ensembles
to be monotonic under the more general class of \textit{stochastic}
operations, i.e., generalized measurements. After all, these are
operations that transform ensembles into ensembles. We will show
that this is not satisfied by the Kantorovich distance and fidelity.
We will also relate this property to the fact that the Kantorovich
fidelity yields two different results in the two `classical' limits
since a necessary condition for a Kantorovich measure to be
monotonic under measurements is that both its classical limits are
the same. This condition, however, is not sufficient, as shown by
the case of the Kantorovich distance.

Note, however, that our definitions of the Kantorovich measures were
based on the trace distance and the square root fidelity. In an
analogous manner, one can define Kantorovich measures based on any
other distance or fidelity between states. Non-monotonicity under
generalized measurements is not a problem \textit{per se} and we
will see that there is no reason why we should expect it,
considering the operational meaning of the Kantorovich measures
based on the trace distance and the square root fidelity.
Nevertheless, it would be useful to have measures such that the
distinguishability between ensembles that they describe cannot be
increased by any possible operation (see Sec.~V). Driven by this
motivation, we derive necessary and sufficient conditions that a
measure of distance or fidelity between states has to satisfy in
order for the corresponding Kantorovich measure to be monotonic
under measurements.

Let us first formulate precisely what we mean by monotonicity under
generalized measurements. As pointed out earlier, under the most
general type of quantum measurement, the state of a system
transforms as in Eq.~\eqref{genmeas}.

\textbf{Definition 3 (Monotonicity under generalized measurements).}
Consider a measurement $\mathbf{M}$ with measurement superoperators
$\{\mathcal{M}_i\}$. Denote the set of distinct density matrices
among all possible outcomes $\frac{\mathcal{M}_i(\rho)}{\textrm{Tr}
\mathcal{M}_i(\rho)}$ over all possible inputs $\rho \in \Omega$ by
$\Omega_{\mathbf{M}}$. If we apply the same generalized measurement
\eqref{genmeas} to every state in an ensemble $P(\rho)$, $\rho\in
\Omega$, we obtain another ensemble $P'(\rho')$, $\rho'\in
\Omega_{\mathbf{M}}$. Thus the generalized measurement
\eqref{genmeas} induces a map from the set of probability
distributions over $\Omega$ to the set of probability distributions
over $\Omega_{\mathbf{M}}$. Denote this map by
$M:\mathcal{P}_{\Omega} \rightarrow
\mathcal{P}_{\Omega_{\mathbf{M}}}$. When we say that a distance
function $D(Q,P)$ between ensembles of states $Q$ and $P$ is
monotonically decreasing (or simply monotonic) under generalized
measurements, we mean that for any generalized measurement
\eqref{genmeas},
\begin{gather}
D(M(P),M(Q))\leq D(P,Q),
\end{gather}
where $M:\mathcal{P}_{\Omega} \rightarrow
\mathcal{P}_{\Omega_{\mathbf{M}}}$ is the map induced by the
measurement. Similarly, a monotonicity of a fidelity $F(Q,P)$ means
\begin{gather}
F(M(P),M(Q))\geq F(P,Q)
\end{gather}
for any generalized measurement.

\textbf{Property.} The Kantorovich distance based on the trace
distance (Eq.~\eqref{definition}) and the Kantorovich fidelity based
on the square root fidelity (Eq.~\eqref{definition2}) are not
monotonic under generalized measurements.

\textbf{Proof.} The proof is presented in Appendix B.

The lack of monotonicity of the Kantorovich measures is something
that should not be surprising considering the operational
interpretations we discussed in the previous subsection. Generally,
monotonicity under certain types of operations means that the type
of distinguishability described by the measures cannot be increased
under these operations. However, from the above game scenarios we
see that the distinguishability concerns Bob's ability do
distinguish which of a pair of states Alice has sent to him, in the
case where Alice has chosen the way she pairs the states in an
optimal way. Certainly, by applying a measurement on the state he
receives, Bob cannot improve his chances of guessing correctly
beyond what he would obtain by doing the optimal measurement.
However, the question of monotonicity we are asking concerns
applying the same measurement to all states in the original
ensembles before Alice has chosen her optimal strategy. There is no
reason to expect that after applying a measurement on all of the
states in the original ensembles, the optimal strategy that Alice
can employ for the resulting ensembles can only be better than her
optimal strategy for the original ensembles. Indeed, as shown in
Appendix B, this is not the case when the figure of merit is based
on the trace distance or the square root fidelity.

We now provide necessary and sufficient conditions that a measure of
distance or fidelity between states has to satisfy in order for the
Kantorovich measure based on it to be monotonic under measurements.
We will denote by $D_d^K$ the Kantorovich distance based on a
distance $d(\rho,\sigma)$ between states, which is defined as in
Eq.~\eqref{definition} with $d$ in the place of $\Delta$. Similarly,
by $F_f^K$ we will denote the Kantorovich fidelity based on a
fidelity $f(\rho,\sigma)$ between states.

\textbf{Theorem 2 (Conditions for monotonicity of the Kantorovich
measures under generalized measurements).} Let $d(\rho,\sigma)$ and
$f(\rho,\sigma)$ be normalized distance and fidelity between states,
which are monotonic under CPTP maps and jointly convex (concave).
The Kantorovich distance $D_d^K(P,Q)$ or fidelity $F_f^K(P,Q)$ based
on $d(\rho,\sigma)$ and $f(\rho,\sigma)$, respectively, is monotonic
under generalized measurements if and only if for every two states
of the form $\sum_i p_i \rho_i \otimes|i\rangle\langle i|$ and
$\sum_i q_i \sigma_i \otimes|i\rangle\langle i|$, where $\{|i\rangle
\}$ is an orthonormal set of states, the distance and fidelity
satisfy
\begin{gather}
d(\sum_i p_i \rho_i \otimes|i\rangle\langle i|, \sum_i q_i \sigma_i \otimes|i\rangle\langle i|)\notag\\
=\sum_i
\left(\textrm{min}(p_i,q_i)\Delta(\rho_i,\sigma_i)+\frac{1}{2}|p_i-q_i|\right),\label{the1}
\end{gather}
and
\begin{gather}
f(\sum_i p_i \rho_i \otimes|i\rangle\langle i|, \sum_i q_i \sigma_i \otimes|i\rangle\langle i|)\notag\\
=\sum_i \textrm{min}(p_i,q_i)F(\rho_i,\sigma_i),\label{the2}
\end{gather}
respectively.

\textbf{Proof.} The proof is presented in Appendix C.

\textbf{Comment 1.} This theorem is a statement regarding the
relation between the values of a given measure (distance or
fidelity) between states over Hilbert spaces of different
dimensions. Note that if a measure has a well-defined operational
interpretation formulated without reference to the dimension of the
Hilbert space (to the best of our knowledge, this is the case for
all known measures of distance and fidelity between states), that
measure is automatically defined for any dimension. The property of
monotonicity that we are interested in is also
dimension-independent. We remark that the above theorem concerns
distance and fidelity measures between states which are monotonic
under CPTP maps without the restriction that the CPTP maps preserve
the dimension of the Hilbert space since we are interested in
proving monotonicity under the most general type of quantum
operations. One can easily see that monotonicity under CPTP maps
that can increase the dimension is equivalent to monotonicity under
dimension-preserving CPTP maps plus the stability condition
$d(\rho,\sigma)=d(\rho\otimes\kappa,\sigma\otimes\kappa)$ and
$f(\rho,\sigma)=f(\rho\otimes\kappa,\sigma\otimes\kappa)$ for all
$\rho,\sigma\in\mathcal{B}(\mathcal{H})$ and $\kappa\in
\mathcal{B}(\mathcal{H}')$ where $\mathcal{H}$ and $\mathcal{H}'$
are arbitrary Hilbert spaces. Similarly, monotonicity under CPTP
maps that can decrease the dimension is equivalent to monotonicity
under dimension-preserving CPTP maps plus monotonicity under partial
tracing.

\textbf{Comment 2.}  The third Jozsa axiom  states that a fidelity
function should satisfy \cite{Jozsa}
\begin{gather}
f(\rho,|\psi\rangle\langle\psi|)=\langle\psi|\rho|\psi\rangle.
\label{Jozsa}
\end{gather}
The square root fidelity we have considered above satisfies a modified version of that axiom, namely,
\begin{gather}
F(\rho,|\psi\rangle\langle\psi|)=\sqrt{\langle\psi|\rho|\psi\rangle}.\label{Jozsa2}
\end{gather}
But one can see that if the fidelity $f$ satisfies Eq.~\eqref{the2},
it must satisfy
\begin{gather}
f(\sum_j p_j\rho_j\otimes |j\rangle\langle j|, |\psi\rangle\langle\psi|\otimes |i\rangle\langle i|)\notag\\
=p_i f(
\rho_i\otimes |i\rangle\langle i|,|\psi\rangle\langle\psi|\otimes |i\rangle\langle i|),
\end{gather}
which can be only consistent with Eq.~\eqref{Jozsa} and not with Eq.~\eqref{Jozsa2}.
This rules out a class of possible fidelity functions.

A natural question to ask is whether there actually exist measures
of distance or fidelity between states that satisfy the conditions
of the theorem and thereby would give rise to Kantorovich measures
that are monotonic under generalized measurements. We leave this
problem open for future investigation. Instead, in the next section
we propose distance and fidelity between ensembles which are based
on the trace distance and the square root fidelity but are not of
the Kantorovich type and satisfy the desired monotonicity.

\section{Distance and fidelity based on the extended-Hilbert-space representation of ensembles}

\subsection{Motivating the definitions}

In this section, we adopt a different approach to defining measures
between ensembles of quantum states, which is based on the
extended-Hilbert-space (EHS) representation of ensembles that we
briefly touched upon in Sec.~II. As we pointed out, an ensemble
describes states occurring randomly according to some probability
distribution, but an indispensable part of the ensemble is the
classical side information about the identity of the given state.
The idea behind the EHS representation is that the classical system
storing that information is ultimately quantum and therefore it must
be possible to describe it in the language of quantum mechanics. In
the original formulation of the EHS representation \cite{DevWin03},
an ensemble of the form $\{(p_x, \rho_x)\}$ is represented in terms
of a state of the form $\widehat{\rho}=\sum_x p_x \rho_x\otimes [x]$
(Eq.~\eqref{state2}). When only a single ensemble is involved, this
representation is sufficient and it is not important what the
pointer (or flag) states $[x]\equiv |x\rangle\langle x|$ are, as
long as they form an orthonormal set and each $[x]$ is unambiguously
associated with $\rho_x$. However, if we want to use the EHS idea to
compare two ensembles, we need to go beyond this simple formulation.
In Sec.~III, we already saw one example where a naive application of
this idea fails. Namely, we argued that if we represent two
ensembles $P(\rho)$ and $Q(\rho)$, $\rho\in\Omega$, by the states
$\sum_{\rho\in\Omega}P(\rho)\rho\otimes[\rho]$ and
$\sum_{\rho\in\Omega}Q(\rho)\rho\otimes[\rho]$, a distance or
fidelity between these EHS representations is equivalent to a
distance or fidelity between the probability distributions $P(\rho)$
and $Q(\rho)$ in which $\rho$ is treated as a classical variable.
Such a measure does not capture the idea of closeness between
different quantum states. In this section, we will provide a
generalized formulation of an EHS representation of an ensemble,
which will allow us to define measures of distance and fidelity
between ensembles that possess all properties that we would like
such measures to have.

For this purpose, it is convenient to introduce the notion of a
`classical' system whose states live in a 'classical' space which we
define to be a fixed set $\Omega^C$ of orthogonal pure states $[c]$,
$\textrm{Tr}([c][c'])=\delta_{cc'}$, where we use the notation
$[c]\equiv |c\rangle\langle c|$ to distinguish the states of the
`classical' system from the states of the quantum system. Generally,
the classical space can consist of infinitely many different states,
but later we will see that it suffices to consider a classical space
of cardinality $|\Omega^C|=|\Omega|^2$, where $|\Omega|$ is the
cardinality of the set $\Omega$ of density matrices participating in
the ensembles.

Given the classical system described by the classical space
$\Omega^C$ and a set $\Omega$ of states of a quantum system, we can
ask what are the most general states of the quantum-classical system
that represent an ensemble $P(\rho)$, $\rho\in\Omega$, consistently
with our notion of ensemble. As we pointed out, the information
about the identity of a quantum state from the ensemble must be
stored in the classical system in a way which allows one to
unambiguously identify the state by measuring the state of the
classical system. If we take this to be the definition of a valid
EHS representation, then we should allow for the possibility that
several flag states $\{ [c_i(\rho)]\}$ point at the same quantum
state as long as every flag state is associated with a single
quantum state and, of course, each quantum state $\rho$ still
appears with the correct total probability. More succinctly, the
most general EHS representation should allow for mixed flag states,
i.e.,
\begin{gather}
\widehat{\rho}_P=\sum_{\rho\in\Omega} P(\rho)\rho \otimes
\left(\sum_i p_i(\rho) [c_i(\rho)]\right).\label{genEHS}
\end{gather}

Having a quantum-classical state of this form is equivalent to
having the ensemble $\{(P(\rho),\rho)\}$ because by measuring the
state of the classical system, we can infer which state from the
ensemble we are given, and given a state drawn randomly from the
ensemble we can always prepare the state \eqref{genEHS} by attaching
the corresponding classical state and discarding any additional
information. Note that in the expression \eqref{genEHS} we have
written the classical states as $[c_i(\rho)]$, explicitly indicating
which classical states are associated with the quantum state $\rho$,
but it is convenient to express the condition that every pointer
state is associated with a unique $\rho\in\Omega$ as a condition on
a general state of the quantum-classical system.

\textbf{Definition 4 (EHS representation of an ensemble).} An EHS
representation of an ensemble $P(\rho)$, $\rho\in\Omega$, is a
quantum-classical state of the form
\begin{gather}
\widehat{\rho}=\sum_{\rho\in\Omega}\sum_{[c]\in\Omega^C}
\widetilde{P}(\rho,[c]) \rho\otimes [c],\label{definition4}
\end{gather}
for which the non-negative quantities $\widetilde{P}(\rho,[c])$
satisfy
\begin{gather}
\sum_{[c]\in\Omega^C}\widetilde{P}(\rho,[c])=P(\rho),\hspace{0.2cm}\forall\rho\in\Omega,\label{PP1}\\
\widetilde{P}(\rho,[c])\widetilde{P}(\sigma,[c])=0,
\hspace{0.2cm}\forall\rho,\sigma\in\Omega| \hspace{0.1cm}
\rho\neq\sigma, \hspace{0.2cm} \forall [c]\in\Omega^C. \label{PP0}
\end{gather}

Equation~\eqref{PP1} ensures that every quantum state
$\rho\in\Omega$ occurs with the correct probability $P(\rho)$ and
Eq.~\eqref{PP0} expresses the fact that a given pointer state $[c]$
in $\Omega^C$ cannot be associated with more than one state in
$\Omega$. In other words, there exists an injective function
$\zeta:\Omega\rightarrow \Omega^C$ which specifies the pointer
states associated with a given $\rho\in\Omega$, and
$\widetilde{P}(\rho,[c])=0$ if $\zeta^{-1}([c])\neq \rho$. It is
important to note that a given ensemble can be encoded using many
different injections. If two ensembles $P$ and $Q$ are encoded using
injections $\zeta_P$ and $\zeta_Q$ which map the space $\Omega$ to
two non-overlapping subsets of $\Omega^C$, the corresponding EHS
representations of the two ensembles would be completely orthogonal
and therefore perfectly distinguishable. However, if the sets of
quantum states participating in the two ensembles are not
orthogonal, one can always chose two EHS representations of the two
ensembles which have a non-zero overlap because one can assign one
and the same pointer to two non-overlapping states from the two
ensembles. At the same time, unless the two ensembles are identical,
their EHS representations cannot be made identical. This suggests a
way of defining distance and fidelity between ensembles based on an
optimal choice of their EHS representations.

\textbf{Definition 5 (EHS distance between ensembles).} The EHS
distance between the ensembles $P(\rho)$ and $Q(\rho)$, $\rho \in
\Omega$, is
\begin{gather}
D^{\textrm{\tiny EHS}}(P,Q)=\min_{\widehat{\rho},
\widehat{\sigma}}\Delta(\widehat{\rho},
\widehat{\sigma}),\label{EHSdist}
\end{gather}
where $\Delta$ is the trace distance (Eq.~\eqref{tracedistance}),
and minimum is taken over all EHS representations $\widehat{\rho}$
and $\widehat{\sigma}$ of $P(\rho)$ and $Q(\rho)$, respectively.

\textbf{Definition 6 (EHS fidelity between ensembles).} The EHS
fidelity between the ensembles $P(\rho)$ and $Q(\rho)$, $\rho \in
\Omega$, is
\begin{gather}
F^{\textrm{\tiny EHS}}(P,Q)=\max_{\widehat{\rho},
\widehat{\sigma}}F(\widehat{\rho}, \widehat{\sigma}),\label{EHSfid}
\end{gather}
where $F$ is the square root fidelity (Eq.~\eqref{fidelity}), and
maximum is taken over all EHS representations $\widehat{\rho}$ and
$\widehat{\sigma}$ of $P(\rho)$ and $Q(\rho)$, respectively.

Before we proceed with studying the properties of these measures, it
is convenient to present two equivalent formulations of the above
definitions.

\textbf{Lemma 2 (Equivalent form of the EHS distance).} The EHS
distance \eqref{EHSdist} is equivalent to
\begin{gather}
D^{\textrm{\tiny EHS}}(P,Q)=\label{EHSdist'}\\
\mathop{\min_{P(\rho,\sigma),}}_{Q(\rho,\sigma)}\Delta(\sum_{\rho,\sigma\in\Omega}P(\rho,\sigma)\rho\otimes[\rho\sigma],
\sum_{\rho,\sigma\in\Omega}Q(\rho,\sigma)\sigma\otimes[\rho\sigma]
),\notag
\end{gather}
where minimum is taken over pairs of joint probability distributions
$P(\rho,\sigma)$ and $Q(\rho,\sigma)$ such that the left marginal of
$P(\rho,\sigma)$ is equal to $P(\rho)$ and the right marginal of
$Q(\rho,\sigma)$ is equal to $Q(\sigma)$. The set of pointer states
$[\rho\sigma]$ is fixed and has cardinality equal to the square of
the cardinality of $\Omega$.

\textbf{Proof.} First, observe that for any two EHS representations
$\widehat{\rho}$ and $\widehat{\sigma}$ of $P$ and $Q$, the distance
$\Delta(\widehat{\rho}, \widehat{\sigma})$ has the form
\begin{gather}
\Delta(\widehat{\rho}, \widehat{\sigma})
=\\\Delta(\sum_{\rho\in\Omega}\sum_{[c]\in\Omega^C}
\widetilde{P}(\rho,[c]) \rho\otimes [c],
\sum_{\rho\in\Omega}\sum_{[c]\in\Omega^C} \widetilde{Q}(\rho,[c])
\rho\otimes [c]),\notag
\end{gather}
where $\widetilde{P}(\rho,[c])$ and $\widetilde{Q}(\rho,[c])$ are
consistent with Definition 4. It can generally happen that one and
the same pointer $[c]$ is attached to a state $\rho$ from the first
ensemble and to a state $\sigma$ from the second ensemble, that is,
$\widetilde{P}(\rho,[c])\neq 0$ and $\widetilde{Q}(\sigma,[c])\neq
0$. However, having a pair of states $\rho$ and $\sigma$ from the
first and second ensembles, respectively, attached simultaneously to
more than one pointer, does not help in attaining the minimum in
Eq.~\eqref{EHSdist}. This follows from the fact that we could
replace the second pointer by the first one, which would result in
valid EHS representations of the two ensembles. But the latter
operation also corresponds to a CPTP map on the states in the
extended Hilbert space, and since $\Delta$ is monotonic under CPTP
maps, the resultant representations will be closer. Therefore,
without loss of generality, we can assume that every pair of states
$\rho$ and $\sigma$ from the first and second ensemble,
respectively, is associated with a single pointer state, which we
will label by $[\rho\sigma]$. This implies that the minimum in
Eq.~\eqref{EHSdist} can be taken over EHS representations of $P$ and
$Q$ of the form
$\sum_{\rho,\sigma\in\Omega}P(\rho,\sigma)\rho\otimes[\rho\sigma]$
and
$\sum_{\rho,\sigma\in\Omega}Q(\rho,\sigma)\sigma\otimes[\rho\sigma]$,
where the condition of consistency with the original distributions
$P$ and $Q$ amounts to conditions on the left and right marginals of
$P(\rho,\sigma)$ and $Q(\rho,\sigma)$, respectively:
\begin{gather}
\sum_{\sigma}P(\rho,\sigma)=P(\rho),\\
\sum_{\rho}Q(\rho,\sigma)=Q(\sigma).
\end{gather}
This completes the proof.

\textbf{Lemma 3 (Equivalent form of the EHS fidelity).} The EHS
fidelity \eqref{EHSfid} is equivalent to
\begin{gather}
F^{\textrm{\tiny EHS}}(P,Q)=\label{EHSfid'}\\
\max_{P(\rho,\sigma),Q(\rho,\sigma)}F(\sum_{\rho,\sigma\in\Omega}P(\rho,\sigma)\rho\otimes[\rho\sigma],
\sum_{\rho,\sigma\in\Omega}Q(\rho,\sigma)\sigma\otimes[\rho\sigma]
),\notag
\end{gather}
where minimum is taken over pairs of joint probability distributions
$P(\rho,\sigma)$ and $Q(\rho,\sigma)$ such that the left marginal of
$P(\rho,\sigma)$ is equal to $P(\rho)$ and the right marginal of
$Q(\rho,\sigma)$ is equal to $Q(\sigma)$. The set of pointer states
$[\rho\sigma]$ is fixed and has cardinality equal to the square of
the cardinality of $\Omega$.

\textbf{Proof.} The proof is analogous to the proof of Lemma 2.

\textbf{Corollary (Formulation without reference to an extended
Hilbert space).} Considering the explicit forms of the trace
distance and the square root fidelity, one can see that
Eqs.~\eqref{EHSdist'} and \eqref{EHSfid'} can be written without
reference to the classical pointer system:
\begin{gather}
D^{\textrm{\tiny
EHS}}(P,Q)=\frac{1}{2}\min_{P(\rho,\sigma),Q(\rho,\sigma)}\sum_{\rho,\sigma\in\Omega}\parallel
P(\rho,\sigma)\rho - Q(\rho,\sigma)\sigma \parallel,
\label{EHSdist''}
\end{gather}
\begin{gather}
F^{\textrm{\tiny
EHS}}(P,Q)=\max_{P(\rho,\sigma),Q(\rho,\sigma)}\sum_{\rho,\sigma\in\Omega}\sqrt{P(\rho,\sigma)Q(\rho,\sigma)}F(\rho,\sigma),
\label{EHSfid''}
\end{gather}
where optimization is taken over all joint distributions
$P(\rho,\sigma)$ with left marginal $P(\rho)$ and $Q(\rho,\sigma)$
with right marginal $Q(\sigma)$.

\subsection{Properties of the EHS distance}

\textbf{Property 1 (Positivity).}
\begin{gather}
D^{\textrm{\tiny EHS}}(P,Q)\geq 0, \\ \forall \hspace{0.1cm}P,Q \in
\mathcal{P}_{\Omega},   \notag
\end{gather}
with equality
\begin{gather}
D^{\textrm{\tiny EHS}}(P,Q)=0 \hspace{0.2cm} \textrm{iff}
\hspace{0.2cm} P(\rho)=Q(\rho), \hspace{0.2cm} \forall \rho \in
\Omega.
\end{gather}

\textbf{Proof.} The EHS distance is obviously non-negative since
$\Delta(\rho,\sigma)\geq 0$. If both ensembles are the same,
$P(\rho)=Q(\rho)$, $\forall \rho \in \Omega $, clearly
$D^{\textrm{\tiny EHS}}(P,Q)=0$, because we can choose identical EHS
representations for both ensembles. Reversely, if $D^{\textrm{\tiny
EHS}}(P,Q)=0$, this means that the EHS representations of $P$ and
$Q$ must be identical, which means that $P$ and $Q$ must be the
same.

\textbf{Property 2 (Normalization).}
\begin{gather}
D^{\textrm{\tiny EHS}}(P,Q)\leq 1,  \\ \forall \hspace{0.1cm}P,Q \in
\mathcal{P}_{\Omega}, \notag
\end{gather}
with equality
\begin{gather}
D^{\textrm{\tiny EHS}}(P,Q)=1
\end{gather}
if and only if the supports of P and Q are orthogonal sets of
states.

\textbf{Proof.} Since $\Delta(\rho,\sigma)\leq 1$, obviously
$D^{\textrm{\tiny EHS}}(P,Q)\leq 1$. If $P$ and $Q$ have supports on
orthogonal sets of states, then all of their EHS representations
will also be orthogonal, which implies $D^{\textrm{\tiny EHS}}(P,Q)=
1$. Reversely, if $D^{\textrm{\tiny EHS}}(P,Q)= 1$, this means that
the EHS states for which the minimum in Eq.~\eqref{EHSdist} is
achieved, must be orthogonal. But unless $P$ and $Q$ have supports
on orthogonal sets of states, it is always possible to find EHS
representations of $P$ and $Q$ which have non-zero overlap because
we can assign one and the same pointer to two non-overlapping states
from the two different ensembles.

\textbf{Property 3 (Symmetry).}
\begin{gather}
D^{\textrm{\tiny EHS}}(P,Q)=D^{\textrm{\tiny EHS}}(Q,P),
\\ \forall \hspace{0.2cm}P,Q \in \mathcal{P}_{\Omega}\notag.
\end{gather}

\textbf{Proof.} The symmetry follows from the definition
\eqref{definition} and the symmetry of $\Delta(\rho,\sigma)$.

\textbf{Property 4 (Triangle inequality).}
\begin{gather}
D^{\textrm{\tiny EHS}}(P,R)\leq D^{\textrm{\tiny
EHS}}(P,Q)+D^{\textrm{\tiny EHS}}(Q,R),\\ \forall
\hspace{0.1cm}P,Q,R \in \mathcal{P}_{\Omega}.\label{triangleEHS}
\end{gather}

\textbf{Proof.} The proof is presented in Appendix D.

\textbf{Property 5 (Joint convexity).}
\begin{gather}
D^{\textrm{\tiny EHS}}(pP_1+(1-p)P_2,pQ_1+(1-p)Q_2)\\\leq pD^{\textrm{\tiny EHS}}(P_1,Q_1)+(1-p)D^{\textrm{\tiny EHS}}(P_2,Q_2),\notag\\
\forall \hspace{0.1cm}P_1,P_2,Q_1,Q_2 \in \mathcal{P}_{\Omega},
\hspace{0.2cm} \forall \hspace{0.1cm}p\in[0,1] \notag.
\end{gather}

\textbf{Proof.} Let
\begin{gather}
D^{\textrm{\tiny EHS}}(P_1,Q_1)=\notag\\
\Delta(\sum_{\rho,\sigma\in\Omega}P_1(\rho,\sigma)\rho\otimes[\rho\sigma],\sum_{\rho,\sigma\in\Omega}Q_1(\rho,\sigma)\sigma\otimes[\rho\sigma]
)
\end{gather}
and
\begin{gather}
D^{\textrm{\tiny EHS}}(P_2,Q_2)=\notag\\
\Delta(\sum_{\rho,\sigma\in\Omega}P_2(\rho,\sigma)\rho\otimes[\rho\sigma],\sum_{\rho,\sigma\in\Omega}Q_2(\rho,\sigma)\sigma\otimes[\rho\sigma]
),
\end{gather}
where the joint distributions $P_1(\rho,\sigma)$ and
$P_2(\rho,\sigma)$ have left marginals $P_1(\rho)$ and $P_2(\rho)$,
respectively, and the joint distributions $Q_1(\rho,\sigma)$ and
$Q_2(\rho,\sigma)$ have right marginals $Q_1(\sigma)$ and
$Q_2(\sigma)$, respectively. Since $\Delta$ is jointly convex, we
have
\begin{gather}
p D^{\textrm{\tiny EHS}}(P_1,Q_1)+(1-p)D^{\textrm{\tiny EHS}}(P_2,Q_2)\geq\notag\\
\Delta(\sum_{\rho,\sigma\in\Omega}(pP_1(\rho,\sigma)+(1-p)P_2(\rho,\sigma)
)\rho\otimes[\rho\sigma],\notag\\
\sum_{\rho,\sigma\in\Omega}(pQ_1(\rho,\sigma)+(1-p)Q_2(\rho,\sigma)
)\sigma\otimes[\rho\sigma] ).\label{conv}
\end{gather}
But obviously $pP_1(\rho,\sigma)+(1-p)P_2(\rho,\sigma)$ is a joint
distribution with left marginal $pP_1(\rho)+(1-p)P_2(\rho)$, and
$pQ_1(\rho,\sigma)+(1-p)Q_2(\rho,\sigma)$ is a joint distribution
with right marginal $pQ_1(\sigma)+(1-p)Q_2(\sigma)$. Therefore, the
quantity on the right-hand side of Eq.~\eqref{conv} is greater than
or equal to $D^{\textrm{\tiny EHS}}(pP_1+(1-p)P_2,pQ_1+(1-p)Q_2)$,
which completes the proof.

\textbf{Property 6 (Monotonicity under generalized measurements).}
$D^{\textrm{\tiny EHS}} (P,Q)$ is monotonic under generalized
measurements in the sense of Definition 3, $D^{\textrm{\tiny
EHS}}(P,Q)\geq D^{\textrm{\tiny EHS}}(M(P),M(Q))$.

\textbf{Proof.}
Let $P(\rho)$ and $Q(\rho)$, $\rho\in\Omega$ be two ensembles of quantum states, and let
\begin{gather}
D^{\textrm{\tiny
EHS}}(P,Q)=\notag\\
\Delta(\sum_{\rho,\sigma\in\Omega}P(\rho,\sigma)\rho\otimes[\rho\sigma],\sum_{\rho,\sigma\in\Omega}Q(\rho,\sigma)\sigma\otimes[\rho\sigma]
).
\end{gather}
Let $\{\mathcal{M}_i\}$, $\mathcal{M}_i(\rho)=\sum_j M_{ij}\rho
M_{ij}^{\dagger}$, be the measurement superoperators of a
generalized measurement $\mathbf{M}$,
$\sum_{i,j}M_{ij}^{\dagger}M_{ij}=I$. Consider the following CPTP
map:
\begin{gather}
\mathcal{M}(\rho)\rightarrow \sum_i\mathcal{M}_i(\rho)\otimes [i],
\end{gather}
where $\{[i]\}$ is an orthonormal set of pure states in the Hilbert
space of some additional system. Since $\Delta$ is monotonic under
CPTP maps, we have
\begin{gather}
D^{\textrm{\tiny EHS}}(P,Q)\geq \notag\\
\Delta(\mathcal{M}(\sum_{\rho,\sigma\in\Omega}P(\rho,\sigma)\rho\otimes[\rho\sigma]),\mathcal{M}(\sum_{\rho,\sigma\in\Omega}Q(\rho,\sigma)\sigma\otimes[\rho\sigma]))\notag\\
=\Delta(\sum_{\rho,\sigma\in\Omega}\sum_iP(\rho,\sigma)\textrm{Tr}(\mathcal{M}_i(\rho)) \frac{\mathcal{M}_i(\rho)}{\textrm{Tr}(\mathcal{M}_i(\rho))}\otimes[\rho\sigma i],\notag\\
\sum_{\rho,\sigma\in\Omega}\sum_iQ(\rho,\sigma)\textrm{Tr}(\mathcal{M}_i(\sigma)) \frac{\mathcal{M}_i(\sigma)}{\textrm{Tr}(\mathcal{M}_i(\sigma))}\otimes[\rho\sigma i]
)\notag\\
\geq D^{\textrm{\tiny EHS}}(M(P),M(Q)),
\end{gather}
where $M:\mathcal{P}_{\Omega} \rightarrow
\mathcal{P}_{\Omega_{\mathbf{M}}}$ is the map induced by the
measurement as explained in Definition 3. The last inequality
follows from the fact that
$\sum_{\rho,\sigma\in\Omega}\sum_iP(\rho,\sigma)\textrm{Tr}(\mathcal{M}_i(\rho))
\frac{\mathcal{M}_i(\rho)}{\textrm{Tr}(\mathcal{M}_i(\rho))}\otimes[\rho\sigma
i]$ and
$\sum_{\rho,\sigma\in\Omega}\sum_iQ(\rho,\sigma)\textrm{Tr}(\mathcal{M}_i(\sigma))
\frac{\mathcal{M}_i(\sigma)}{\textrm{Tr}(\mathcal{M}_i(\sigma))}\otimes[\rho\sigma
i]$ are EHS representations of the new ensembles $M(P)$ and $M(Q)$.

\textbf{Corollary (Monotonicity under CPTP maps and invariance under
unitary maps).} Property 6 obviously implies monotonicity under CPTP
maps, which can be regarded as a special type of generalized
measurements. This in turn implies invariance under unitary maps
since the latter are reversible CPTP maps.

\textbf{Property 7 (Monotonicity under averaging).} Let
$\overline{P}$ denote the singleton ensemble consisting of the
average state of $P(\rho)$,
$\overline{\rho}_P=\underset{\rho\in\Omega}{\sum}P(\rho)\rho$. Then
\begin{equation}
D^{\textrm{\tiny EHS}}(P,Q)\geq D^{\textrm{\tiny EHS}}(\overline{P},
\overline{Q}).
\end{equation}

\textbf{Proof.} Let
\begin{gather}
D^{\textrm{\tiny
EHS}}(P,Q)=\notag\\
\Delta(\sum_{\rho,\sigma\in\Omega}P(\rho,\sigma)\rho\otimes[\rho\sigma],\sum_{\rho,\sigma\in\Omega}Q(\rho,\sigma)\sigma\otimes[\rho\sigma]
).
\end{gather}
Observe that
\begin{gather}
\overline{\rho}_P=\textrm{Tr}_C(\sum_{\rho,\sigma\in\Omega}P(\rho,\sigma)\rho\otimes[\rho\sigma])
\end{gather}
and
\begin{gather}
\overline{\rho}_Q=\textrm{Tr}_C(\sum_{\rho,\sigma\in\Omega}Q(\rho,\sigma)\sigma\otimes[\rho\sigma]),
\end{gather}
where $\textrm{Tr}_C$ denotes partial tracing over the subsystem
containing the classical pointers $\{[\rho\sigma]\}$. On the other
hand, $\Delta(\overline{\rho},\overline{\sigma})=D^{\textrm{\tiny
EHS}}(\overline{P}, \overline{Q})$ (see Eq.~\eqref{2sinEHS} below).
Since $\Delta(\rho,\sigma)$ is monotonic under partial tracing
(which is a CPTP map), the property follows.

\textbf{Corollary.} If two distributions are close, their average
states are also close, i.e.,
\begin{equation}
\textrm{if}\hspace{0.2cm}D^{\textrm{\tiny EHS}}(P,Q)\leq
\varepsilon,\hspace{0.2cm}\textrm{then}\hspace{0.2cm}
\Delta(\overline{\rho}_P, \overline{\rho}_Q)\leq\varepsilon.
\end{equation}

\textbf{Property 8 (Continuity of the average of a continuous
function)}. Let $h(\rho)$ be a bounded function, which is continuous
with respect to the distance $\Delta$. Then the ensemble average of
$h(\rho)$,
$\overline{h}_P=\underset{\rho\in\Omega}{\sum}P(\rho)h(\rho)$, is
continuous with respect to $D^{\textrm{\tiny EHS}}$.

\textbf{Proof.} The proof is presented in Appendix E.

\textbf{Comment.} Again, as we pointed out in relation to the
Kantorovich distance, Property 8 naturally reflects the idea of
states as resources---if a resource is a continuous function of the
state, when two ensembles are close, their average resources must
also be close.

\textbf{Property 9 (The EHS distance is upper bounded by the
Kantorovich distance)}.
\begin{equation}
D^{\textrm{\tiny EHS}}(P,Q)\leq D^K(P,Q).
\end{equation}

\textbf{Proof.} Let $\Pi(\rho,\sigma)$ be a joint probability
distribution with left and right marginals $P(\rho)$ and $Q(\sigma)$
for which the minimum in the definition \eqref{definition} of
$D^K(P,Q)$ is attained. Obviously, the minimum in
Eq.~\eqref{EHSdist'} satisfies
\begin{gather}
D^{\textrm{\tiny
EHS}}(P,Q)\leq\notag\\\Delta(\sum_{\rho,\sigma\in\Omega}\Pi(\rho,\sigma)\rho\otimes[\rho\sigma],\sum_{\rho,\sigma\in\Omega}\Pi(\rho,\sigma)\sigma\otimes[\rho\sigma]
)\notag\\
=\sum_{\rho,\sigma\in\Omega}\Pi(\rho,\sigma)\Delta(\rho,\sigma)=D^K(P,Q).
\end{gather}

\textbf{Property 10 (Stability).} Let $P(\rho)$ and $Q(\rho)$ be two
ensembles of states in $\Omega$ and $R(\sigma')$ be an ensemble of
states in $\Omega'$, where $\Omega$ and $\Omega'$ are sets of states
of two different systems. Then,
\begin{gather}
D^{\textrm{\tiny EHS}}(P\otimes R,Q\otimes R)=D^{\textrm{\tiny
EHS}}(P,Q).\label{stabi'}
\end{gather}

\textbf{Proof.} Let
\begin{gather}
D^{\textrm{\tiny EHS}}(P\otimes R,Q\otimes
R)=\notag\\
\Delta(\sum_{\rho,\sigma\in\Omega}\sum_{\tau',\kappa'\in\Omega'}\Pi(\rho\otimes
\tau',\sigma \otimes \kappa')\rho\otimes
\tau'\otimes[\rho\tau'\sigma\kappa'],\notag\\
\sum_{\rho,\sigma\in\Omega}\sum_{\tau',\kappa'\in\Omega'}J(\rho\otimes
\tau',\sigma \otimes
\kappa')\sigma\otimes\kappa'\otimes[\rho\tau'\sigma\kappa']),
\end{gather}
where $\Pi(\rho\otimes \tau',\sigma \otimes \kappa')$ has left
marginal $P(\rho)R(\tau')$ and $J(\rho\otimes \tau',\sigma \otimes
\kappa')$ has right marginal $Q(\sigma)R(\tau')$.

One can readily see that the monotonicity of $\Delta$ under partial
tracing implies
\begin{gather}
D^{\textrm{\tiny EHS}}(P\otimes R,Q\otimes R)\geq D^{\textrm{\tiny
EHS}}(P,Q).\label{1}
\end{gather}
Using the stability of $\Delta$, we see that if we choose
$\Pi(\rho\otimes \tau',\sigma \otimes
\kappa')=P(\rho,\sigma)R(\tau')\delta_{\tau'\kappa'}$ and
$J(\rho\otimes \tau',\sigma \otimes
\kappa')=Q(\rho,\sigma)R(\tau')\delta_{\tau'\kappa'}$, where
$P(\rho,\sigma)$ and $Q(\rho,\sigma)$ are two joint distributions
for which the minimum in Eq.~\eqref{EHSdist'} is attained, we obtain
\begin{gather}
D^{\textrm{\tiny EHS}}(P\otimes R,Q\otimes R)\leq D^{\textrm{\tiny
EHS}}(P,Q),\label{2}
\end{gather}
which together with Eq.~\eqref{1} implies Eq.~\eqref{stabi'}.

This property can also be seen to follow from Property 6 because one
can go from $P$ and $Q$ to $P\otimes R$ and $Q\otimes R$,
respectively, and vice versa, via stochastic operations.

\textbf{Property 11 (Convex optimization).} The task of finding the
optimal $P(\rho,\sigma)$ and $Q(\rho,\sigma)$ in Eq.\eqref{EHSdist'}
is a convex optimization problem.

\textbf{Proof.} We can think of $P(\rho,\sigma)$ and
$Q(\rho,\sigma)$ as the components of a vector $x$ of dimension
$2N^2$, where $N$ is the cardinality of the set $\Omega$. The first
$N^2$ components of the vector are equal to $P(\rho,\sigma)$ and the
second $N^2$ components are equal to $Q(\rho,\sigma)$. The convexity
of the function
\begin{gather}
\xi(x)\equiv\Delta(\sum_{\rho,\sigma\in\Omega}P(\rho,\sigma)\rho\otimes[\rho\sigma],\sum_{\rho,\sigma\in\Omega}Q(\rho,\sigma)\sigma\otimes[\rho\sigma]
)
\end{gather}
can be seen from the fact that for any $x_1$, $x_2$, and $t$, $0\leq
t\leq 1$, we have
\begin{gather}
\xi(tx_1+(1-t)x_2)=\notag\\
\Delta(\sum_{\rho,\sigma\in\Omega}(tP_1(\rho,\sigma)+(1-t)P_2(\rho,\sigma))\rho\otimes[\rho\sigma],\notag\\
\sum_{\rho,\sigma\in\Omega}
(tQ_1(\rho,\sigma)+(1-t)Q_2(\rho,\sigma))\sigma\otimes[\rho\sigma]
)\leq\notag\\
t\Delta(\sum_{\rho,\sigma\in\Omega}P_1(\rho,\sigma)\rho\otimes[\rho\sigma],\sum_{\rho,\sigma\in\Omega}Q_1(\rho,\sigma)\sigma\otimes[\rho\sigma]
)+\notag\\
(1-t)\Delta(\sum_{\rho,\sigma\in\Omega}P_2(\rho,\sigma)\rho\otimes[\rho\sigma],\sum_{\rho,\sigma\in\Omega}Q_2(\rho,\sigma)\sigma\otimes[\rho\sigma]
)\notag\\
=t\xi(x_1)+(1-t)\xi(x_2),
\end{gather}
due to the joint convexity of $\Delta$. Notice that if
$P_1(\rho,\sigma)$ and $P_2(\rho,\sigma)$ have left marginals equal
to $P(\rho)$, so does $tP_1(\rho,\sigma)+(1-t)P_2(\rho,\sigma)$.
Similarly, if $Q_1(\rho,\sigma)$ and $Q_2(\rho,\sigma)$ have right
marginals equal to $Q(\rho)$, so does
$tQ_1(\rho,\sigma)+(1-t)Q_2(\rho,\sigma)$. Since the marginal
conditions on $x$ are linear, the problem of finding $x$ which
minimizes $\xi(x)$ subject to these constraints is a convex
optimization problem, for which efficient numerical techniques
exist.

\textbf{Limiting case 1 (Two singleton ensembles).} If
$P(\rho)=\delta_{\rho\tau}$, $\rho, \tau\in \Omega$ and
$Q(\rho)=\delta_{\rho\sigma}$, $\rho, \sigma\in \Omega$, i.e., each
of the ensembles $P$ and $Q$ consists of only a single state, then
the distance between the ensembles is equal to the distance between
the respective states,
\begin{equation}
D^{\textrm{\tiny EHS}}(P,Q)=\Delta(\tau,\sigma).\label{2sinEHS}
\end{equation}

\textbf{Proof.} Due to the monotonicity of $\Delta$ under partial
tracing over the pointer system, we have that
$D(P,Q)\geq\Delta(\tau,\sigma)$. But clearly, equality is achievable
because we can choose the probability distributions in
Eq.~\eqref{EHSdist'}
$P(\kappa,\rho)=Q(\kappa,\rho)=\delta_{\kappa\tau}\delta_{\rho\sigma}$.

\textbf{Limiting case 2 (One singleton ensemble)}. Unlike the
Kantorovich distance, when the ensemble $Q(\rho)$ consists of only
one state $\sigma$, i.e., $Q(\rho)=\delta_{\rho\sigma}$, $\rho,
\sigma\in \Omega$, the EHS distance between $P(\rho)$ and $Q(\rho)$
is generally \textit{not} equal to the average distance between a
state drawn from the ensemble $P(\rho)$ and the state $\sigma$,
\begin{equation}
D^{\textrm{\tiny
EHS}}(P,Q)\neq\underset{\rho\in\Omega}{\sum}P(\rho)\Delta(\rho,\sigma).
\end{equation}

\textbf{Proof.} We provide a proof by counterexample. Let the
singleton ensemble consist of the sate
$\sigma_0=\tilde{\sigma}\otimes|0\rangle\langle 0|$ and let the
other ensemble consist of two states, $\rho_0=\tilde{\rho}_0\otimes
|0\rangle\langle 0|$ and $\rho_1=\tilde{\rho}_1\otimes
|1\rangle\langle 1|$, with probabilities $p_0$ and $p_1=1-p_0$,
respectively. The average distance between the state $\sigma_0$ and
the states from the other ensemble is
\begin{gather}
\Delta_{ave}=p_0\Delta(\rho_0, \sigma_0)+p_1\Delta(\rho_1, \sigma_0)\notag\\
=p_0\Delta(\tilde{\rho}_0,\tilde{\sigma}_0)+p_1.
\end{gather}
However, if we choose the joint distributions
$P(\rho,\sigma)=p_0\delta_{\rho\rho_0}+p_1\delta_{\rho\rho_1}$ and
$Q(\rho,\sigma)=\delta_{\rho_0\sigma_0}$, we see from
Eq.~\eqref{EHSdist''} that
\begin{gather}
D^{\textrm{\tiny EHS}}(P,Q)\leq\frac{1}{2}\parallel p_0\rho_0-\sigma_0\parallel +\frac{1}{2}p_1\leq\notag\\
\frac{p_0}{2}\parallel \rho_0-\sigma_0\parallel+\frac{1}{2}(1-p_0)\parallel\sigma_0\parallel+\frac{1}{2}p_1=\notag\\
p_0\frac{1}{2}\parallel\tilde{\rho}_0-\tilde{\sigma}_0\parallel+p_1=\Delta_{ave}.
\end{gather}
For an appropriate choice of $\tilde{\rho}_0$ and
$\tilde{\sigma_0}$, the second inequality can be made strict, which
completes the proof.

\textbf{Limiting case 3 (Classical distributions).} If the set
$\Omega$ consists of perfectly distinguishable density matrices, i.e., $\Delta
(\rho, \sigma)=1-\delta_{\rho\sigma}$, $\forall \rho, \sigma \in
\Omega$, then $D^{\textrm{\tiny EHS}}(P,Q)$ reduces to the trace
distance $\Delta(\overline{\rho}_P,\overline{\rho}_Q)$ between the
density matrices $\overline{\rho}_P=\sum_{\rho\in\Omega}P(\rho)\rho$
and $\overline{\rho}_Q=\sum_{\rho\in\Omega}Q(\rho)\rho$, which is
equal to the Kolmogorov distance between the classical probability
distributions $P$ and $Q$, $D^{\textrm{\tiny
EHS}}(P,Q)=\frac{1}{2}\underset{\rho\in\Omega}{\sum}|P(\rho)-Q(\rho)|$.

\textbf{Proof.} The property follows from the fact that via CPTP
maps one can go back and forth between any EHS representations of
the ensembles $P(\rho)$ and $Q(\rho)$, $\rho\in\Omega$, and the
states $\overline{\rho}_P$ and $\overline{\rho}_Q$.

\subsection{Properties of the EHS fidelity}

The properties of the EHS fidelity \eqref{EHSfid'} can be proven
analogously to the properties of the EHS distance, which is why we
present them without proof.

\textbf{Property 1 (Positivity and normalization).}
\begin{gather}
0\leq F^{\textrm{\tiny EHS}}(P,Q)\leq 1, \\ \forall
\hspace{0.1cm}P,Q \in \mathcal{P}_{\Omega},   \notag
\end{gather}
with
\begin{gather}
F^{\textrm{\tiny EHS}}(P,Q)=1 \hspace{0.2cm} \textrm{iff}
\hspace{0.2cm} P(\rho)=Q(\rho), \hspace{0.2cm} \forall \rho \in
\Omega,
\end{gather}
and
\begin{gather}
F^{\textrm{\tiny EHS}}(P,Q)=0
\end{gather}
if and only if the supports of P and Q are orthogonal sets of
states.

\textbf{Property 2 (Symmetry).}
\begin{gather}
F^{\textrm{\tiny EHS}}(P,Q)=F^{\textrm{\tiny EHS}}(Q,P),
\\ \forall \hspace{0.2cm}P,Q \in \mathcal{P}_{\Omega}\notag.
\end{gather}

\textbf{Property 3 (Strong concavity).}
\begin{gather}
F^{\textrm{\tiny EHS}}(pP_1+(1-p)P_2,qQ_1+(1-q)Q_2)\\\geq \sqrt{pq}F^{\textrm{\tiny EHS}}(P_1,Q_1)+\sqrt{(1-q)(1-p)}F^{\textrm{\tiny EHS}}(P_2,Q_2),\notag\\
\forall \hspace{0.1cm}P_1,P_2,Q_1,Q_2 \in \mathcal{P}_{\Omega},
\hspace{0.2cm} \forall \hspace{0.1cm}p,q\in[0,1] \notag.
\end{gather}

\textbf{Property 4 (Monotonicity under generalized measurements).}
$F^{\textrm{\tiny EHS}}(P,Q)$ is monotonic under generalized
measurements in the sense of Definition 3, $F^{\textrm{\tiny
EHS}}(P,Q)\leq F^{\textrm{\tiny EHS}}(M(P),M(Q))$.

\textbf{Corollary (Monotonicity under CPTP maps and invariance under
unitary maps).} $F^{\textrm{\tiny EHS}}(P,Q)$ is monotonic under
CPTP maps and invariant under unitary maps.

\textbf{Property 5 (Monotonicity under averaging).} Let
$\overline{P}$ denote the singleton ensemble consisting of the
average state of $P(\rho)$,
$\overline{\rho}_P=\underset{\rho\in\Omega}{\sum}P(\rho)\rho$. Then
\begin{equation}
F^{\textrm{\tiny EHS}}(P,Q)\leq F^{\textrm{\tiny EHS}}(\overline{P},
\overline{Q}).
\end{equation}

\textbf{Corollary.} If two distributions are close, their average
states are also close, i.e.,
\begin{equation}
\textrm{if}\hspace{0.2cm}F^{\textrm{\tiny EHS}}(P,Q)\geq
1-\varepsilon,\hspace{0.2cm}\textrm{then}\hspace{0.2cm}
F(\overline{\rho}_P, \overline{\rho}_Q)\geq 1-\varepsilon.
\end{equation}

\textbf{Property 6 (The EHS fidelity is lower bounded by the
Kantorovich fidelity).}
\begin{gather}
F^{\textrm{\tiny EHS}}(P,Q)\geq F^K(P,Q).
\end{gather}

\textbf{Property 7 (Stability).} Let $P(\rho)$ and $Q(\rho)$ be two
ensembles of states in $\Omega$ and $R(\sigma')$ be an ensemble of
states in $\Omega'$, where $\Omega$ and $\Omega'$ are sets of states
of two different systems. Then,
\begin{gather}
F^{\textrm{\tiny EHS}}(P\otimes R,Q\otimes R)=F^{\textrm{\tiny
EHS}}(P,Q).
\end{gather}

\textbf{Property 8 (Convex optimization).} The task of finding the
optimal $P(\rho,\sigma)$ and $Q(\rho,\sigma)$ in Eq.\eqref{EHSfid'}
is a convex optimization problem.

\textbf{Limiting case 1 (Two singleton ensembles).} Let
$P(\rho)=\delta_{\rho\tau}$, $\rho, \tau\in \Omega$ and
$Q(\rho)=\delta_{\rho\sigma}$, $\rho, \sigma\in \Omega$, i.e., each
of the ensembles $P$ and $Q$ consists of only a single state. Then
the fidelity between the ensembles is equal to the fidelity between
the respective states,
\begin{equation}
F^{\textrm{\tiny EHS}}(P,Q)=F(\tau,\sigma).
\end{equation}

\textbf{Limiting case 2 (One singleton ensemble)}. Unlike the
Kantorovich fidelity, when the ensemble $Q(\rho)$ consists of only
one state $\sigma$, i.e., $Q(\rho)=\delta_{\rho\sigma}$, $\rho,
\sigma\in \Omega$, the EHS fidelity between $P(\rho)$ and $Q(\rho)$
is generally \textit{not} equal to the average fidelity between a
state drawn from the ensemble $P(\rho)$ and the state $\sigma$,
\begin{equation}
F^{\textrm{\tiny
EHS}}(P,Q)\neq\underset{\rho\in\Omega}{\sum}P(\rho)F(\rho,\sigma).
\end{equation}

\textbf{Limiting case 3 (Classical distributions).} If the set
$\Omega$ consists of perfectly distinguishable density matrices, i.e.,
$F(\rho, \sigma)=\delta_{\rho\sigma}$, $\forall \rho, \sigma \in
\Omega$, then $F^{\textrm{\tiny EHS}}(P,Q)$ reduces to the fidelity
$F(\overline{\rho}_P,\overline{\rho}_Q)$ between the density
matrices $\overline{\rho}_P=\sum_{\rho\in\Omega}P(\rho)\rho$ and
$\overline{\rho}_Q=\sum_{\rho\in\Omega}Q(\rho)\rho$, which is equal
to the Bhattacharyya overlap between the classical probability
distributions $P$ and $Q$, $F^{\textrm{\tiny
EHS}}(P,Q)=\underset{\rho\in\Omega}{\sum}\sqrt{P(\rho)Q(\rho)}$.

\textbf{Comment.} Unlike the Kantorovich fidelity, here both
`classical' limits are the same.

\subsection{Operational interpretations of the EHS measures}

Similarly to the Kantorovich measures, we can understand the meaning
of the EHS measures from an operational point of view. However, we
present an interpretation in the spirit of Sec.~IV.D only for the
EHS distance. For the EHS fidelity, we present an interpretation of
a different type, in which an ensemble of density matrices is looked
upon as the output of a stochastic quantum channel with a pure-state
input.

\subsubsection{The EHS distance}

Observe that Eq.~\eqref{EHSdist''} can be written as
\begin{gather}
D^{\textrm{\tiny
EHS}}(P,Q)=\min_{P(\rho,\sigma),Q(\rho,\sigma)}\sum_{\rho,\sigma\in\Omega}\frac{P(\rho,\sigma)+Q(\rho,\sigma)
}{2}\times\notag\\
\parallel \frac{P(\rho,\sigma)}{P(\rho,\sigma)+Q(\rho,\sigma) }\rho
- \frac{Q(\rho,\sigma)}{P(\rho,\sigma)+Q(\rho,\sigma) }\sigma
\parallel. \label{EHSdist''2}
\end{gather}
It is not difficult to see that
\begin{gather}
\parallel \frac{P(\rho,\sigma)}{P(\rho,\sigma)+Q(\rho,\sigma) }\rho
- \frac{Q(\rho,\sigma)}{P(\rho,\sigma)+Q(\rho,\sigma) }\sigma
\parallel\notag\\
=2p_{\mathrm{max}}(\rho,\sigma)-1,\label{pmaxbiased}
\end{gather}
where $p_{\mathrm{max}}(\sigma,\rho)$ is the maximum average probability with
which the two states $\sigma$ and $\rho$, each occurring with prior
probability $\frac{P(\rho,\sigma)}{P(\rho,\sigma)+Q(\rho,\sigma)}$
and $\frac{Q(\rho,\sigma)}{P(\rho,\sigma)+Q(\rho,\sigma)}$,
respectively, can be distinguished by a measurement \cite{Helstrom}.
In the case when each of the states $\rho$ and $\sigma$ is equally
likely, the quantity \eqref{pmaxbiased} reduces to
$\frac{1}{2}\parallel \rho-\sigma\parallel$.

Imagine that Alice is given two ensembles $P(\rho)$ and $Q(\rho)$,
$\rho\in\Omega$, which are also known to Bob. With probability
${1}/{2}$, she chooses one of the two ensembles and draws a random
state from it. Let us say that she draws the state $\rho$ from the
first ensemble. She then sends this state to Bob but tells him that
she is sending either the state $\rho$ drawn from the first ensemble
or the state $\sigma$ drawn from the second ensemble, where Alice
can choose to say a particular $\sigma$ depending on the $\rho$ she
actually got. Bob's task is to distinguish from which ensemble the
state he receives has been drawn, and the figure of merit of his
success is the average number of times he guesses correctly. Alice's
goal is to make Bob's task as difficult as possible, with the caveat
that, although she is free to choose her strategy, she has to reveal
it to Bob. Alice's strategy is described by the probabilities
$T_1(\rho|\sigma)$ with which, when having drawn state $\rho$ from
the first ensemble, she will tell Bob that the state is either
$\rho$ from the first ensemble or $\sigma$ from the second ensemble,
and the probabilities $T_2(\rho|\sigma)$ with which, when having
drawn state $\sigma$ from the second ensemble, she will say that the
sate is either $\sigma$ from the second ensemble or $\rho$ from the
first ensemble. In other words, Bob is aware of the joint
probabilities $P(\rho,\sigma)=P(\rho)T_1(\rho|\sigma)$ and
$Q(\rho,\sigma)=T_2(\rho|\sigma)Q(\sigma)$. Obviously, the
probability that Bob will be told that the state he receives is
either $\rho$ from the first ensemble or $\sigma$ from the second
ensemble is equal to $\frac{P(\rho,\sigma)+Q(\rho,\sigma) }{2}$, and
the prior probability that in such a case the state is $\rho$ is
$\frac{P(\rho,\sigma)}{P(\rho,\sigma)+Q(\rho,\sigma)}$, while the
prior probability that the state is $\sigma$ is
$\frac{Q(\rho,\sigma)}{P(\rho,\sigma)+Q(\rho,\sigma)}$. Then
assuming that Bob performs the optimal measurement to distinguish
these states with these prior probabilities, the optimal strategy
for Alice is to choose $T_1(\rho|\sigma)$ and $T_2(\rho|\sigma)$ (or
equivalently, $P(\rho,\sigma)$ and $Q(\rho,\sigma)$) that minimize
the quantity \eqref{EHSdist''2}. The EHS distance can then be
understood as
\begin{gather}
D^{\textrm{\tiny EHS}}(P,Q)=2p^{\mathrm{Bob}}_{\mathrm{max}}(P,Q)-{1},
\end{gather}
where $p^{\mathrm{Bob}}_{\mathrm{max}}(P,Q)$ is Bob's maximal probability of success
when Alice chooses her strategy optimally.

\subsubsection{The EHS fidelity}

For the EHS fidelity, we propose an interpretation which is similar
to the one proposed for the square root fidelity in
Ref.~\cite{DoddNielsen},
\begin{gather}
F(\rho,\sigma) = \max |\langle\psi|\phi\rangle|,\label{DoddNielsen}
\end{gather}
where maximization is taken over all pure states $|\psi\rangle$ and
$|\phi\rangle$ such that
$\rho=\mathcal{E}(|\psi\rangle\langle\psi|)$ and
$\sigma=\mathcal{E}(|\phi\rangle\langle\phi|)$ for some CPTP map
$\mathcal{E}$. According to this interpretation, if $\rho$ and
$\sigma$ are the outputs of a deterministic quantum channel with
pure-state inputs, the square root fidelity is an upper bound on the
overlap between the input states. It turns out that the EHS fidelity
provides a generalization of this idea to stochastic quantum
channels.

When a generalized measurement $\mathbf{M}$ with measurement
superoperators $\{\mathcal{M}_i\}$ is applied to a given state
$\sigma$, it gives rise to an ensemble $P(\rho)$, $\rho\in\Omega$,
with $P(\rho)=\sum_{i:\textrm{ }\sigma_i=\rho}p_i$, where
$p_i=\textrm{Tr}(\mathcal{M}_i(\sigma))$ are the probabilities for
the different measurement outcomes, and
$\sigma_i=\mathcal{M}_i(\sigma)/p_i$ are their corresponding output
states. In other words, $\mathbf{M}$ can be viewed as a stochastic
quantum channel which for a given input state outputs an ensemble of
states. We will use the short-cut notation $\mathbf{M}(\sigma)$ to
denote the ensemble of states resulting from the action of the
channel $\mathbf{M}$ on the state $\sigma$.

\textbf{Theorem 3 (Channel-based interpretation of the EHS
fidelity).} Let $P(\rho)$ and $Q(\rho)$, $\rho\in\Omega$, be two
ensembles of density matrices on $\mathcal{H}^S$. Then,
\begin{gather}
F^{\textrm{\tiny EHS}}(P,Q) = \max |\langle\psi|\phi\rangle|,
\end{gather}
where maximization is taken over all pure states
$|\psi\rangle\in\mathcal{H}^S$ and $|\phi\rangle\in\mathcal{H}^S$
such that $\mathbf{M}(|\psi\rangle\langle\psi|)=\{(P(\rho),\rho)\}$,
$\rho\in\Omega$, and
$\mathbf{M}(|\phi\rangle\langle\phi|)=\{(Q(\rho),\rho)\}$,
$\rho\in\Omega$, for some stochastic channel $\mathbf{M}$.

\textbf{Proof.} From the monotonicity of the EHS fidelity under
generalized measurements it follows that for any generalized
measurement $\mathbf{M}$ and two states $|\psi\rangle$ and
$|\phi\rangle$,
\begin{gather}
F^{\textrm{\tiny
EHS}}(\mathbf{M}(|\psi\rangle\langle\psi|),\mathbf{M}(|\phi\rangle\langle\phi|))
\geq |\langle\psi|\phi\rangle|.
\end{gather}
Therefore, we only have to show that there exist states
$|\psi\rangle, |\phi\rangle\in\mathcal{H}^S$ and a generalized
measurement $\mathbf{M}$, for which equality is attained.

Let $P(\rho,\sigma)$ and $Q(\rho,\sigma)$ be two joint probability
distributions which achieve the maximum in Eq.~\eqref{EHSfid''} for
the pair of probability distribution $P(\rho)$ and $Q(\rho)$. From
Uhlmann's theorem \cite{Uhl76} we know that for any pair
$(\rho,\sigma)\in\Omega\times\Omega$, there exist purifications
$|\psi_{\rho,\sigma}\rangle^{SB}\in\mathcal{H}^S\otimes\mathcal{H}^B$
and
$|\phi_{\rho,\sigma}\rangle^{SB}\in\mathcal{H}^S\otimes\mathcal{H}^B$
of $\rho$ and $\sigma$, respectively, such that
$F(\rho,\sigma)=\langle
\psi_{\rho,\sigma}|\phi_{\rho,\sigma}\rangle^{SB}$. The second
system $B$ can be chosen to have the same dimension as that of $S$.
Let us introduce a third system with a Hilbert space $\mathcal{H}^E$
of dimension $N^2$, where $N$ is the cardinality of the set
$\Omega$. Let $\{|(\rho,\sigma)\rangle^E\}$,
$(\rho,\sigma)\in\Omega\times\Omega$, be an orthonormal basis of
$\mathcal{H}^E$. From Eq.~\eqref{EHSfid''} one can readily see that
the pure states
\begin{gather}
|P\rangle^{SBE}=\sum_{\rho,\sigma\in\Omega}\sqrt{P(\rho,\sigma)}|\psi_{\rho,\sigma}\rangle^{SB}|(\rho,\sigma)\rangle^E,\\
|Q\rangle^{SBE}=\sum_{\rho,\sigma\in\Omega}\sqrt{Q(\rho,\sigma)}|\phi_{\rho,\sigma}\rangle^{SB}|(\rho,\sigma)\rangle^E,
\end{gather}
by construction satisfy
\begin{gather}
\langle P|Q \rangle^{SBE}=F^{\textrm{\tiny EHS}}(P,Q).
\end{gather}
Notice that there exists a unitary transformation
$U\in\mathcal{B}(\mathcal{H}^S\otimes\mathcal{H}^B\otimes\mathcal{H}^E)$
such that
\begin{gather}
U|\psi\rangle^S|0\rangle^{BE}=|P\rangle^{SBE}, \\
U|\phi\rangle^S|0\rangle^{BE}=|Q\rangle^{SBE},
\end{gather}
where $|0\rangle^{BE}$ is some state in
$\mathcal{H}^B\otimes\mathcal{H}^E$, and $|\psi\rangle^S$ and
$|\phi\rangle^S$ are states in $\mathcal{H}^S$. Since unitary
operations preserve the overlap between states,
\begin{gather}
\langle\psi|\phi\rangle^S=\langle P|Q\rangle^{SBE}=F^{\textrm{\tiny
EHS}}(P,Q).
\end{gather}
But from the states $|P\rangle^{SBE}$ and $|Q\rangle^{SBE}$ we can
obtain the ensembles $\{(P(\rho),\rho)\}$ and $\{(Q(\rho),\rho)\}$,
respectively, by performing a destructive measurement on subsystem
$\mathcal{H}^E$ in the basis $\{|(\rho,\sigma)\rangle^E\}$ and
tracing out subsystem $\mathcal{H}^B$. Therefore, starting from the
two states $|\psi\rangle^S$ and $|\phi\rangle^S$ we can obtain the
ensembles $\{(P(\rho),\rho)\}$ and $\{(Q(\rho),\rho)\}$ by appending
the state $|0\rangle^{BE}$, applying the unitary operation $U$,
measuring in the basis $\{|(\rho,\sigma)\rangle^E\}$ and discarding
system $B$. This operation is equivalent to a generalized
measurement $\mathbf{M}$ on system $S$. This completes the proof.

\section{An ensemble-based interpretation of the square root fidelity}

As we pointed out in Sec.~V, the EHS fidelity can be formulated
without reference to an extended Hilbert space
(Eq.~\eqref{EHSfid''}). In the case when the set $\Omega$ consists
of pure states, the quantity \eqref{EHSfid''} can be written as
\begin{gather}
F^{\textrm{\tiny
EHS}}(P,Q)=\max_{P(\psi,\phi),Q(\psi,\phi)}\sum_{\psi,\phi\in\Omega}\sqrt{P(\psi,\phi)Q(\psi,\phi)}|\langle\psi|\phi\rangle|,
\label{EHSfid''3}
\end{gather}
where optimization is taken over all joint distributions
$P(\psi,\phi)$ with left marginal $P(\psi)$, and $Q(\psi,\phi)$ with
right marginal $Q(\phi)$. Notice that for fixed $P(\psi,\phi)$ and
$Q(\psi,\phi)$, the quantity
$\sum_{\psi,\phi\in\Omega}\sqrt{P(\psi,\phi)Q(\psi,\phi)}|\langle\psi|\phi\rangle|$
can be thought of as a generalization of the Bhattacharyya overlap
between classical probability distributions over the variable
$(\psi,\phi)$, where the overlap $\sqrt{P(\psi,\phi)Q(\psi,\phi)}$
between the probabilities $P(\psi,\phi)$ and $Q(\psi,\phi)$ is
modified by the factor $|\langle\psi|\phi\rangle|$. Heuristically,
we could think that the probabilities of the two distributions are
of a quantum nature, i.e., instead of $P(\psi,\phi)$ and
$Q(\psi,\phi)$ at a given point $(\psi,\phi)$, we have
$P(\psi,\phi)|\psi\rangle\langle\psi|$ and
$Q(\psi,\phi)|\phi\rangle\langle\phi|$, whose overlap is given by
$\sqrt{P(\psi,\phi)Q(\psi,\phi)}|\langle\psi|\phi\rangle|$. Note
that expression \eqref{EHSfid''3} is formulated without any
reference to mixed-state fidelity.

\textbf{Theorem 4.} The square root fidelity
$F(\rho,\sigma)=\textrm{Tr}\sqrt{\sqrt{\sigma}\rho\sqrt{\sigma}}$ is
equal to the maximum of the fidelity \eqref{EHSfid''3} between all
possible pure-state ensembles whose average density matrices are
equal to $\rho$ and $\sigma$, i.e.,
\begin{gather}
F(\rho,\sigma)=\max_{P,Q}F^{\textrm{\tiny EHS}}(P,Q),\label{newint}
\end{gather}
where maximization is taken over all
$P=\{(P(\psi),|\psi\rangle\langle\psi|)\}$ and
$Q=\{(Q(\phi),|\phi\rangle\langle\phi|)\}$, such that
\begin{gather}
\sum_{\psi}P(\psi)|\psi\rangle\langle\psi|=\rho,\\
\sum_{\phi}Q(\phi)|\phi\rangle\langle\phi|=\sigma.
\end{gather}
More succinctly,
\begin{gather}
F(\rho,\sigma)=\mathop{\max_{P(\psi,\phi),Q(\psi,\phi),}}_{\Omega}\sum_{\psi,\phi\in\Omega}\sqrt{P(\psi,\phi)Q(\psi,\phi)}|\langle\psi|\phi\rangle|,\label{newinterp}
\end{gather}
where maximization is taken over all sets of pure states $\Omega$
and joint distributions $P(\psi,\phi)$ and $Q(\psi,\phi)$,
$\psi,\phi\in\Omega$, such that
\begin{gather}
\sum_{\psi,\phi\in\Omega}P(\psi,\phi)|\psi\rangle\langle\psi|=\rho,\\
\sum_{\psi,\phi\in\Omega}Q(\psi,\phi)|\phi\rangle\langle\phi|=\sigma.
\end{gather}

\textbf{Proof.} From the monotonicity of $F^{\textrm{\tiny
EHS}}(P,Q)$ under averaging, it follows that
\begin{gather}
F(\rho,\sigma)\geq\mathop{\max_{P(\psi,\phi),Q(\psi,\phi),}}_{\Omega}\sum_{\psi,\phi\in\Omega}\sqrt{P(\psi,\phi)Q(\psi,\phi)}|\langle\psi|\phi\rangle|.\label{newinterpin}
\end{gather}
To prove that there are pure-state ensembles for which equality is
achieved, we will make use of Uhlmann's theorem \cite{Uhl76}
according to which
\begin{gather}
F(\rho,\sigma)=\max_{|\widetilde{\psi}\rangle,|\widetilde{\phi}\rangle}|\langle\widetilde{\psi}|\widetilde{\phi}\rangle|,\label{Uhlman}
\end{gather}
where maximization is taken over all possible purifications
$|\widetilde{\psi}\rangle$ and $|\widetilde{\phi}\rangle$ of $\rho$
and $\sigma$, respectively. Let $|\widetilde{\psi}_0\rangle$ and
$|\widetilde{\phi}_0\rangle$ be two purifications for which the
maximum in Eq.~\eqref{Uhlman} is attained. Choose an orthonormal
basis $\{|i\rangle\}$ in the auxiliary system needed for the
purification. We can write
\begin{gather}
|\widetilde{\psi}_0\rangle=\sum_i\alpha_i|\psi_i\rangle|i\rangle,\label{alphai}
\end{gather}
\begin{gather}
|\widetilde{\phi}_0\rangle=\sum_i\beta_i|\phi_i\rangle|i\rangle.\label{betai}
\end{gather}
The overlap between these states can be written as
\begin{gather}
|\langle\widetilde{\psi}_0|\widetilde{\phi}_0\rangle|=|\sum_i\alpha_i^*\beta_i\langle\psi_i|\phi_i\rangle|\leq
\sum_i|\alpha_i^*\beta_i\langle\psi_i|\phi_i\rangle|.\label{inequhl}
\end{gather}
Notice that if we change arbitrarily the phases of $\alpha_i$ and
$\beta_i$ in Eqs.~\eqref{alphai} and \eqref{betai}, we obtain valid
(although not necessarily optimal) purifications of $\rho$ and
$\sigma$. If we choose the phases such that each of the quantities
$\alpha_i^*\beta_i\langle\psi_i|\phi_i\rangle$ have the same phase,
then equality in Eq.~\eqref{inequhl} is attained. Therefore, for
optimal purifications we have
\begin{gather}
|\langle\widetilde{\psi}_0|\widetilde{\phi}_0\rangle|=\sum_i|\alpha_i||\beta_i||\langle\psi_i|\phi_i\rangle|.
\end{gather}
Notice that the ensembles
$\{|\alpha_i|^2,|\psi_i\rangle\langle\psi_i|\}$ and
$\{|\beta_i|^2,|\phi_i\rangle\langle\phi_i|\}$ are such that their
averages give rise to $\rho$ and $\sigma$, i.e., they are among
those ensembles over which maximization in Eq.~\eqref{newint} is
taken. But $\sum_i|\alpha_i||\beta_i||\langle\psi_i|\phi_i\rangle|$
is exactly of the form on the right-hand side of
Eq.~\eqref{newinterp}, i.e., equality in Eq.~\eqref{newinterpin} is
attained by $\{|\alpha_i|^2,|\psi_i\rangle\langle\psi_i|\}$ and
$\{|\beta_i|^2,|\phi_i\rangle\langle\phi_i|\}$. This completes the
proof.

Clearly, all interpretations of the fidelity must be equivalent, but
they provide different intuitive ways of understanding the same
quantity. Theorem 4 gives an interpretation based on the pure-state
ensembles from which a mixed state can be prepared by averaging and
thus reflects the common intuition of mixed states as describing
mixtures of pure states.

\section{Distance and fidelity between stochastic quantum operations}

In practice, it often makes sense to ask how close two quantum
processes are. For example, we may want to compare an ideal quantum
operation which we would like to implement, with an imperfect
operation that we are able to implement. Distance measures between
deterministic quantum operations (CPTP maps) have been defined,
e.g., in Ref.~\cite{GLN05}. However, a similar treatment for
stochastic quantum operations (generalized measurements) has been
missing. Stochastic operations are an important tool for quantum
information processing with applications in various areas, such as
quantum control, state estimation, entanglement manipulation, and
error correction, to name a few. Identifying such measures could
thus be very useful.

Before we propose distinguishability measures between stochastic
quantum operations, let us discuss what we mean when we say that two
such operations are different. For the purposes of the present
paper, we will identify a stochastic quantum operation
${\mathbf{M}}$ (or a {generalized measurement}) with an ensemble
$\{(m_i, \bar{\mathcal{M}}_i)\}$, $m_i\geq 0$, of \textit{different}
completely positive measurement superoperators
$\bar{\mathcal{M}}_i(\cdot)=\sum_j\bar{M}_{ij}(\cdot)\bar{M}_{ij}^{\dagger}$
which are \textit{normalized} as
\begin{equation}
\textrm{Tr}(\sum_j\bar{M}_{ij}^{\dagger}\bar{M}_{ij})=d,
\hspace{0.1cm}\forall i,
\end{equation}
and satisfy
\begin{gather}
\sum_{i,j}m_i\bar{M}_{ij}^{\dagger}\bar{M}_{ij}=I.
\end{gather}
The unnormalized measurement superoperators $\mathcal{M}_i$ which
appear in the usual description of a measurement
(Eq.~\eqref{genmeas}) are related to the normalized ones via
\begin{gather}
\bar{\mathcal{M}}_i=\mathcal{M}_i/m_i,\\
m_i=\textrm{Tr}(\sum_jM_{ij}^{\dagger}M_{ij})/d.
\end{gather}

Notice that the weights $m_i$ satisfy $\sum_i m_i=1$, i.e., they can
be thought of as `probabilities' and $\{(m_i,
\bar{\mathcal{M}}_i)\}$ can be thought of as a `probabilistic'
ensemble of normalized superoperators $\bar{\mathcal{M}}_i$. Note,
however, that $m_i$ are \textit{not} equal to the probabilities of
the measurement outcomes which generally depend on the input state
$\rho$ and are given by
$p_i=m_i\textrm{Tr}(\bar{\mathcal{M}}_i(\rho))$.

The reason why we associate different outcomes with
\textit{normalized} superoperators is that we want our description
to explicitly emphasize the fact that measurement outcomes whose
measurement superoperators differ from each other only by a factor
are not considered different. This is because for us a generalized
measurement is not a characterization of a particular physical
device (which could produce classical readings not necessarily
related to the quantum system of interest), but the most abstract
characterization of an operation on the state of the quantum system,
which includes information extraction as well as state
transformation. Clearly, two measurement superoperators which differ
from each other by a factor do not provide any different information
about the state of the system prior to the measurement (according to
Bayes's rule) nor give rise to different post-measurement states.
Note that when we say that two normalized measurement superoperators
$\bar{\mathcal{M}}_i(\cdot)=\sum_{j}\bar{M}_{ij}(\cdot)\bar{M}_{ij}^{\dagger}$
and
$\bar{\mathcal{N}}_k(\cdot)=\sum_{k}\bar{N}_{kl}(\cdot)\bar{N}_{kl}^{\dagger}$
are the same, we compare them as completely positive maps, i.e.,
irrespectively of their operator-sum representations. In other
words, $\bar{\mathcal{M}}_i= \bar{\mathcal{N}}_k$ if and only if
there exists a unitary matrix with components $U_{jl}$, such that
$\bar{M}_{ij}=\sum_l U_{jl}\bar{N}_{kl}$, $\forall j$
\cite{NieChu00}. In that sense, if two measurements are described by
identical ensembles of normalized measurement superoperators, they
are the same measurement. Conversely, if two measurements are
described by different ensembles of normalized measurement
superoperators, they should be considered different because they
either give rise to different output ensembles for some input, or
provide different information about the input state, or both.
Therefore, we will specify a generalized measurement $\mathbf{M}$ by
the correspondence
\begin{gather}
\mathbf{M}\leftrightarrow
\{(m_i,\bar{\mathcal{M}}_i)\}.\label{measidentity}
\end{gather}

There are many possible ways in which one can define distance
between two quantum operations. The following desirable properties
for a distance $D$ between deterministic quantum operations
$\mathcal{E}$ and $\mathcal{F}$ were pointed out and discussed in
Ref.~\cite{GLN05}: \textbf{(1) \textit{metric}}---the measure should
be positive, symmetric, satisfy the triangle inequality, and vanish
if and only if the two operations are identical; \textbf{(2)
\textit{computability}}---it should be possible to evaluate $D$ in a
direct manner; \textbf{(3) \textit{measurability}}---there should be
an achievable experimental procedure for determining $D$;
\textbf{(4) \textit{physical interpretation}}---the distance should
have a well motivated physical interpretation; \textbf{(5)
\textit{stability}}---$D(\mathcal{I}\otimes\mathcal{E},\mathcal{I}\otimes\mathcal{F})=D(\mathcal{E},\mathcal{F})$,
which means that unrelated physical systems should not affect the
value of $D$; \textbf{(6)
\textit{chaining}}---$D(\mathcal{E}_2\otimes\mathcal{E}_1,\mathcal{F}_2\otimes\mathcal{F}_1)\leq
D(\mathcal{E}_1,\mathcal{F}_1)+D(\mathcal{E}_2,\mathcal{F}_2)$,
i.e., for a process composed of several steps, the total error
should be less than the sum of the errors in the individual steps.
We will consider the same requirements for a distance between
stochastic quantum operations. In the deterministic case, in view of
the above desiderata, two main approaches to distinguishing quantum
operations stand out---comparison based on the Jamio{\l}kowski
isomorphism and worst-case comparison. We will adopt the same
approaches here.

Since many of the properties for the following measures and their
proofs are similar to those discussed in Ref.~\cite{GLN05}, we will
only comment on them briefly. In what follows, we will use $D$ and
$F$ to denote distance and fidelity between ensembles, which can be
either of the Kantorovich or of the EHS type. We will use
$\mathbf{M}(\rho)$ to denote the ensemble of output states that
results from the action of a stochastic quantum operation
$\mathbf{M}$ on an input state $\rho$.

\subsection{Measures based on the Jamio{\l}kowski isomorphism}

The Jamio{\l}kowski isomorphism \cite{Jamiolkowski} is a one-to-one
correspondence between completely positive maps (superoperators)
$\mathcal{M}:\mathcal{B}(\mathcal{H}^S)\rightarrow
\mathcal{B}(\mathcal{H}^S)$ and positive operators
$\rho_{\mathcal{M}}\in\mathcal{B}(\mathcal{H}^A\otimes\mathcal{H}^S)$,
where $\textrm{dim}(\mathcal{H}^A)=\textrm{dim}(\mathcal{H}^S)=d$.
The correspondence is established via
\begin{equation}
\rho_{\mathcal{M}}=\mathcal{I}^A\otimes\mathcal{M}^S(|\Phi\rangle\langle\Phi|^{AS}),
\end{equation}
where $|\Phi\rangle^{AS}=\sum_j|j\rangle^A|j\rangle^S/\sqrt{d}$ is a
maximally entangled state on $\mathcal{H}^A\otimes\mathcal{H}^S$
(here $\{|j\rangle^A\}$ and $\{|j\rangle^S\}$ are orthonormal bases
of $\mathcal{H}^A$ and $\mathcal{H}^S$, respectively). Notice that
if the completely positive map $\mathcal{M}$ is trace-preserving,
the corresponding positive operator $\rho_{\mathcal{M}}$ is a
density matrix, i.e, $\textrm{Tr}(\rho_{\mathcal{M}})=1$. However,
not all density matrices on $\mathcal{H}^A\otimes\mathcal{H}^S$
correspond to CPTP maps, but only those whose reduced density matrix
on subsystem $A$ is the maximally mixed state $I/d$. It is easy to
see that most generally, a density matrix on
$\mathcal{H}^A\otimes\mathcal{H}^S$ corresponds to a completely
positive superoperator
$\bar{\mathcal{M}}(\cdot)=\sum_i\bar{M}_i(\cdot)\bar{M}_i^{\dagger}$,
which is normalized as
\begin{equation}
\textrm{Tr}(\sum_i\bar{M}_i^{\dagger}\bar{M}_i)=d.\label{normalized}
\end{equation}
The reverse is also true---every completely positive superoperator
on $\mathcal{B}(\mathcal{H}^S)$, which satisfies
Eq.~\eqref{normalized}, gives rise to a density matrix when applied
to $|\Phi\rangle\langle\Phi|^{AS}$. We therefore see that there is
an isomorphism
\begin{gather}
\{(m_i, \bar{\mathcal{M}}_i)\}\leftrightarrow
\{(m_i,\rho_{\bar{\mathcal{M}}_i})\}
\end{gather}
between ensembles of normalized completely positive superoperators
and ensembles of density matrices. Of course, just like not every
completely positive map corresponding to a density matrix is trace
preserving, not every ensemble $\{(m_i, \mathcal{M}_i)\}$,
$\sum_im_i=1$, forms a generalized measurement
($\sum_{i,j}m_i\bar{M}_{ij}^{\dagger}\bar{M}_{ij}=I$). But since the
reverse is true, we can use the isomorphism to define distance and
fidelity between generalized measurements through the distance and
fidelity between ensembles of states.

\textbf{Definition 7 (Distance between generalized measurements
based on the Jamio{\l}kowski isomorphism).} Let $\mathbf{M}$ and
$\mathbf{N}$ be two generalized measurements acting on
$\mathcal{B}(\mathcal{H}^S)$. Then,
\begin{gather}
D_{\textrm{iso}}(\mathbf{M},\mathbf{N})\equiv\notag\\
D\left(\mathcal{I}^A\otimes\mathbf{M}^S(|\Phi\rangle\langle\Phi|^{AS}),
\mathcal{I}^A\otimes\mathbf{N}^S(|\Phi\rangle\langle\Phi|^{AS})\right),\label{Diso}
\end{gather}
where $\mathcal{I}^A\otimes\mathbf{M}^S$ and
$\mathcal{I}^A\otimes\mathbf{N}^S$ denote the generalized
measurements $\mathbf{M}$ and $\mathbf{N}$ applied locally on
subsystem $S$ and  $|\Phi\rangle^{AS}=\sum_j|j\rangle^A|j\rangle^S/\sqrt{d}$ is a
maximally entangled state on $\mathcal{H}^A\otimes\mathcal{H}^S$.

\textbf{Property 1 (Metric).} It follows from the metric properties
of $D$.

\textbf{Property 2 (Computability).} It follows from the
computability of $D$ which is either a linear program (in the
Kantorovich case) or a convex-optimization problem (in the EHS
case).

\textbf{Property 3 (Measurability).} As in the deterministic case,
$D_{\textrm{iso}}$ can be determined by doing full process
tomography \cite{QPT1, QPT2}.

\textbf{Property 4 (Physical interpretation).} In addition to the
obvious meaning of $D_{\textrm{iso}}$ following from its definition,
it was pointed out in Ref.~\cite{GLN05} that in the deterministic
case, $D_{\textrm{iso}}(\mathcal{E},\mathcal{F})\geq
\frac{1}{d}\sum_x\Delta (\mathcal{E}(|x\rangle\langle x|),
\mathcal{F}(|x\rangle\langle x|))$, where the sum is over a set of
orthonormal basis states $|x\rangle$ which can be thought of as the
different instances of a computational problem. In a similar manner,
it can be seen that $D_{\textrm{iso}}(\mathbf{M},\mathbf{N})\geq
\frac{1}{d}\sum_x\Delta (\mathbf{M}(|x\rangle\langle x|),
\mathbf{N}(|x\rangle\langle x|))$.

\textbf{Property 5 (Stability).} It follows from the stability of
$D$.

\textbf{Property 6 (Chaining).} The proof of this property assumes
monotonicity of $D$ under generalized measurements and therefore it
holds for the EHS distance. Similarly to the deterministic case
\cite{GLN05}, it can be shown that $D_{\textrm{iso}}$ satisfies
$D_{\textrm{iso}}(\mathbf{M}_2\circ\mathbf{M}_1,\mathbf{N}_2\circ\mathbf{N}_1)\leq
D_{\textrm{iso}}(\mathbf{M}_2,\mathbf{N}_2)+D_{\textrm{iso}}(\mathbf{M}_1,\mathbf{N}_1)$,
provided that $\mathbf{N}_1$ is a \textit{unital measurement}, i.e.,
$\sum_jn_{1j}\bar{\mathcal{N}}_{1j}(I)=I$, where
$\{(n_{1j},\bar{\mathcal{N}}_{1j})\}$ is the ensemble of normalized
measurement superoperators corresponding to $\mathbf{N}_1$.

\textbf{Definition 8 (Fidelity between generalized measurements
based on the Jamio{\l}kowski isomorphism).} Let $\mathbf{M}$ and
$\mathbf{N}$ be two generalized measurements acting on
$\mathcal{B}(\mathcal{H}^S)$. Then,
\begin{gather}
F_{\textrm{iso}}(\mathbf{M},\mathbf{N})\equiv\notag\\
F\left(\mathcal{I}^A\otimes\mathbf{M}^S(|\Phi\rangle\langle\Phi|^{AS}),
\mathcal{I}^A\otimes\mathbf{N}^S(|\Phi\rangle\langle\Phi|^{AS})\right).\label{Fiso}
\end{gather}

The fidelity satisfies similar properties to those of the distance,
except for the triangle inequality.

\subsection{Measures based on worst-case comparison}

\textbf{Definition 9 (Distance between generalized measurements
based on the worst case).} Let $\mathbf{M}$ and $\mathbf{N}$ be two
generalized measurements acting on $\mathcal{B}(\mathcal{H}^S)$,
$\textrm{dim}(\mathcal{H}^S)=d$. Introduce an ancillary system $A$
with a Hilbert space $\mathcal{H}^A$,
$\textrm{dim}(\mathcal{H}^A)=d$. Then,
\begin{gather}
D_{\max}(\mathbf{M},\mathbf{N})\equiv\notag\\
\max_{|\psi\rangle}D\left(\mathcal{I}^A\otimes\mathbf{M}^S(|\psi\rangle\langle\psi|),
\mathcal{I}^A\otimes\mathbf{N}^S(|\psi\rangle\langle\psi|)\right),\label{Dworst}
\end{gather}
where maximum is taken over all $|\psi\rangle\in\mathcal{H}^A\otimes\mathcal{H}^S$.

The definition is based on a maximization over states in an extended
Hilbert space in order to guarantee stability of the distance, as it
is known that without this extension even the analogously defined
distance between CPTP maps based on the trace distance is not stable
\cite{Aharonov98}. Note that this definition takes maximum over
pure-state inputs. As we saw in Sec.~IV.E, a generalized measurement
can be defined to act on ensembles of mixed states so that it most
generally transforms ensembles of density matrices into ensembles of
density matrices. However, it is easy to see that one cannot obtain
a larger value by maximizing over mixed states or ensembles of mixed
states. This follows from the joint convexity of $D$ with respect to
ensembles and from the joint convexity of $\Delta$ with respect to
mixed states.

\textbf{Property 1 (Metric).} It follows from the metric properties
of $D$. (The fact that the distance between different measurements
is non-zero follows from the fact that for the input state
$|\Phi\rangle^{SA}$, different measurements yield different output
ensembles.)

\textbf{Property 2 (Computability).} We already pointed out that the
measure $D$ for any particular pair of ensembles is computable. In
Ref.~\cite{GLN05} it was argued that in the case of deterministic
operations, the corresponding optimization in Eq.~\eqref{Dworst} is
a convex optimization problem and therefore computable. By a similar
argument it can be seen that for stochastic quantum operations,
finding the maximum in Eq.~\eqref{Dworst} is also a convex
optimization problem.

\textbf{Property 3 (Measurability).} Here too, the value of
$D_{\max}$ can be determined using quantum process tomography
\cite{QPT1, QPT2}.

\textbf{Property 4 (Physical interpretation).} The physical meaning
of $D_{\max}$ follows directly from its definition and the physical
meaning of $D$.

\textbf{Property 5 (Stability).} The proof goes along the same lines
as the proof for the deterministic case (Ref.~\cite{GLN05})---all
one needs to show is that the quantity \eqref{Dworst} is independent
of the dimension of system $A$, as long as this dimension is greater
than or equal to $d$. This follows from the observation that an
input state which achieves the maximum in Eq.~\eqref{Dworst} can
have at most $d$ Schmidt coefficients, which implies that there is a
subspace of $\mathcal{H}^A$ with dimension $d$ such that the maximum
can be achieved by maximization inside that subspace.

\textbf{Property 6 (Chaining).} The chaining property follows from
the triangle inequality and the monotonicity of $D$ under
generalized measurements, i.e., it holds for the EHS distance.

\textbf{Definition 10 (Fidelity between generalized measurements
based on the worst case).} Let $\mathbf{M}$ and $\mathbf{N}$ be two
generalized measurements acting on $\mathcal{B}(\mathcal{H}^S)$,
$\textrm{dim}(\mathcal{H}^S)=d$. Introduce an ancillary system $A$
with a Hilbert space $\mathcal{H}^A$,
$\textrm{dim}(\mathcal{H}^A)=d$. Then,
\begin{gather}
F_{\min}(\mathbf{M},\mathbf{N})\equiv\notag\\
\min_{|\psi\rangle}F\left(\mathcal{I}^A\otimes\mathbf{M}^S(|\psi\rangle\langle\psi|),
\mathcal{I}^A\otimes\mathbf{N}^S(|\psi\rangle\langle\psi|)\right),\label{Fworst}
\end{gather}
where minimum is taken over all
$|\psi\rangle\in\mathcal{H}^A\otimes\mathcal{H}^S$.

The fidelity $F_{\min}$ satisfies properties analogous to those of
$D_{\max}$ with the exception of the triangle inequality.

\subsection{Distance and fidelity between POVMs}

A very useful concept in quantum information is that of a positive
operator-valued measure (POVM)---a set of positive operators
$\{E_i\}$, $E_i>0$, which sum up to the identity, $\sum_iE_i=I$. A
POVM provides the most general description of a quantum measurement
in situations where one is not interested in the post-measurement
state. In terms of the measurement superoperators $\mathcal{M}_i$,
the POVM elements are given by $E_i=\sum_jM_{ij}^{\dagger}M_{ij}$,
i.e., there is no unique generalized measurement which corresponds
to a given POVM. Similarly to the case of generalized measurements,
we can express a POVM as an ensemble of \textit{normalized} POVM
elements, $\{(m_i,\bar{E}_i)\}$, where $m_i=\textrm{Tr}(E_i)/d$,
$\bar{E}_i=E_i/m_i$. Notice that the operators
\begin{equation}
\rho_{\bar{E}_i}\equiv\bar{E}_i/d
\end{equation}
are density matrices ($\textrm{Tr}(\rho_{\bar{E}_i})=1$), i.e.,
there is a one-to-one correspondence between POVMs and ensembles of
density matrices $\{(m_i,\rho_{\bar{E}_i})\}$ which satisfy
$\sum_im_i\rho_{\bar{E}_i}=I/d$. Therefore, we can compare POVMs
directly using the distinguishability measures between ensembles of
states.

\textbf{Definition 11 (Distance between POVMs).} Let $\{E_i\}$ and
$\{G_j\}$ be two POVMs and let $P_E\equiv\{(m_i,\rho_{\bar{E}_i})\}$
($m_i=\textrm{Tr}(E_i)/d$, $\rho_{\bar{E}_i}=E_i/(m_id)$) and
$P_G\equiv\{(n_j,\rho_{\bar{G}_j})\}$ ($n_j=\textrm{Tr}(G_j)/d$,
$\rho_{\bar{G}_j}=G_j/(n_jd)$) be the ensembles of density matrices
that correspond to them. Then,
\begin{gather}
D_{\textrm{POVM}}(\{E_i\}, \{G_j\} )\equiv D (P_E,P_G).
\end{gather}

\textbf{Definition 12 (Fidelity between POVMs).} Let $\{E_i\}$ and
$\{G_j\}$ be two POVMs and let $P_E\equiv\{(m_i,\rho_{\bar{E}_i})\}$
($m_i=\textrm{Tr}(E_i)/d$, $\rho_{\bar{E}_i}=E_i/(m_id)$) and
$P_G\equiv\{(n_j,\rho_{\bar{G}_j})\}$ ($n_j=\textrm{Tr}(G_j)/d$,
$\rho_{\bar{G}_j}=G_j/(n_jd)$) be the ensembles of density matrices
that correspond to them. Then,
\begin{gather}
F_{\textrm{POVM}}(\{E_i\}, \{G_j\} )\equiv F (P_E,P_G).
\end{gather}

The properties of these measures can be obtained in a
straightforward manner from the properties of the distance and
fidelity between states. We only remark that the ensemble of states
$P_E=\{(m_i,\rho_{\bar{E}_i})\}$ that corresponds to a given POVM
$\{E_i\}$ has the following operational meaning---it is the ensemble
of states of system $A$ that we obtain from the maximally entangled
state $|\Phi\rangle^{AS}$ if we perform the destructive POVM
$\{E_i\}$ on subsystem $S$,
\begin{gather}
|\Phi\rangle\langle\Phi|^{AS}\rightarrow\rho_{\bar{E}_i}^A=\textrm{Tr}_S(I^A\otimes
E^S_i|\Phi\rangle\langle\Phi|^{AS})/m_i,\\
\textrm{with probability} \hspace{0.2cm}m_i=\textrm{Tr}(I^A\otimes
E^S_i|\Phi\rangle\langle\Phi|^{AS})=\textrm{Tr}(E^S_i).\notag
\end{gather}

As quantum detector tomography is now within the reach of
experimental technology \cite{TQD}, it becomes relevant to ask how
much a real quantum detector differs from an ideal one. The distance
and fidelity between POVMs introduced in this section provide
rigorous means of quantifying such difference.

\section{Conclusion}

In this paper we defined measures of distance and fidelity between
probabilistic ensembles of quantum states and used them to define
measures of distance and fidelity between stochastic quantum
operations. We proposed two types of measures between ensembles.

The first one is based on the ability of one ensemble to mimic
another and leads to measures of a Kantorovich type, which appear in
the context of optimal transportation and can be computed as linear
programs. However, when based on the trace distance or the square
root fidelity, these measures are not monotonic under generalized
measurements. We derived necessary and sufficient conditions that
the basic measures of distance and fidelity between states have to
satisfy in order for the corresponding Kantorovich distance and
fidelity to be monotonic under measurements (Theorem 2). An
interesting open problem is whether measures of distance and
fidelity that satisfy the conditions of Theorem 2 exist.

The second type of measures is based on the notion of an
extended-Hilbert-space (EHS) representation of an ensemble. We
showed that for every ensemble there is a class of valid EHS
representations and defined the measures as a minimum (maximum) of
the trace distance (square root fidelity) between all EHS
representations of the ensembles being compared. These measures,
which are monotonic under generalized measurements, can be computed
as convex optimization problems. We provided operational
interpretations for the measures and showed that the EHS fidelity is
an upper bound of the overlap between all possible pure-state inputs
that could give rise to the two ensembles being compared under the
action of a stochastic quantum operation. We also used the EHS
fidelity between ensembles to provide a novel interpretation of the
square root fidelity between density matrices. We showed that the
square root fidelity is equal to the minimum fidelity between all
possible pure-state ensembles from which the density matrices being
compared can be obtained.

An interesting question is whether any of the measures between
ensembles that we introduced can be used to define a Riemannian
metric on the space of ensembles, which endows the space with
geometrical notions such as volume or geodesics. Clearly, the
measures based on the trace distance would not induce a Riemannian
metric because the trace distance is known not to be Riemannian
\cite{SomZyc03}. The Kantorovich fidelity is not a good candidate
either because in one of the classical limits it reduces to a
function of the Kolmogorov distance. However, we can define an EHS
distance which is a generalization of the Bures distance between
density matrices, $B^{\textrm{\tiny
EHS}}(P,Q)=\sqrt{1-F^{\textrm{\tiny EHS}}(P,Q)}$, or an EHS angle
which is a generalization of the Bures angle, $A^{\textrm{\tiny
EHS}}(P,Q)=\arccos{F^{\textrm{\tiny EHS}}(P,Q)}$. It is known that
the Bures distance and angle induce a Riemannian metric, and it
would be interesting to see if their EHS generalizations induce such
a metric on the space of ensembles. This problem is left open for
future investigation.

Finally, based on the measures between ensembles, we defined two
types of distinguishability measures between generalized
measurements. The first one is based on the Jamio{\l}kowski
isomorphism and the second one on the worst-case comparison. These
measures are generalizations of the distance and fidelity between
CPTP maps proposed in Ref.~\cite{GLN05} and similarly to them
satisfy the desiderata outlined in Ref.~\cite{GLN05}. One of the
desired properties---the chaining property---is satisfied only by
the measures based on the EHS distance and fidelity since this
property requires monotonicity under generalized measurements of the
corresponding measures between ensembles of states. In addition to
generalized measurements, we also defined distinguishability
measures between POVMs. The proposed measures may find various
applications as they provide a rigorous general tool for assessing
the performance of non-destructive and destructive measurement
schemes.

\appendix

\section{Continuity of the average of a
continuous function with respect to the Kantorovich distance}

\label{app:0}

Let $h(\rho)$ be a bounded function which is continuous with respect
to the distance $\Delta$, i.e., for every $\delta>0$, there exists
$\varepsilon>0$, such that for all $\rho$ and $\sigma$ for which
\begin{gather}
\Delta(\rho,\sigma)\leq\varepsilon,
\end{gather}
we have
\begin{gather}
|h(\rho)-h(\sigma)|\leq\frac{1}{2}\delta.\label{con}
\end{gather}
(The factor of $\frac{1}{2}$ in front of $\delta$ is chosen for
convenience.) Let $\overline{h}_{P}$ denote the average of the
function $h(\rho)$ over the ensemble $P(\rho)$, $\rho\in\Omega$,
\begin{equation}
\overline{h}_P=\underset{\rho\in\Omega}{\sum}P(\rho)h(\rho).
\end{equation}
We will prove that for every $\delta>0$, there exists
$\varepsilon'>0$, such that for all $P,Q\in\mathcal{P}_{\Omega}$ for
which
\begin{gather}
D^K(P,Q)\leq \varepsilon',
\end{gather}
we have
\begin{equation}
|\overline{h}_P-\overline{h}_Q|\leq \delta.\label{con2}
\end{equation}

Assume that $D^K(P,Q)\leq \varepsilon'$. Let $\Pi(\rho,\sigma)$ be a
joint distribution for which the minimum in the definition
\eqref{definition} of $D^K(P,Q)$ is achieved, i.e.,
\begin{gather}
D^K(P,Q)=\sum_{\rho,\sigma\in\Omega}\Pi(\rho,\sigma)\Delta(\rho,\sigma)\leq\varepsilon'.\label{edno}
\end{gather}
Define the sets $\Omega_{>\varepsilon}$ and
$\Omega_{\leq\varepsilon}$ as the sets of all pairs of states
$(\rho,\sigma)$ for which $\Delta(\rho,\sigma)>\varepsilon$ and
$\Delta(\rho,\sigma)\leq\varepsilon$, respectively. The sum in
Eq.~\eqref{edno} can then be split in two sums,
\begin{gather}
\sum_{\Omega_{>\varepsilon}}\Pi(\rho,\sigma)\Delta(\rho,\sigma)+\sum_{\Omega_{\leq\varepsilon}}\Pi(\rho,\sigma)\Delta(\rho,\sigma)\leq
\varepsilon'.
\end{gather}
The first sum obviously can be bounded as follows,
\begin{gather}
\sum_{\Omega_{>\varepsilon}}\Pi(\rho,\sigma)\varepsilon\leq\sum_{\Omega_{>\varepsilon}}\Pi(\rho,\sigma)\Delta(\rho,\sigma)\leq\varepsilon',
\end{gather}
which implies that
\begin{gather}
\sum_{\Omega_{>\varepsilon}}\Pi(\rho,\sigma)\leq
\frac{\varepsilon'}{\varepsilon}.\label{con3}
\end{gather}
On the other hand, we have
\begin{gather}
|\overline{h}_P-\overline{h}_Q|=|\underset{\rho\in\Omega}{\sum}P(\rho)h(\rho) -\underset{\sigma\in\Omega}{\sum}Q(\sigma)h(\sigma)|\notag\\
=|\sum_{\rho,\sigma\in \Omega}\Pi(\rho,\sigma)h(\rho) -\sum_{\rho,\sigma\in \Omega}\Pi(\rho,\sigma)h(\sigma)|\notag\\
\leq \sum_{\rho,\sigma\in
\Omega}\Pi(\rho,\sigma)|h(\rho)-h(\sigma)|=\notag\\
\sum_{\Omega_{>\varepsilon}}\Pi(\rho,\sigma)|h(\rho)-h(\sigma)|+\sum_{\Omega_{\leq\varepsilon}}\Pi(\rho,\sigma)|h(\rho)-h(\sigma)|
.\label{conti}
\end{gather}
Since $h(\rho)$ is bounded, there exists a constant $h_{\mathrm{max}}>0$ such
that $|h(\rho)-h(\sigma)|\leq h_{\mathrm{max}}$ for all $\rho$ and $\sigma$.
Using this fact, together with Eq.~\eqref{con3} and the assumption
that for all $(\rho,\sigma)\in\Omega_{\leq\varepsilon}$,
$|h(\rho)-h(\sigma)|\leq\frac{1}{2}\delta$, we can upper bound the
last line in Eq.~\eqref{conti} as follows:
\begin{gather}
\sum_{\Omega_{>\varepsilon}}\Pi(\rho,\sigma)|h(\rho)-h(\sigma)|+\sum_{\Omega_{\leq\varepsilon}}\Pi(\rho,\sigma)|h(\rho)-h(\sigma)|\leq\notag\\
\frac{\varepsilon'}{\varepsilon}h_{\mathrm{max}}
+\sum_{\Omega_{\leq\varepsilon}}\Pi(\rho,\sigma)\frac{1}{2}\delta\leq\notag\\\frac{\varepsilon'}{\varepsilon}h_{\mathrm{max}}+\frac{1}{2}\delta.
\end{gather}
Therefore, we see that by choosing
\begin{equation}
\varepsilon'\leq\frac{\delta\varepsilon}{2h_{\mathrm{max}}},
\end{equation}
we obtain
\begin{gather}
|\overline{h}_P-\overline{h}_Q|\leq \delta.
\end{gather}
Since $\delta$ was arbitrarily chosen, the property follows.

\section{Non-monotonicity under generalized measurements of the
Kantorovich measures}

\label{app:A}

To show that the Kantorovich distance is not monotonic under
measurements, let us look at a particular example. Consider the case
of two singleton ensembles consisting of the states
$\sum_ip_i\rho_i\otimes|i\rangle\langle i|$ and
$\sum_iq_i\sigma_i\otimes|i\rangle\langle i|$, respectively, where
the states $\{|i\rangle \} $ are an orthonormal set, $\langle i |
j\rangle = \delta_{ij}$. Imagine that we apply a nondestructive
projective measurement on the second subsystem in the basis
$\{|i\rangle \} $. This measurement yields the ensembles
$\{(p_i,\rho_i\otimes|i\rangle\langle i| )\}$ and $\{(
q_i,\sigma_i\otimes|i\rangle\langle i| )\}$, which we will denote by
$p$ and $q$ for short. Observe that the Kantorovich distance between
the resulting ensembles, as defined in Eq.~\eqref{definition}, is
equal to
\begin{gather}
D^K(p,q)=\frac{1}{2}\sum_i (\textrm{min}(p_i, q_i) \parallel \rho_i
- \sigma_i
\parallel + |p_i-q_i|).\label{DKopt}
\end{gather}
This follows from the observation that for any joint probability
distribution $\Pi(\rho_i\otimes|i\rangle\langle i|,
\sigma_j\otimes|j\rangle\langle j|)$, the quantity in
Eq.~\eqref{quantity} reads
\begin{gather}
D_{\Pi}(p,q)=\frac{1}{2}\sum_i \Pi( \rho_i\otimes|i\rangle\langle
i|, \sigma_i\otimes|i\rangle\langle i|)\parallel \rho_i-\sigma_i
\parallel\notag\\+ \sum_{i\neq j}\Pi(\rho_i\otimes|i\rangle\langle i|, \sigma_j\otimes|j\rangle\langle
j|)
\end{gather}
because $\langle i | j\rangle = \delta_{ij}$. Since $ \sum_i \Pi(
\rho_i\otimes|i\rangle\langle i|, \sigma_i\otimes|i\rangle\langle
i|)+ \sum_{i\neq j}\Pi(\rho_i\otimes|i\rangle\langle i|,
\sigma_j\otimes|j\rangle\langle j|)=1$, and $\parallel
\rho_i-\sigma_i
\parallel\leq 1$, if each of the terms $\Pi( \rho_i\otimes|i\rangle\langle i|,
\sigma_i\otimes|i\rangle\langle i|)$ is equal to its maximal
possible value consistent with the marginal conditions, then the
value of $D_{\Pi}(p,q)$ would be minimal and it would be equal to
the Kantorovich distance $D^K(p,q)$. The maximum possible value of
$\Pi( \rho_i\otimes|i\rangle\langle i|,
\sigma_i\otimes|i\rangle\langle i|)$ consistent with the marginal
probability distributions is $\textrm{min}(p_i,q_i)$ because if,
say, $\textrm{min}(p_i,q_i)=p_i$ and $\Pi(
\rho_i\otimes|i\rangle\langle i|, \sigma_i\otimes|i\rangle\langle
i|)>p_i$, then $\sum_j \Pi( \rho_i\otimes|i\rangle\langle i|,
\sigma_j\otimes|j\rangle\langle j|)$ would be strictly larger than
$p_i$, while by definition it has to be equal to $p_i$. Each of
these values is achievable because there exist joint probability
distributions $\Pi( \rho_i\otimes|i\rangle\langle i|,
\sigma_j\otimes|j\rangle\langle j|)$ with the correct marginals that
satisfy
\begin{gather}
\Pi( \rho_i\otimes|i\rangle\langle i|,
\sigma_i\otimes|i\rangle\langle i|)=\textrm{min}(p_i,q_i),
\hspace{0.1cm} \forall i.\label{diagonalT}
\end{gather}
The latter can be seen from the fact that
$\Pi(\rho_i\otimes|i\rangle\langle i|,
\sigma_j\otimes|j\rangle\langle j|)$ describes a transportation plan
which tells us what probability weights taken from $p_i$ and $q_j$
come together as we transport one distribution on top of the other.
The condition $\Pi( \rho_i\otimes|i\rangle\langle i|,
\sigma_i\otimes|i\rangle\langle i|)=\textrm{min}(p_i,q_i)$ simply
specifies how to pair certain parts of the two distributions, each
having a total weight of $\sum_i \textrm{min}(p_i,q_i)$. Since the
remaining parts of the two distributions have equal weights, $
1-\sum_i \textrm{min}(p_i,q_i)$, there certainly exists a
transportation plan according to which one can be mapped on top of
the other. Therefore, the Kantorovich distance between $p$ and $q$
is given by Eq.~\eqref{DKopt}.

However, the Kantorovich distance between the original singleton
ensembles is equal to the trace distance between the two states,
\begin{gather}
\frac{1}{2}\parallel \sum_i p_i\sigma_i\otimes |i\rangle\langle i| -
\sum_j q_j\sigma_j\otimes |j\rangle\langle j|
\parallel \notag\\
=\frac{1}{2}\sum_i
\parallel p_i\rho_i- q_i\sigma_i\parallel.\label{trD}
\end{gather}
Assume that for a given $i$, $\textrm{min}(p_i,q_i)=p_i$. We can
write
\begin{gather}
\parallel p_i\rho_i- q_i\sigma_i\parallel=p_i\parallel \rho_i -
\frac{q_i}{p_i}\sigma_i
\parallel.
\end{gather}
But from the triangle inequality we have
\begin{gather}
\parallel \rho_i - \frac{q_i}{p_i}\sigma_i
\parallel\leq \parallel \rho_i - \sigma_i
\parallel+ \parallel \sigma_i - \frac{q_i}{p_i}\sigma_i
\parallel \notag\\
=\parallel \rho_i - \sigma_i
\parallel+ (\frac{q_i}{p_i}-1),\label{ineqtri}
\end{gather}
i.e.,
\begin{gather}
\parallel p_i\rho_i- q_i\sigma_i\parallel\leq p_i (\parallel \rho_i - \sigma_i
\parallel+ (\frac{q_i}{p_i}-1))\notag\\
=p_i \parallel \rho_i - \sigma_i
\parallel+ ({q_i}-{p_i})\notag\\
=\textrm{min}(p_i,q_i)\parallel \rho_i - \sigma_i
\parallel + |p_i-q_i|.\label{ineq}
\end{gather}
Since we arbitrarily assumed which is the smaller of the two values
$p_i$ and $q_i$, the inequality \eqref{ineq} must hold for every
$i$. Comparing Eq.~\eqref{DKopt} and Eq.~\eqref{trD}, we see that
\begin{equation}
\frac{1}{2}\parallel \sum_i p_i\sigma_i\otimes |i\rangle\langle i| -
\sum_j q_j\sigma_j\otimes |j\rangle\langle j|
\parallel \leq
D^K(p,q).\label{lastineq}
\end{equation}
For most choices of $\rho_i$ and $\sigma_i$, the inequality
\eqref{lastineq} is strict since the triangle inequality used in
Eq.~\eqref{ineqtri} is generally strict. Thus we see that the
Kantorovich distance is not monotonically decreasing under
measurements. Obviously it is not monotonically increasing either
because it decreases under CPTP maps (Property 6, Sec.~IV.B).

For the Kantorovich fidelity, we already observed that its values in
the two classical limits are not the same: the fidelity between
the two singleton distributions consisting of states of the form
$\rho=\sum_i p_i |i\rangle\langle i|$ and $\sigma=\sum_i q_i
|i\rangle\langle i|$, where $\{|i\rangle\}$ is an orthonormal set,
is equal to
\begin{gather}
F^K(P,Q)=F(\rho, \sigma)=\underset{i}{\sum}\sqrt{p_iq_i},
\end{gather}
whereas the fidelity between the ensembles $ \{(p_i,
|i\rangle\langle i|)\}$ and $\{(q_i, |i\rangle\langle i|)\}$ is
equal to $\sum_i\textrm{min}(p_i,q_i)$, which is strictly smaller
than $F(\rho,\sigma)$ unless $p_i=q_i$, $\forall i$. The latter pair
of ensembles are exactly the ensembles that result from a
measurement in the $\{|i\rangle \}$ basis applied to the states
$\rho$ and $\sigma$. Therefore, the Kantorovich fidelity can
decrease under measurements. Clearly, it is not always decreasing
because it increases under CPTP maps (Property 4, Sec.~IV.C).

We can now see that the difference of the values of the Kantorovich
fidelity in the two `classical' limits discussed earlier can be
linked to its lack of monotonicity under measurements. Obviously,
through a projective measurement and averaging, we can go back and
forth between these two limits. Since the Kantorovich fidelity is
monotonic under averaging, if it were also monotonic under
measurements, it would have to remain invariant under these
operations since they are reversible. By the same token, any measure
of distinguishability between ensembles, which is monotonic both
under measurements and averaging of the ensembles, would have to
have the same values in the two classical limits. As we saw for the
case of the Kantorovich distance, however, the latter property by
itself is not a guarantee for monotonicity.

\section{Proof of Theorem 2}

\label{app:B}

From the proof of Property 7 in Sec.~IV.B, it can be seen that if the
distance (fidelity) between states is jointly convex (concave), the
corresponding Kantorovich measure would be monotonic under averaging
of the ensembles. The necessity of the conditions in Theorem 2
follows from the observation that if we apply a measurement on the
second subsystem in the basis $\{|i\rangle\}$, we obtain the
ensembles $\{(p_i,\rho_i\otimes |i\rangle\langle i|)\}$ and
$\{(q_i,\sigma_i\otimes |i\rangle\langle i|)\}$, and if we follow
the measurement by an averaging of the ensembles, we obtain the
original states. If the Kantorovich measures are monotonic both
under measurements and averaging, they must be invariant during the
process. By an argument analogous to the one following
Eq.~\eqref{DKopt}, it can be seen that a Kantorovich distance
$D^K_d$ between ensembles of the form $\{(p_i,\rho_i\otimes
|i\rangle\langle i|)\}$ and $\{(q_i,\sigma_i\otimes |i\rangle\langle
i|)\}$ is equal to $\sum_i
(\textrm{min}(p_i,q_i)d(\rho_i,\sigma_i)+\frac{1}{2}|p_i-q_i|)$.
Similarly, a Kantorovich fidelity $F^K_f$ between ensembles of the
form $\{(p_i,\rho_i\otimes |i\rangle\langle i|)\}$ and
$\{(q_i,\sigma_i\otimes |i\rangle\langle i|)\}$ is equal to $\sum_i
\textrm{min}(p_i,q_i)f(\rho_i,\sigma_i)$.

To prove the sufficiency of condition \eqref{the1}, consider two
ensembles of states $P(\rho)$ and $Q(\sigma)$. Let
$\Pi(\rho,\sigma)$ be a joint probability distribution that attains
the minimum in Eq.~\eqref{definition}, i.e.,
\begin{gather}
D_d^K(P,Q)=\sum_{\rho,\sigma\in\Omega}\Pi(\rho,\sigma)d(\rho,\sigma).
\end{gather}
According to condition \eqref{the1},
\begin{gather}
D_d^K(P,Q)=\notag\\
d(\sum_{\rho,\sigma\in\Omega}\Pi(\rho,\sigma)\rho\otimes|\rho\sigma\rangle\langle
\rho\sigma| ,
\sum_{\rho,\sigma\in\Omega}\Pi(\rho,\sigma)\sigma\otimes|\rho\sigma\rangle\langle
\rho\sigma|),
\end{gather}
where $|\rho\sigma\rangle$ is a set of orthonormal states,
$\langle\rho\sigma|\rho'\sigma'\rangle=\delta_{\rho\rho'}\delta_{\sigma\sigma'}$.
Let $\{\mathcal{M}_i\}$, $\mathcal{M}_i(\rho)=\sum_j M_{ij}\rho
M_{ij}^{\dagger}$, be a set of completely positive maps that form a
generalized measurement, $\sum_{i,j}M_{ij}^{\dagger}M_{ij}=I$.
Consider the following CPTP map:
\begin{gather}
\mathcal{M}(\rho)=\sum_i\mathcal{M}_i(\rho)\otimes |i\rangle\langle
i|,
\end{gather}
where $\{|i\rangle\}$ is an orthonormal set of states in the Hilbert
space of some additional system (this map is not
dimension-preserving). From the monotonicity of $d(\rho,\sigma)$
under CPTP maps and property \eqref{the1}, it follows that
\begin{gather}
D_d^K(P,Q)=\notag\\
d(\sum_{\rho,\sigma\in\Omega}\Pi(\rho,\sigma)\rho\otimes|\rho\sigma\rangle\langle
\rho\sigma| ,
\sum_{\rho,\sigma\in\Omega}\Pi(\rho,\sigma)\sigma\otimes|\rho\sigma\rangle\langle
\rho\sigma|)\notag\\
\geq
d(\sum_{\rho,\sigma\in\Omega}\sum_i\Pi(\rho,\sigma)\mathcal{M}_i(\rho)\otimes|\rho\sigma\rangle\langle
\rho\sigma| \otimes |i\rangle\langle i|,\notag\\
\sum_{\rho,\sigma\in\Omega}\sum_i\Pi(\rho,\sigma)\mathcal{M}_i(\sigma)\otimes|\rho\sigma\rangle\langle
\rho\sigma|\otimes |i\rangle\langle i|)\notag\\
=\sum_{\rho,\sigma\in\Omega}\sum_i
\textrm{min}\left(\Pi(\rho,\sigma)p_i(\rho),
\Pi(\rho,\sigma)p_i(\sigma) \right) d\left(\rho_i,\sigma_i
\right)+\notag\\\frac{1}{2}\sum_{\rho,\sigma\in\Omega}\sum_i|\Pi(\rho,\sigma)p_i(\rho)
- \Pi(\rho,\sigma)p_i(\sigma) |,\label{longineq}
\end{gather}
where $p_i(\rho)=\textrm{Tr}(\mathcal{M}_i(\rho))$,
$\rho_i=\mathcal{M}_i(\rho)/p_i(\rho)$. Now observe that there
exists a joint probability distribution $\breve{\Pi}(\rho_i,
\sigma_j )$ that satisfies
\begin{gather}
\breve{\Pi}(\sigma_i, \rho_i )=
\textrm{min}\left(\Pi(\rho,\sigma)p_i(\rho),
\Pi(\rho,\sigma)p_i(\sigma) \right)\label{prop1}
\end{gather}
and has marginals
\begin{gather}
\sum_{\sigma\in\Omega}\sum_j\breve{\Pi}(\rho_i, \sigma_j
)=P(\rho)p_i(\rho),\label{prop2}
\end{gather}
\begin{gather}
\sum_{\rho\in\Omega}\sum_i\breve{\Pi}(\rho_i, \sigma_j
)=Q(\sigma)p_j(\sigma).\label{prop3}
\end{gather}
This is because condition \eqref{prop1} is compatible with the
marginal conditions \eqref{prop2} and \eqref{prop3}, which follows
from an argument analogous to the one in the paragraph after
Eq.\eqref{diagonalT}. For this distribution, we can write
\begin{gather}
\sum_{\rho,\sigma\in\Omega}\sum_{i,j} \breve{\Pi}\left(\rho_i,
\sigma_j \right) d \left(\rho_i,
\sigma_j\right)=\notag\\
\sum_{\rho,\sigma\in\Omega}\sum_i
\textrm{min}\left(\Pi(\rho,\sigma)p_i(\rho),
\Pi(\rho,\sigma)p_i(\sigma) \right) d\left(\rho_i,\sigma_i
\right)+\notag\\
\sum_{\rho,\sigma\in\Omega}\sum_{i\neq j} \breve{\Pi}\left(\rho_i,
\sigma_j \right) d \left(\rho_i, \sigma_j\right).\label{longineq2}
\end{gather}
But we have that
\begin{gather}
d \left(\rho_i, \sigma_j\right)\leq 1
\end{gather}
and
\begin{gather}
\sum_{\rho,\sigma\in\Omega}\sum_{i\neq j} \breve{\Pi}\left(\rho_i,
\sigma_j\right)=\notag\\
1-\sum_{\rho,\sigma\in\Omega}\sum_i
\textrm{min}\left(\Pi(\rho,\sigma)p_i(\rho),
\Pi(\rho,\sigma)p_i(\sigma)\right),
\end{gather}
from which we obtain that the second sum on the right-hand side of
Eq.~\eqref{longineq2} satisfies
\begin{gather}
\sum_{\rho,\sigma\in\Omega}\sum_{i\neq j} \breve{\Pi}\left(\rho_i,
\sigma_j \right) d \left(\rho_i,
\sigma_j\right)\leq \notag\\
 1-\sum_{\rho,\sigma\in\Omega}\sum_i
\textrm{min}\left(\Pi(\rho,\sigma)p_i(\rho),
\Pi(\rho,\sigma)p_i(\sigma) \right)\notag\\
=\frac{1}{2}\sum_{\rho,\sigma\in\Omega}\sum_i|\Pi(\rho,\sigma)p_i(\rho)
-  \Pi(\rho,\sigma)p_i(\sigma) |.\label{longineq3}
\end{gather}
Combining Eqs.~\eqref{longineq2} and \eqref{longineq3}, we see that
the expression on the right-hand side of the last equality in
Eq.~\eqref{longineq} is greater than or equal to
\begin{gather}
\sum_{\rho,\sigma\in\Omega}\sum_{i,j} \breve{\Pi}\left(\rho_i,
\sigma_j \right) d \left(\rho_i, \sigma_j \right)\equiv
D_{d,\breve{\Pi}}(M(P),M(Q)).\label{DbreveT}
\end{gather}
But notice that the quantity \eqref{DbreveT} is greater than or
equal to $D_d^K(M(P),M(Q))$, where $M:
\mathcal{P}_{\Omega}\rightarrow \mathcal{P}_{ \Omega_{\mathbf{M}}}$
is the map on the original probability distributions induced by the
measurement $\mathbf{M}$ with measurement superoperators
$\{\mathcal{M}_i \}$. This is because $\breve{\Pi}\left(\rho_i,
\sigma_j \right)$ is a joint probability distribution with marginals
$P(\rho)p_i(\rho)$ and $Q(\sigma)p_j(\sigma)$, which are consistent
with the distributions $M(P)$ and $M(Q)$ over $\Omega_{\mathbf{M}}$,
and therefore the quantity Eq.~\eqref{DbreveT} is among those
quantities over which the minimum in the definition of
$D_d^K(M(P),M(Q))$ is taken. Therefore, we have shown that for an
arbitrary generalized measurement,
\begin{equation}
D_d^{K}(P,Q)\geq D_d^K(M(P),M(Q)).
\end{equation}
This completes the proof of the sufficiency of Eq.~\eqref{the1}. The
proof of the sufficiency of Eq.~\eqref{the2} follows in a similar
manner, and we do not present it here.

\section{Triangle inequality for the EHS distance}

\label{app:C}

Let
\begin{gather}
D^{\textrm{\tiny
EHS}}(P,Q)=\notag\\
\Delta(\sum_{\rho,\sigma\in\Omega}P(\rho,\sigma)\rho\otimes[\rho\sigma],\sum_{\rho,\sigma\in\Omega}Q(\rho,\sigma)\sigma\otimes[\rho\sigma]
)
\end{gather}
and
\begin{gather}
D^{\textrm{\tiny
EHS}}(Q,R)=\notag\\
\Delta(\sum_{\kappa,\sigma\in\Omega}Q'(\kappa,\sigma)\sigma\otimes[\kappa\sigma],
\sum_{\kappa,\sigma\in\Omega}R'(\kappa,\sigma)\kappa\otimes[\kappa\sigma]
).
\end{gather}
Here, the joint probability distributions $P(\rho,\sigma)$,
$Q(\rho,\sigma)$, $Q'(\rho,\sigma)$, $R'(\rho,\sigma)$ are such that
the maxima for $D^{\textrm{\tiny EHS}}(P,Q)$ and $D^{\textrm{\tiny
EHS}}(Q,R)$ in Eq.~\eqref{EHSdist'} are achieved. (The left
marginals of $P(\rho,\sigma)$ and $R'(\rho,\sigma)$ are $P(\rho)$
and $R(\rho)$, respectively, and the right marginals of
$Q(\rho,\sigma)$ and $Q'(\rho,\sigma)$ are equal to $Q(\sigma)$.)

Note that $Q(\rho,\sigma)$ and $Q'(\rho,\sigma)$ are generally
different, and we cannot use directly the triangle inequality of
$\Delta$ to prove Eq.~\eqref{triangleEHS}. This is why, we will
construct two CPTP maps, $\mathcal{M}$ and $\mathcal{M'}$, which map
the states
$\sum_{\rho,\sigma\in\Omega}Q(\rho,\sigma)\sigma\otimes[\rho\sigma]$
and
$\sum_{\kappa,\sigma\in\Omega}Q'(\kappa,\sigma)\sigma\otimes[\kappa\sigma]$,
respectively, to the same state, while at the same time transform
the states
$\sum_{\rho,\sigma\in\Omega}P(\rho,\sigma)\rho\otimes[\rho\sigma]$
and
$\sum_{\kappa,\sigma\in\Omega}R'(\kappa,\sigma)\kappa\otimes[\kappa\sigma]$,
respectively, to valid EHS representations of the ensembles
$P(\rho)$ and $R(\rho)$. Then using the monotonicity under CPTP maps
of $\Delta$, it will follow that
\begin{gather}
D^{\textrm{\tiny EHS}}(P,Q)+D^{\textrm{\tiny
EHS}}(Q,R)\geq\notag\\
\Delta(\mathcal{M}(\sum_{\rho,\sigma\in\Omega}P(\rho,\sigma)\rho\otimes[\rho\sigma]),\mathcal{M}(\sum_{\rho,\sigma\in\Omega}Q(\rho,\sigma)\sigma\otimes[\rho\sigma]
))+\notag\\
  \Delta(\mathcal{M'}(\sum_{\kappa,\sigma\in\Omega}Q'(\kappa,\sigma)\sigma\otimes[\kappa\sigma]),\mathcal{M'}(\sum_{\kappa,\sigma\in\Omega}R'(\kappa,\sigma)\kappa\otimes[\kappa\sigma]
))\notag\\
\geq
\Delta(\mathcal{M}(\sum_{\rho,\sigma\in\Omega}P(\rho,\sigma)\rho\otimes[\rho\sigma]),\mathcal{M'}(\sum_{\kappa,\sigma\in\Omega}R'(\kappa,\sigma)\kappa\otimes[\kappa\sigma]
))\notag\\
=\Delta(\widehat{\rho}, \widehat{\kappa})\geq D^{\textrm{\tiny
EHS}}(P,R),
\end{gather}
where $\widehat{\rho}$ and $\widehat{\kappa}$ are EHS
representations of $P(\rho)$ and $R(\rho)$. What remains to be shown
is that maps $\mathcal{M}$ and $\mathcal{M'}$ with the above
properties exist.

The maps that we propose act on the pointer space as follows:
\begin{gather}
\mathcal{M}([\rho\sigma])=T_{\sigma}(\kappa|\rho)[\kappa\rho\sigma],\label{M}
\end{gather}
\begin{gather}
\mathcal{M'}([\kappa\sigma])=T'_{\sigma}(\rho|\kappa)[\kappa\rho\sigma],\label{M'}
\end{gather}
where for every $\sigma$, $T_{\sigma}(\kappa|\rho)$ and
$T'_{\sigma}(\rho|\kappa)$ describe transition probabilities from
$\rho$ to $\kappa$ and from $\kappa$ to $\rho$, respectively, such
that
\begin{gather}
T_{\sigma}(\kappa|\rho)Q(\rho,\sigma)=T'_{\sigma}(\rho|\kappa)Q'(\kappa,\sigma)\equiv
J_{\sigma}(\kappa,\rho).
\end{gather}
The fact that such transition probabilities exist follows from the
fact that for every $\sigma$,
$\sum_{\rho}Q(\rho,\sigma)=\sum_{\kappa}Q'(\kappa,\sigma)=Q(\sigma)$,
i.e., for every fixed $\sigma$, $Q(\rho,\sigma)$ and
$Q'(\kappa,\sigma)$ describe (unnormalized) distributions of $\rho$
and $\kappa$ that have the same weight and therefore can be mapped
one on top of each other via stochastic matrices that map $\rho$ to
$\kappa$ or $\kappa$ to $\rho$.

By construction, we have
\begin{gather}
\mathcal{M}(\sum_{\rho,\sigma\in\Omega}Q(\rho,\sigma)\sigma\otimes[\rho\sigma]
)=
\mathcal{M'}(\sum_{\kappa,\sigma\in\Omega}Q'(\kappa,\sigma)\sigma\otimes[\kappa\sigma])\notag\\
=\sum_{\kappa,\rho,\sigma}J_{\sigma}(\kappa,\rho)\sigma\otimes
[\kappa\rho\sigma].
\end{gather}
Let us now verify that $\mathcal{M}$ and $\mathcal{M'}$ applied to
$\sum_{\rho,\sigma\in\Omega}P(\rho,\sigma)\rho\otimes[\rho\sigma]$
and
$\sum_{\kappa,\sigma\in\Omega}R'(\kappa,\sigma)\kappa\otimes[\kappa\sigma]$,
respectively, give rise to valid EHS representations of $P$ and $R$.
From the definition of the maps \eqref{M} and \eqref{M'}, one
immediately obtains
\begin{gather}
\mathcal{M}(\sum_{\rho,\sigma\in\Omega}P(\rho,\sigma)\rho\otimes[\rho\sigma])=\notag\\
\sum_{\kappa,\rho,\sigma}T_{\sigma}(\kappa|\rho)P(\rho,\sigma)\rho\otimes[\kappa\rho\sigma]
\end{gather}
and
\begin{gather}
\mathcal{M'}(\sum_{\kappa,\sigma\in\Omega}R'(\kappa,\sigma)\kappa\otimes[\kappa\sigma])=\notag\\
\sum_{\kappa,\rho,\sigma}T'_{\sigma}(\rho|\kappa)R'(\kappa,\sigma)\kappa\otimes[\kappa\rho\sigma].
\end{gather}
The fact that these are EHS representations of the ensembles $P$ and
$R$ follows from two observations. The first one is that from the
pointer $[\kappa\rho\sigma]$ one can unambiguously determine the
state $\rho$ or $\kappa$ in the ensemble $P$ or $R$. The second one
is that the joint probability distributions
$T_{\sigma}(\kappa|\rho)P(\rho,\sigma)$ and
$T'_{\sigma}(\rho|\kappa)R'(\kappa,\sigma)$ have the correct
marginals,
\begin{gather}
\sum_{\kappa,\sigma}T_{\sigma}(\kappa|\rho)P(\rho,\sigma)=\notag\\
\sum_{\sigma}(\sum_{\kappa}T_{\sigma}(\kappa|\rho))P(\rho,\sigma)=\sum_{\sigma}P(\rho,\sigma)=P(\rho),
\end{gather}
\begin{gather}
\sum_{\rho,\sigma}T'_{\sigma}(\rho|\kappa)R'(\kappa,\sigma)=\notag\\
\sum_{\sigma}(\sum_{\rho}T'_{\sigma}(\rho|\kappa))R'(\kappa,\sigma)=\sum_{\sigma}R'(\kappa,\sigma)=R(\kappa).
\end{gather}
This completes the proof.

\section{Continuity of the average of a continuous function with
respect to the EHS distance}

\label{app:D}

Let $h(\rho)$ be a bounded function, which is continuous with
respect to the distance $\Delta$, i.e., for every $\delta>0$, there
exists $\varepsilon>0$, such that for all $\rho$ and $\sigma$ for
which
\begin{gather}
\Delta(\rho,\sigma)\leq\varepsilon,
\end{gather}
we have
\begin{gather}
|h(\rho)-h(\sigma)|\leq\frac{1}{2}\delta.\label{con'}
\end{gather}
Let $\overline{h}_{P}$ denote the average of the function $h(\rho)$
over the ensemble $P(\rho)$, $\rho\in\Omega$,
\begin{equation}
\overline{h}_P=\underset{\rho\in\Omega}{\sum}P(\rho)h(\rho).
\end{equation}
We will show that for every $\delta>0$, there exists
$\varepsilon'>0$, such that for all $P,Q\in\mathcal{P}_{\Omega}$ for
which
\begin{gather}
D^{\textrm{\tiny EHS}}(P,Q)\leq \varepsilon',
\end{gather}
we have
\begin{equation}
|\overline{h}_P-\overline{h}_Q|\leq \delta.\label{con2'}
\end{equation}

Assume that $D^{\textrm{\tiny EHS}}(P,Q)\leq \varepsilon'$. Let
$P(\rho,\sigma)$ and $Q(\rho,\sigma)$ be two joint distributions for
which the minimum in Eq.~\eqref{EHSdist''} is attained. We then have
\begin{gather}
D^{\textrm{\tiny
EHS}}(P,Q)=\frac{1}{2}\sum_{\rho,\sigma\in\Omega}\parallel
P(\rho,\sigma)\rho-Q(\rho,\sigma)\sigma\parallel\leq\varepsilon'.\label{edno'}
\end{gather}
Define the sets $\Omega_{>\varepsilon}$ and
$\Omega_{\leq\varepsilon}$ as the sets of all pairs of states
$(\rho,\sigma)$ for which $\Delta(\rho,\sigma)>\varepsilon$ and
$\Delta(\rho,\sigma)\leq\varepsilon$, respectively. The sum in
Eq.~\eqref{edno'} can then be split in two sums,
\begin{gather}
\frac{1}{2}\sum_{\Omega_{>\varepsilon}}\parallel
P(\rho,\sigma)\rho-Q(\rho,\sigma)\sigma\parallel
+\notag\\\frac{1}{2}\sum_{\Omega_{\leq\varepsilon}}\parallel
P(\rho,\sigma)\rho-Q(\rho,\sigma)\sigma\parallel\leq \varepsilon'.
\end{gather}
The first sum obviously can be bounded from above as
\begin{equation}
\frac{1}{2}\sum_{\Omega_{>\varepsilon}}\parallel P(\rho,\sigma)\rho
- Q(\rho,\sigma)\sigma\parallel\leq \varepsilon'.\label{firstin}
\end{equation}
Notice also that since the trace distance is monotonic under
tracing, we have
\begin{gather}
\frac{1}{2}\sum_{\rho,\sigma\in\Omega}|P(\rho,\sigma)-Q(\rho,\sigma)
|\leq\notag\\
\frac{1}{2}\sum_{\rho,\sigma\in\Omega}\parallel
P(\rho,\sigma)\rho-Q(\rho,\sigma)\sigma\parallel \leq
\varepsilon'.\label{zeroin}
\end{gather}
Therefore,
\begin{gather}
\frac{1}{2}\sum_{\Omega_{>\varepsilon}}|P(\rho,\sigma)-Q(\rho,\sigma)|\leq\varepsilon',\label{secondin}
\end{gather}
and
\begin{gather}
\frac{1}{2}\sum_{\Omega_{\leq\varepsilon}}|
P(\rho,\sigma)-Q(\rho,\sigma)|\leq \varepsilon'.
\end{gather}
On the other hand, we have
\begin{gather}
\sum_{\Omega_{>\varepsilon}}P(\rho,\sigma)\varepsilon\leq
\frac{1}{2}\sum_{\Omega_{>\varepsilon}}P(\rho,\sigma)\parallel
\rho-\sigma\parallel\leq\notag\\
\frac{1}{2}\sum_{\Omega_{>\varepsilon}}\parallel P(\rho,\sigma)\rho
-
Q(\rho,\sigma)\sigma\parallel+\frac{1}{2}\sum_{\Omega_{>\varepsilon}}|Q(\rho,\sigma)-P(\rho,\sigma)|\notag\\
\leq \varepsilon'+\varepsilon'=2\varepsilon',\label{thirdin}
\end{gather}
where the second inequality follows from the triangle inequality for
the trace distance and the third inequality follows from
Eqs.~\eqref{firstin} and \eqref{secondin}. This implies
\begin{gather}
\sum_{\Omega_{>\varepsilon}}P(\rho,\sigma)\leq\frac{2\varepsilon'}{\varepsilon}.\label{thirdin'}
\end{gather}

Let us now look at the difference between the average functions over
the two ensembles.
\begin{gather}
|\overline{h}_P-\overline{h}_Q|=|\underset{\rho\in\Omega}{\sum}P(\rho)h(\rho) -\underset{\sigma\in\Omega}{\sum}Q(\sigma)h(\sigma)|\notag\\
=|\sum_{\rho,\sigma\in \Omega}P(\rho,\sigma)h(\rho) -\sum_{\rho,\sigma\in \Omega}Q(\rho,\sigma)h(\sigma)|\notag\\
 \leq \sum_{\rho,\sigma\in
\Omega}|P(\rho,\sigma)h(\rho)-Q(\rho,\sigma)h(\sigma)|\leq\notag\\
\sum_{\rho,\sigma\in
\Omega}(P(\rho,\sigma)|h(\rho)-h(\sigma)|+|Q(\rho,\sigma)-P(\rho,\sigma)||h(\sigma)|
\notag\\
=\sum_{\Omega_{>\varepsilon}}P(\rho,\sigma)|h(\rho)-h(\sigma)|+
\sum_{\Omega_{\leq\varepsilon}}P(\rho,\sigma)|h(\rho)-h(\sigma)|+\notag\\
\sum_{\rho,\sigma\in
\Omega}|Q(\rho,\sigma)-P(\rho,\sigma)||h(\sigma)| .  \label{conti'}
\end{gather}
Since $h(\rho)$ is bounded, there exists a constant $h_{\mathrm{max}}>0$ such
that $|h(\rho)-h(\sigma)|\leq h_{\mathrm{max}}$ and $|h(\rho)|\leq h_{\mathrm{max}}$
for all $\rho$ and $\sigma$. Using this fact, together with
Eqs.~\eqref{thirdin'} and \eqref{zeroin} and the assumption that for
all $(\rho,\sigma)\in\Omega_{\leq\varepsilon}$,
$|h(\rho)-h(\sigma)|\leq\frac{1}{2}\delta$, we can upper bound the
last line in Eq.~\eqref{conti'} as follows:
\begin{gather}
\sum_{\Omega_{>\varepsilon}}P(\rho,\sigma)|h(\rho)-h(\sigma)|+
\sum_{\Omega_{\leq\varepsilon}}P(\rho,\sigma)|h(\rho)-h(\sigma)|+\notag\\
\sum_{\rho,\sigma\in
\Omega}|Q(\rho,\sigma)-P(\rho,\sigma)||h(\sigma)|\leq\notag\\
\frac{2\varepsilon'}{\varepsilon}h_{\mathrm{max}}+\sum_{\Omega_{\leq\varepsilon}}P(\rho,\sigma)\frac{1}{2}\delta+2\varepsilon'
h_{\mathrm{max}}\leq\notag\\
\frac{2\varepsilon'}{\varepsilon}h_{\mathrm{max}}+\frac{1}{2}\delta+2\varepsilon'
h_{\mathrm{max}}.
\end{gather}
Therefore, we see that by choosing
\begin{equation}
\varepsilon'\leq\frac{\delta\varepsilon}{4h_{\mathrm{max}}(1+\varepsilon)},
\end{equation}
we obtain
\begin{gather}
|\overline{h}_P-\overline{h}_Q|\leq \delta.
\end{gather}
Since $\delta$ was arbitrarily chosen, the property follows.

\section*{Acknowledgments}
The authors thank Emili Bagan, Ram\'{o}n Mu\~{n}oz-Tapia, Oriol
Romero-Isart, Igor Devetak and Nathan K. Langford for helpful
discussions. This work was supported by the Spanish MICINN through
the Ram\'{o}n y Cajal program (JC), contract FIS2008-01236/FIS, and
project QOIT (CONSOLIDER2006-00019), and by the Generalitat de
Catalunya through CIRIT 2005SGR-00994.

\end{document}